\documentclass[aps,twocolumn,superscriptaddress]{revtex4-1}

\usepackage{breqn}
\usepackage{graphicx}
\usepackage{dcolumn}   
\usepackage{bm}        
\usepackage{amssymb}   
\usepackage{soul}
\usepackage{gensymb}
\usepackage{enumerate}

\usepackage[breaklinks=true,colorlinks=true,linkcolor=blue,urlcolor=blue,citecolor=blue]{hyperref}

\usepackage{float}

\begin{document}

\title{Core transport barriers induced by fast ions in global gyrokinetic GENE simulations}

\author{A.~Di Siena} 
\affiliation{The University of Texas at Austin 201 E 24th St 78712 Austin Texas USA}
\author{R.~Bilato}
\affiliation{Max Planck Institute for Plasma Physics Boltzmannstr 2 85748 Garching Germany}
\author{T.~G\"orler}
\affiliation{Max Planck Institute for Plasma Physics Boltzmannstr 2 85748 Garching Germany}
\author{E.~Poli}
\affiliation{Max Planck Institute for Plasma Physics Boltzmannstr 2 85748 Garching Germany}
\author{A.~Ba\~n\'on~Navarro}
\affiliation{Max Planck Institute for Plasma Physics Boltzmannstr 2 85748 Garching Germany}
\author{D.~Jarema}
\affiliation{Max Planck Institute for Plasma Physics Boltzmannstr 2 85748 Garching Germany}
\author{F.~Jenko} 
\affiliation{Max Planck Institute for Plasma Physics Boltzmannstr 2 85748 Garching Germany}

\begin{abstract}

A novel type of internal transport barrier (ITB) called F-ATB (fast ion-induced anomalous transport barrier) has been recently observed in state-of-the-art global gyrokinetic simulations on a properly optimized ASDEX Upgrade experiment [A. Di Siena et al. Phys. Rev. Lett. {\bf 127} 025002 (2021)]. Unlike the transport barriers previously reported in literature, the trigger mechanism for the F-ATB is a basically electrostatic wave-particle resonant interaction between supra-thermal particles - generated via ion cyclotron resonance heating (ICRH) - and ion scale plasma turbulence. This resonant effect strongly depends on the particular shape of the fast ion temperature and density profiles. Therefore, to further improve our theoretical understanding of this transport barrier, we present results exploring the parameter space and physical conditions for the F-ATB generation by performing a systematic study with global GENE simulations. Particular emphasis is given to the transport barrier width and its localization by scanning over different energetic particle temperature profiles. The latter are varied in amplitude, half-width, and radial localization of an ad-hoc Gaussian-like energetic particle logarithmic temperature gradient profile. For the reference parameters at hand, a threshold in the amplitude of the fast ion logarithmic temperature gradient is identified to trigger the transport barrier effectively.

\end{abstract}


\maketitle


\section{Introduction}

One of the primary goals of fusion research is identifying possible experimental actuators to control turbulent transport in magnetic confinement devices. This is particularly important in view of future fusion reactors to maximize the fusion triple product by increasing the on-axis plasma pressure and hence the fusion output. Promising results along these lines are related to the formation of narrow regions in the core of fusion devices with reduced turbulent transport, commonly called internal transport barriers (ITBs) \cite{Connor_NF_2004} (and reference therein). Internal transport barriers have been widely observed in magnetic confinement devices under different experimental conditions in both tokamaks and stellarators \cite{Rice_PoP_1996,Gruber_PPCF_2000,Tardini_NF_2007,Fujita_PRL_1997,Levinton_PRL_1995,Litaudon_PPCF_1999,Ernst_PoP_2004,Zhurovich_NF_2007,Gormezano_PPCF_1999,Ida_PRL_2003,Fujisawa_PRL_1999,Castejon_NF_2002,Stroth_PRL_2001}. Their formation is typically associated with nearly flat or reversed shear configurations $s = \rho/q$ $(d q / d\rho) <0$ (with $q$ safety factor and $\rho$ the radial coordinate) \cite{Joffrin_NF_2002,Joffrin_PPCF_2002,Garbet_PoP_2001}, MHD activity in proximity of low-integer rational surfaces in the safety factor \cite{Joffrin_2002_q1,Joffrin_NF_2003} and fast ion effects in highly electromagnetic regimes \cite{Wong_NF_2004,Romanelli_PPCF_2010}.

While a large variety of ITBs has been reported in the literature from experimental studies, only a limited number of transport barriers has been observed with global gyrokinetic codes up-to-date, e.g., Refs.~\cite{Dif-Pradalier_PRL_2015,Strugarek_PPCF_2013} for reduced adiabatic electron setups. Recently, a new type of internal transport barrier has been first predicted and then observed via global gyrokinetic GENE \cite{Jenko_PoP2000,Goerler_JCP2011} simulations on a properly designed ASDEX Upgrade discharge, revealing signatures of improved plasma confinement consistently with the numerical results \cite{DiSiena_PRL_2021}. As shown by state-of-the-art GENE simulations (including kinetic electrons with realistic mass ratio, supra-thermal particles modeled with realistic distributions, electromagnetic effects and collisions), this transport barrier, called F-ATB (fast ion-induced anomalous transport barrier), leads to a full turbulence suppression within its radial domain and an increase in the neoclassical counterpart. The trigger mechanism responsible for its generation is an essentially-electrostatic wave-particle resonance interaction between ion-scale plasma turbulence and supra-thermal particles generated via ion cyclotron resonance heating (ICRH), recently identified in both tokamak and stellarator devices \cite{DiSiena_NF2018,DiSiena_PoP2019,W7X}. These findings hence suggest an intriguing possibility to access fusion relevant physical conditions more easily. 

According to analytic theory and gyrokinetic results, ion-temperature-gradient (ITG) micro-instabilities can interact with certain plasma species whenever their magnetic drift frequency $(\omega_{df})$ is comparable with the ITG frequencies $(\omega_{ITG})$ for some relevant wave-numbers. Whilst this condition is hardly matched by thermal species (typically having $\omega_{ITG} \gg \omega_{df}$), it is well fulfilled by mildly supra-thermal particles, commonly generated via auxiliary heating schemes in nowadays experiments, e.g.~ neutral-beam-injection (NBI) and ICRH. As discussed in detail in Ref.~\cite{DiSiena_NF2018}, an additional constraint on the fast particle logarithmic gradients has to be fulfilled to achieve an effective ITG stabilization. More specifically, its logarithmic temperature gradient has to largely overcome the corresponding density gradient to allow an effective energy transfer from the ITGs to the fast particles, thus making this wave-particle resonant effect accessible only for ICRH schemes. Further details can be found in Refs.~\cite{DiSiena_NF2018,DiSiena_PoP2019}. Signatures of this beneficial fast ion effect on turbulent transport have been observed at ASDEX Upgrade \cite{DiSiena_PRL_2021}, JET \cite{Bonanomi_2018} and are also expected for W7-X \cite{W7X} and during the ramp-up phase of an ITER standard scenario \cite{DiSiena_NF2018}.

Given the high relevance of this wave-particle resonant interaction in (possibly) improving plasma confinement by triggering this novel type of transport barriers, this contribution explores the parameter space and physical conditions for the F-ATB formation by performing a systematic study with global GENE simulations. Each of the GENE simulations required approximately 0.8 million CPU·h on the Marconi Skylake partition. Particular attention is given to analyze the transport barrier width and its localization by scanning over different energetic particle temperature profiles. The physics inputs are derived from the reference ASDEX Upgrade discharge $\#36637$ at $t = 4.1$s, which showed an improvement in the plasma performances possibly linked to the presence of the F-ATB. 

It is important to mention here, that radially-global gyrokinetic simulations are essential to fully capture the transport barrier properties, such as global $E \times B_0$ flows, their effect on turbulent transport and the dynamics of turbulent avalanches in the proximity of the transport barrier. Whilst flux-tube linear and nonlinear simulations can locate with good approximation the center of the F-ATB, they fail in fully capturing the transport barrier width and the turbulent levels at the radial boundaries of the transport barrier \cite{DiSiena_PRL_2021}.

The remainder of this paper is organized as follows. Section~\ref{sec1} briefly presents the code GENE and the numerical setup used for carrying out the simulations discussed throughout this paper. The physical parameters, plasma profiles and geometry - taken from the ASDEX Upgrade discharge $\#36637$ at $t = 4.1$s - are discussed in Section~\ref{sec2}, together with the analytic Gaussian-like approximation used for modeling the energetic particle logarithmic temperature gradient profile. The latter defines different fast particle temperatures that vary one to the other in the amplitude, half-width, and radial localization of the Gaussian-like energetic particle logarithmic temperature gradient profile. The impact of each of these different parameters on the F-ATB properties is studied in details in Sec.~\ref{sec3}-\ref{sec6}. Conclusions are drawn in Sec.~\ref{sec7}, while convergence studies are provided in the appendix~\ref{appendix:a}.

\section{Numerical model and setup} \label{sec1}

\subsection{Model description}

The numerical simulations presented in this paper have been performed with the Eulerian gyrokinetic code GENE \cite{Jenko_PoP2000,Goerler_JCP2011}. GENE solves the coupled Vlasov-Maxwell system of equations on the field-aligned coordinate grid $\left(x,y,z\right)$ in configuration space and $\left(v_\shortparallel, \mu\right)$ in velocity. Here, $x = \rho_{tor}a = \sqrt{\Phi_{tor}/\pi B_0}$ denotes the radial coordinate, with $\Phi_{tor}$ toroidal flux, $a$ the minor radius and $B_0$ the on-axis magnetic field; $y = x_0\left(q\chi -\varphi \right)/q(x_0)$ the bi-normal direction with $x_0$ the center of the radial domain, $\chi$ the straight-field-line poloidal angle, $\varphi$ the toroidal angle and $q$ the safety factor; and $z$ the distance along a field line. Moreover, $v_\shortparallel$ represents the velocity component parallel to the background magnetic field $B_0$ and $\mu$ the magnetic moment. While fixed equidistant grids are used for the radial (x), parallel field (z) and $v_\shortparallel$ directions, the magnetic moment ($\mu$) is discretized with Gauss-Laguerre integration points. The bi-normal direction ($y$) is represented in Fourier space. The underlying gyrokinetic equations are solved in GENE for the perturbed part of distribution function $f_1$ assuming a stationary background distribution $f_0$ (so called $\delta$f approach). Whilst different choices of $f_0$ (both analytic and numerical) are currently supported in GENE \cite{DiSiena_2016,DiSiena_PoP_2018,DiSiena_NF_2018_egam}, an equivalent Maxwellian distribution function is employed throughout this paper for simplicity unless stated otherwise. The level of approximation in modeling supra-thermal particle distributions with a Maxwellian background instead of more realistic distribution functions will be discussed in Section \ref{sec3}. 
Kinetic electrons with realistic ion-to-electron mass ratio, collisions modeled with a linearized Landau operator with energy and momentum conserving terms \cite{Crandall_CPC_2020} and a realistic magnetic geometry are retained for the present study. To keep the numerical cost of the simulations at a reasonable level, electromagnetic fluctuations are neglected unless stated otherwise. This approximation is not expected to alter qualitatively the global structure of the F-ATB \cite{DiSiena_PRL_2021}.

In radially global simulations, GENE allows the use of block-structured grids, e.g., radially dependent velocity grids that are properly optimized (based on the temperature profile of a selected plasma species) to reduce the resolution requirements in the velocity space without losing in accuracy \cite{Jarema_CPC_2016}. This specific feature is of critical importance for the numerical simulations presented in this paper, otherwise requiring a prohibitive velocity space resolution to capture the extreme changes in the energetic particle temperature profiles. Although GENE can be run either in a single flux-tube (radially-local) or a broad radial domain (radially-global), only the latter version of the code is used for the nonlinear simulations in the present work. This is particularly important to retain global effects correctly, such as turbulence avalanches, turbulence spreading and radially-global zonal flow structures, which are important in the present study to describe the global features of the F-ATB.

There are two distinct ways of running GENE in radially global simulations, the so-called flux-driven and gradient-driven setups. The flux-driven approach consists in employing particle and energy sources which mimic the external sources applied in experiments and allow the pressure profile to freely adjust until reaching a steady-state solution. Whilst Neumann boundary conditions are employed in the inner boundary, Dirichet boundary conditions are used on the outer boundary. Dirichlet boundary conditions require radial buffer regions (typically few percent of the whole radial domain) to smoothly damp the fluctuation levels at the radial boundary. The steady-state solution is reached when the turbulent fluxes match the volume integral of the external sources in the simulated radial domain. This condition is typically achieved when running the turbulence simulations up to the confinement time-scale (several order of magnitude larger than the turbulence time-scale), making flux-driven simulations extremely demanding numerically. The gradient-driven approach, on the other hand, consists in applying a Krook-type operator to maintain the averaged profiles of each species fixed to the initial ones. Dirichlet boundary conditions are applied for both inner and outer radial boundaries. The main advantage of running in gradient-driven mode is that the turbulence simulation does not need to be run up to the confinement time scale but only on the turbulence time scale, thus significantly reducing the numerical cost of each simulations. Further details on the different numerical schemes and implementation of flux-driven and gradient-driven setups can be found in Ref.~\cite{Goerler_PoP_2011}.

\subsection{Numerical setup and resolution}

In the following, the main parameters and resolutions employed throughout this work are summarized. The numerical simulations are performed (unless stated otherwise) in the radial domain $\rho_{tor} = [0.05-0.55]$. The grid resolution employed in the radial $(x)$, bi-normal $(y)$ and field-aligned $(z)$ directions is respectively $(nx0 \times ny0 \times nz0) = (256\times 48 \times 32$). The discretized toroidal mode number is given by $n = n_{0,min} \cdot j$ with $j$ being integer-valued in the range $j = [0,1,2, ..., ny0-1]$ and $n_{0,min} = 2$.

\begin{figure}
\begin{center}
\includegraphics[scale=0.60]{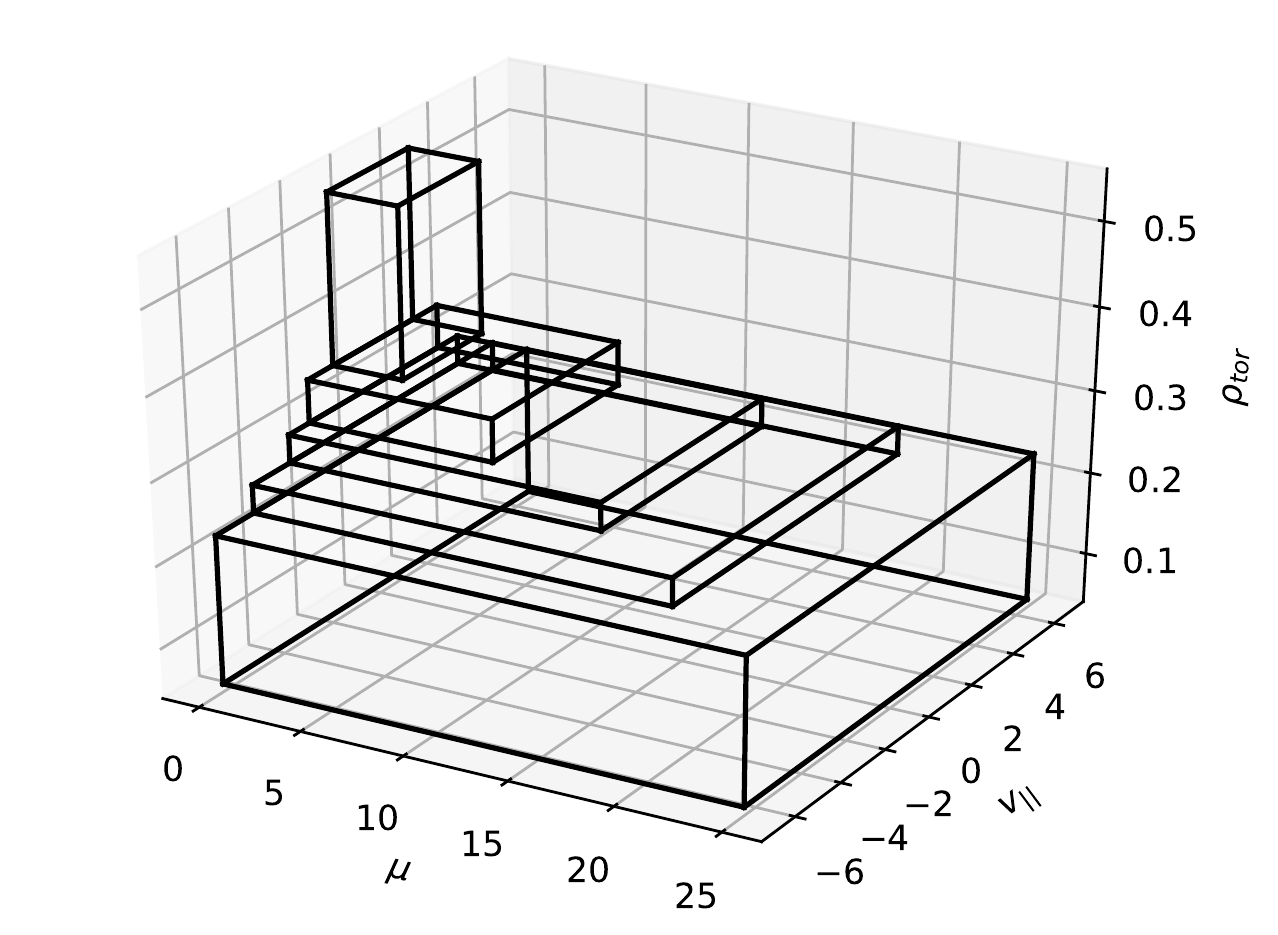}
\includegraphics[scale=0.50]{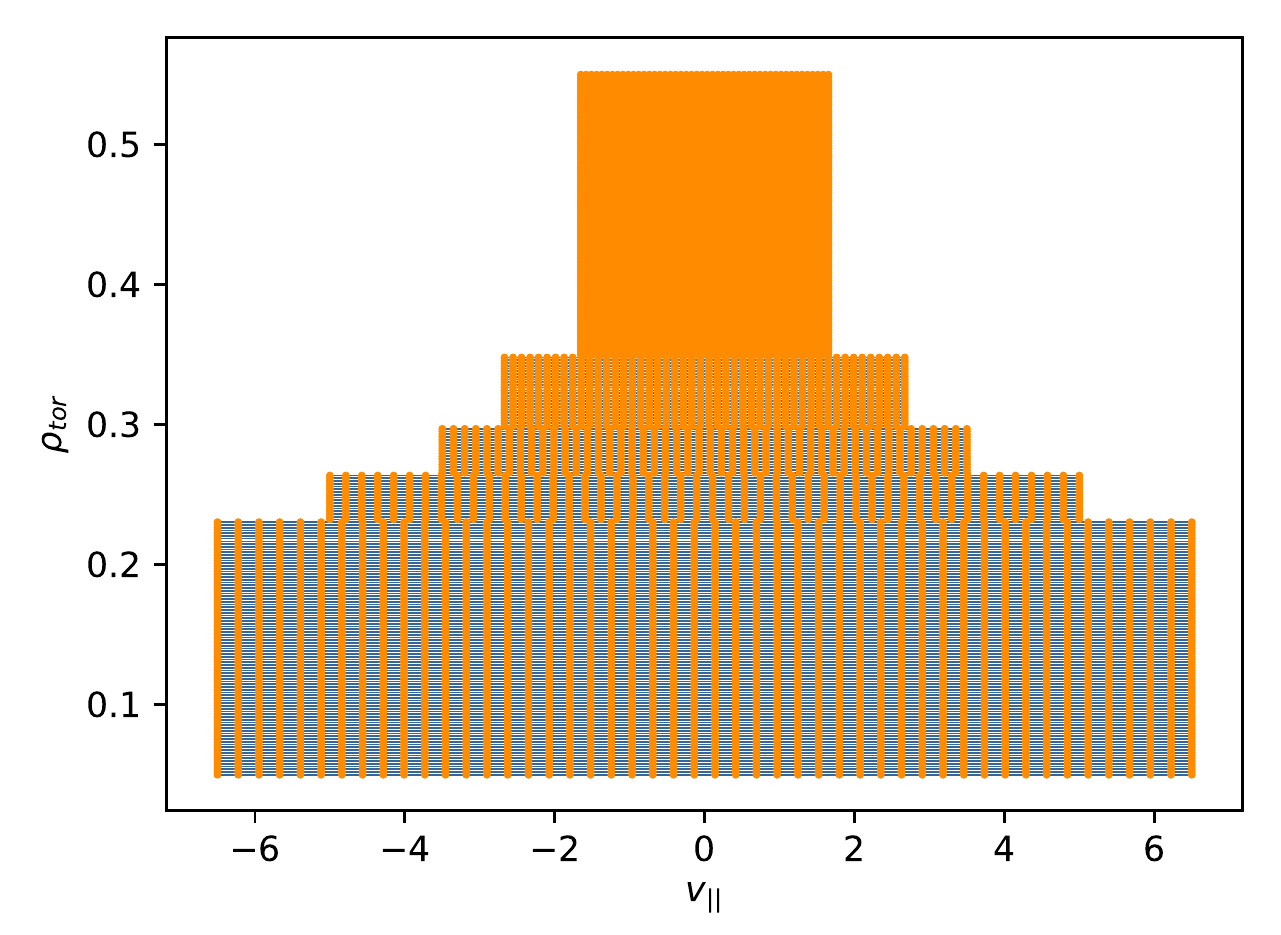}
\includegraphics[scale=0.50]{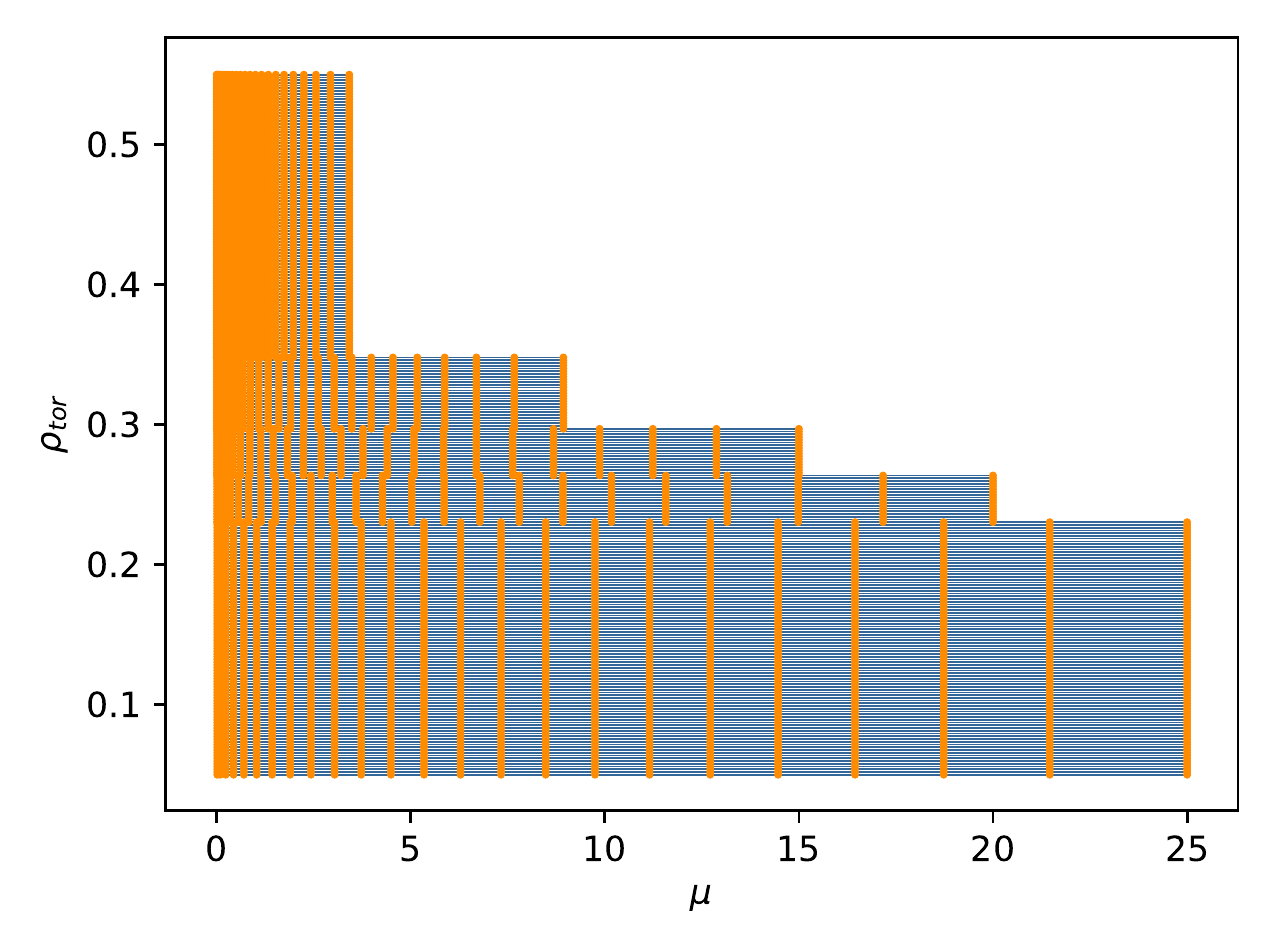}
\par\end{center}
\caption{Reference block-structured grid with five blocks in a) $(\rho_{tor}, v_\shortparallel,\mu)$ used throughout this paper (unless stated otherwise) and its slices in b) $(\rho_{tor}, v_\shortparallel)$ and c) $(\rho_{tor},\mu)$. The vertical orange lines in each blocks of b) and c) represent the $(v_\shortparallel,\mu)$ grid points used in each directions.}
\label{fig:fig1}
\end{figure}

As previously mentioned, a radially dependent block-structured grid is found to be essential to reduce the velocity resolution requirements needed to capture the sharp changes in the fast ion temperature profile. In the present paper, we employed a velocity grid divided into five different blocks. It has been properly designed to resolve correctly the $(v_\shortparallel,\mu)$-dynamics for each different plasma species with 48 points along the parallel velocity and 24 along the magnetic moment. This block-structured grid is shown in Fig.~\ref{fig:fig1}a in the $(v_\shortparallel,\mu,\rho_{tor})$-space, in Fig.~\ref{fig:fig1}b for its slice at $\mu=0$ and in Fig.~\ref{fig:fig1}c at the plane at $v_\shortparallel = 0$.
Dedicated convergence studies on the specify structure of the velocity space grid are shown in the Appendix \ref{appendix:a}. All the numerical simulations presented within this paper have been performed running GENE in gradient-driven mode. Therefore, Krook-type particle and heat operators have been applied to keep the plasma profiles (on average) fixed to the initial ones. The amplitude of the Krook heat ($\gamma_k$) and particle ($\gamma_p$) coefficients used are, respectively, $\gamma_k = 0.085 c_s / a$ and $\gamma_p = 0.85 c_s /a$. Here, $c_s = (T_e / m_i)^{1/2}$ represents the sound speed, with $T_e$ the electron temperature at the reference radial position and $m_i$ the bulk ion mass in proton units.

Buffer regions covering $10\%$ of the main simulation radial domain have been employed in this study near the domain boundaries to ensure consistency with the Dirichlet boundary conditions. In these buffer areas an artificial Krook damping operator $\gamma_b$ is applied with an amplitude $\gamma_b = 1.0 c_s/a$. Finally, to reduce the otherwise prohibitive computational cost of these simulations, numerical fourth order hyperdiffusion is used to damp fluctuations at large toroidal mode number due to electron temperature gradient (ETG) modes.

\section{Physical conditions for transport barrier formation} \label{sec2}

\begin{figure*}
\begin{center}
\includegraphics[scale=0.38]{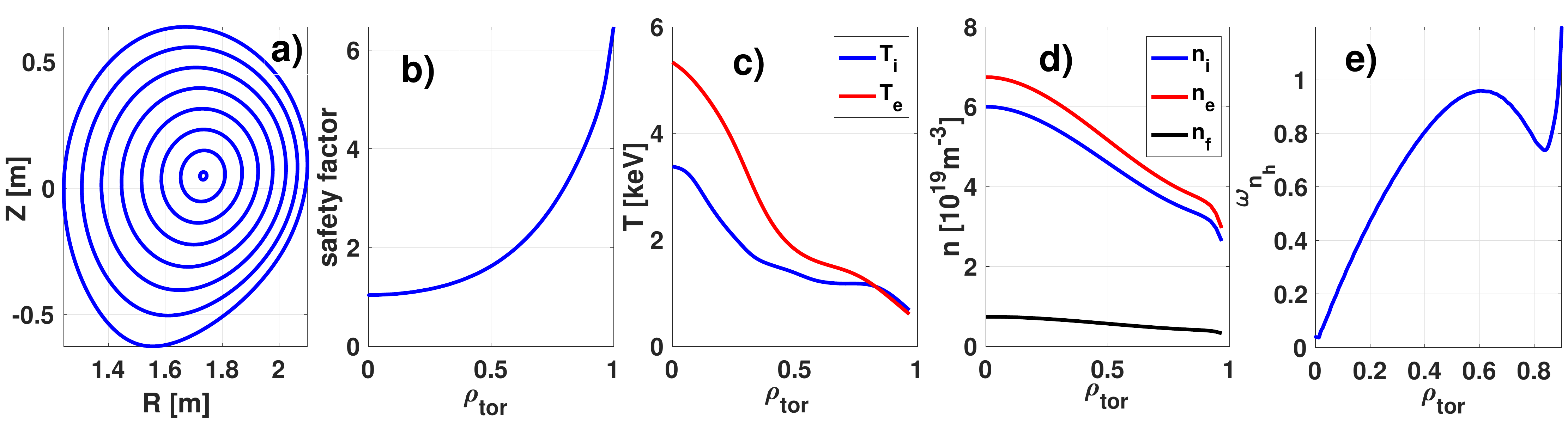}
\par\end{center}
\caption{a) Contours of constant poloidal flux of magnetic equilibrium reconstructed by CLISTE, b) safety factor profile, c) bulk ion and electron temperature profiles, d) corresponding density profiles of bulk ion, electron and hydrogen minority and e) logarithmic density gradient of the hydrogen minority in ASDEX Upgrade discharge $\#36637$ at $ t = 4.1$s used as input to GENE.}
\label{fig:fig2}
\end{figure*}

The parameter space and physical conditions possibly leading to the F-ATB generation are here explored by performing a series of (gradient-driven) global gyrokinetic GENE simulations inspired by the reference ASDEX Upgrade discharge $\#36637$ \cite{DiSiena_PRL_2021}. This is an H-mode deuterium plasma heated with constant $P = 2.5$MW electron-cyclotron-resonance-heating (ECRH) and $P = 2.5$MW neutral-beam-injected (NBI) - source 8 at $93.5$keV. Moreover, a ramp-up of the ion-cyclotron-resonance-heating (ICRH) power is performed going from $0$ to $3.5$MW. The ICRH frequency is $36.5$MHz and the antenna is operating in dipole phasing. For this study we focus our analyses on the time slice $t = 4.1$s, namely when the maximum ICRH power is applied ($P = 3.5$MW). This phase corresponds to the physical conditions where the ion temperature profile reaches the maximum value on-axis with no degradation of the plasma confinement (despite the increase in the auxiliary ICRH heating power). Ref.~\cite{DiSiena_PRL_2021} presented compelling numerical evidence - based on radially global GENE simulations - that, at this time-slice of the discharge, an F-ATB is triggered. The magnetic equilibrium is shown in Fig.~\ref{fig:fig2}a-b together with the safety factor profile. It has been reconstructed by CLISTE \cite{Carthy_PoP_1999} and read into GENE via numerical field line tracing provided by the TRACER-EFIT interface \cite{Xanthopoulos_PoP_2009}. The thermal ion and electron temperature profiles are illustrated in Fig.~\ref{fig:fig2}c while the thermal ion, electron and ICRH-hydrogen minority density profiles in Fig.~\ref{fig:fig2}d. The logarithmic density gradient ($\omega_{n,h} = -(1 / n_h)$ d$n_h /$d$ \rho_{tor}$) of the ICRH-hydrogen minority is shown in Fig.~\ref{fig:fig2}e. This quantity is particularly important in the dynamics of the resonant interaction between supra-thermal particles and ITGs, since an effective turbulence stabilization requires $\omega_{T,h} \gg \omega_{n,h}$ \cite{DiSiena_NF2018,DiSiena_PoP2019}. Here, $\omega_{T,h} = -(1 / T_h)$ d$T_h /$d$ \rho_{tor}$ represents the energetic particle (hydrogen minority) logarithmic temperature gradient. Throughout this paper $\omega_{n,h}$ is kept fixed to the nominal profile of Fig.~\ref{fig:fig2}e.

In the present paper, we investigate the F-ATB properties for different shapes of the energetic particle profiles. As demonstrated in details in Refs.~\cite{DiSiena_NF2018,DiSiena_PoP2019} via analytic theory and numerical simulations, changes in the fast ion temperatures strongly affects the wave-particle resonant dynamics via modifications on the energetic particle magnetic-drift frequency. Therefore, a combination of ten different fast particle temperature profiles has been selected for these analyses. To simplify the definition of the energetic particle temperature profiles, an analytic expression is used
\begin{equation}
T_h = T_{h,0} \cdot {\rm exp}(-\omega_{T_{h,0}}\cdot \delta_x\cdot {\rm tanh}((\rho_{tor}-\rho_0)/\delta_x)).
\label{eq:T0}
\end{equation}
Eq.~\ref{eq:T0} leads to the Gaussian-like function for the logarithmic temperature gradients of the supra-thermal particle species, 
\begin{equation}
\omega_{T,h} =  -(1/T_h) dT_h / d \rho_{tor} = \omega_{T_{h,0}} \cdot {\rm sech^2}((\rho_{tor}-\rho_0)/\delta_x).
\label{eq:omt}
\end{equation}
Here, $T_0 = 8.3 {\rm keV}$.
The three different free parameters of Eqs.~\ref{eq:T0},~\ref{eq:omt} are $\rho_0$, $\delta_x$ and $\omega_{T_{h,0}}$. They represent, respectively, the position (in $\rho_{tor}$) of the peak of the fast ions logarithmic temperature gradient, the full width at half maximum of the Gaussian-like function and its amplitude. Throughout this paper, the impact of each of these different parameters on the F-ATB proprieties is investigated in details.

It is worth mentioning here that the reference profile used in the following studies (called profile A) is constructed to approximate the realistic ICRH-hydrogen minority profile computed by TORIC/SSFPQL \cite{Brambilla_NF2009,Bilato_2011} for the reference AUG discharge $\#36637$ at $t = 4.1$s.
\begin{figure}
\begin{center}
\includegraphics[scale=0.28]{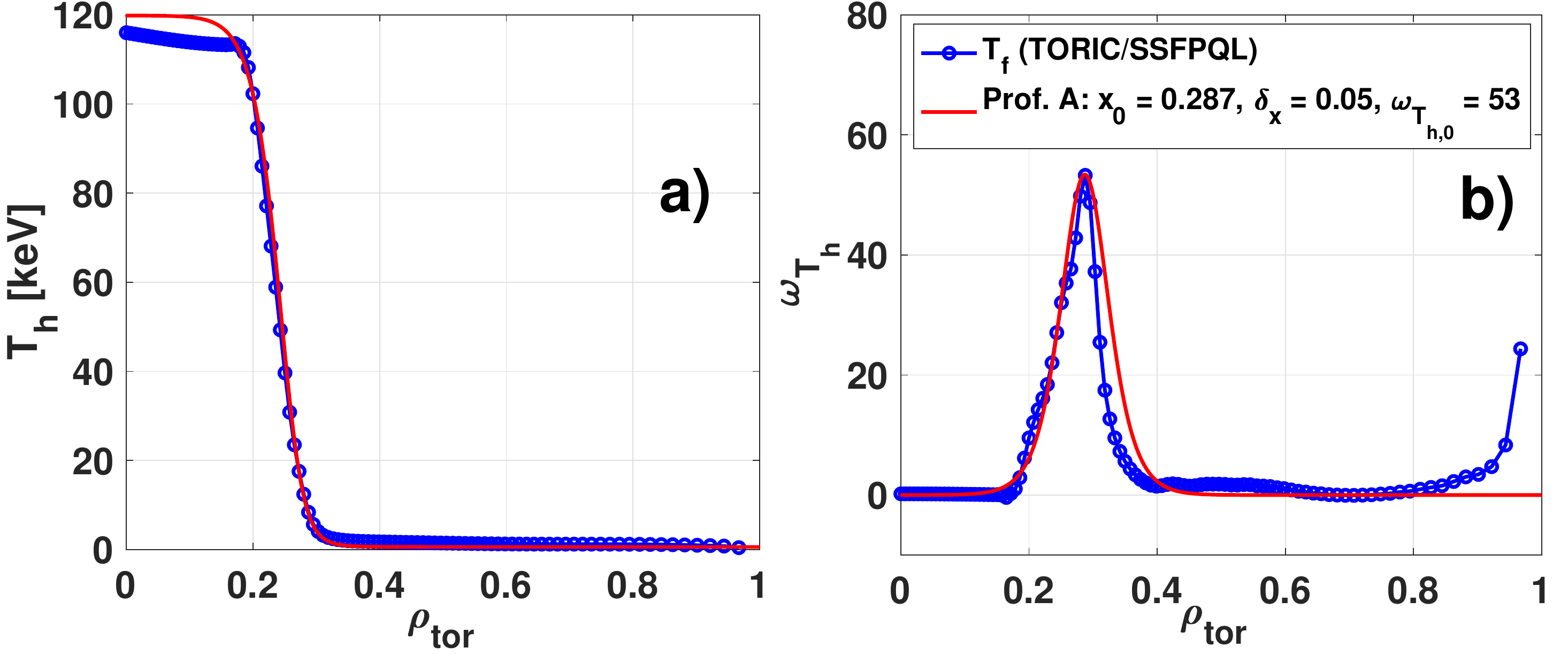}
\par\end{center}
\caption{Radial profile of the hydrogen minority a) temperature and b) logarithmic temperature gradient. The blue line in each plots denotes the temeprature profile computed by TORIC/SSFPQL for the ASDEX Upgrade discharge $\#36637$ at $t = 4.1$s, whilst the red line its simplified analytic approximation.}
\label{fig:fig4}
\end{figure}
A comparison of the realistic fast ion temperature profile and its logarithmic gradient (used in Ref.~\cite{DiSiena_PRL_2021}) with their simplified expression constructed by Eqs.~\ref{eq:T0},~\ref{eq:omt} is illustrated in Fig.~\ref{fig:fig4}.

\begin{figure}
\begin{center}
\includegraphics[scale=0.35]{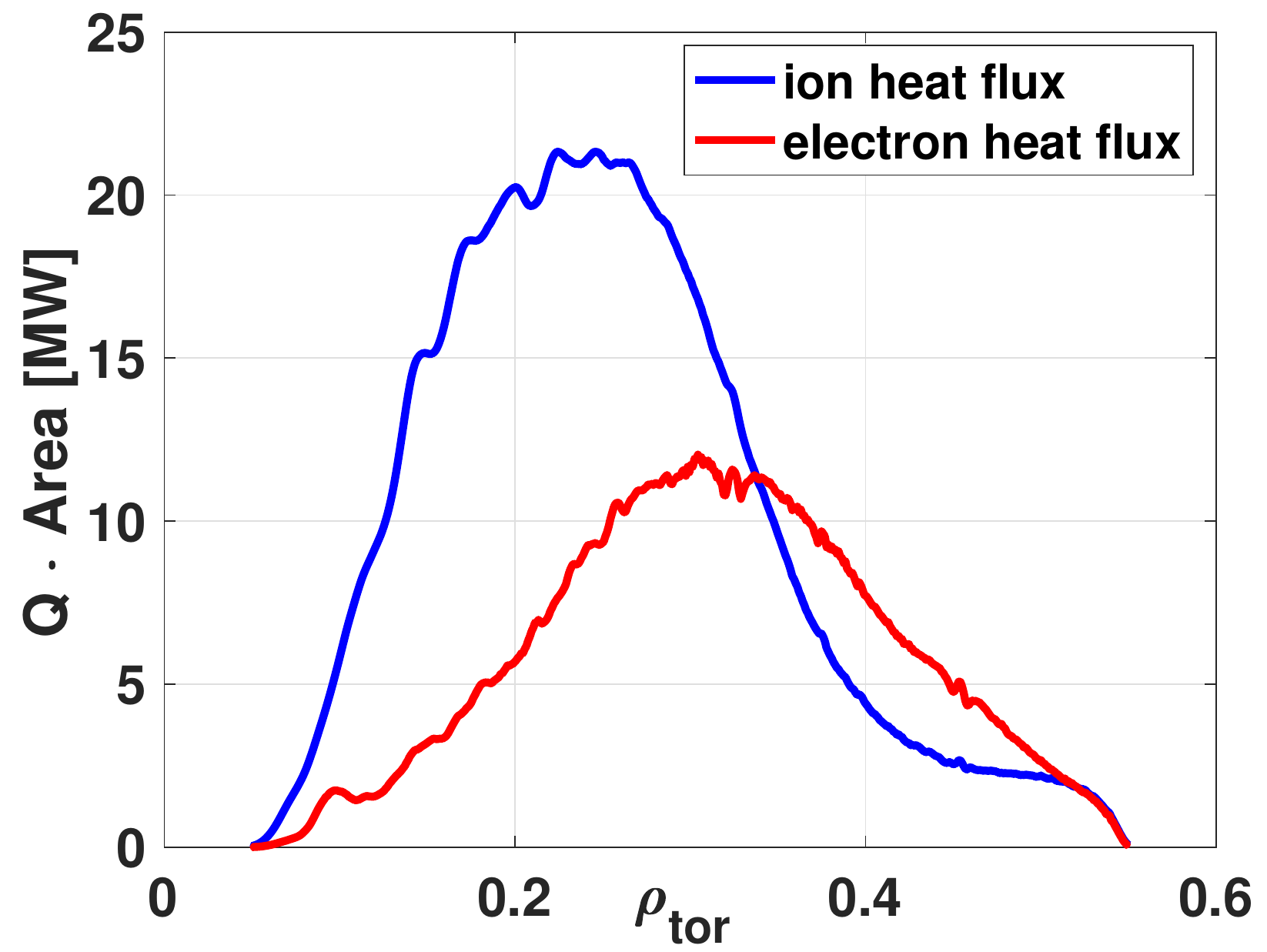}
\par\end{center}
\caption{Radial profile of the heat flux of the bulk ions and electrons obtained from GENE averaged over the time domain $t [c_s / a]= [400 - 600]$. The supra-thermal particle species are neglected.}
\label{fig:fig3}
\end{figure}

To evaluate how the specific shapes of the energetic particle temperature profiles affect the F-ATB properties, we present below a baseline scenario obtained by neglecting the fast hydrogen minority in the GENE simulations. These reference results will be employed in the following Sections to estimate the most effective fast particle temperature profiles maximizing the turbulence suppression via the wave-particle resonant interaction, namely the profiles leading to the strongest turbulence suppression in comparison with the case without fast particles. When neglecting the supra-thermal particles, the time-averaged (over the nonlinear saturated phase) heat flux profiles reach large amplitudes, as shown in Fig.~\ref{fig:fig3}. These simulations have been obtained by enforcing quasi-neutrality on the thermal ion density profile and setting it equal to the electron one. The magnetic geometry is set equal to the experimental one retaining the effect of fast ions.
The heat flux ($Q$) of the generic species s is computed as 
\begin{equation}
    Q_s = \langle {\bf Q_s} \cdot \nabla x \rangle_{S} = \langle \int \frac{1}{2}m_sv^2 \delta f_{1,s} \left({\bf v_{E \times B}} \cdot {\bf \nabla} x \right) d^3v \rangle_{S}.
    \label{eq:}
\end{equation}
Here, $m_s$ and $\delta f_{1,s}$ represent, respectively, the mass and the perturbed part of the distribution function of the species $s$ ($s = i,e,h$ stand respectively for thermal ions, electrons and fast hydrogen minority), while ${\bf v_{E \times B}}$ the ${\bf E \times B}$ velocity and $\langle \cdot \rangle_{S}$ the surface average.

\section{Impact of the amplitude of the logarithmic fast ion temperature gradient} \label{sec3}

\begin{figure}
\begin{center}
\includegraphics[scale=0.28]{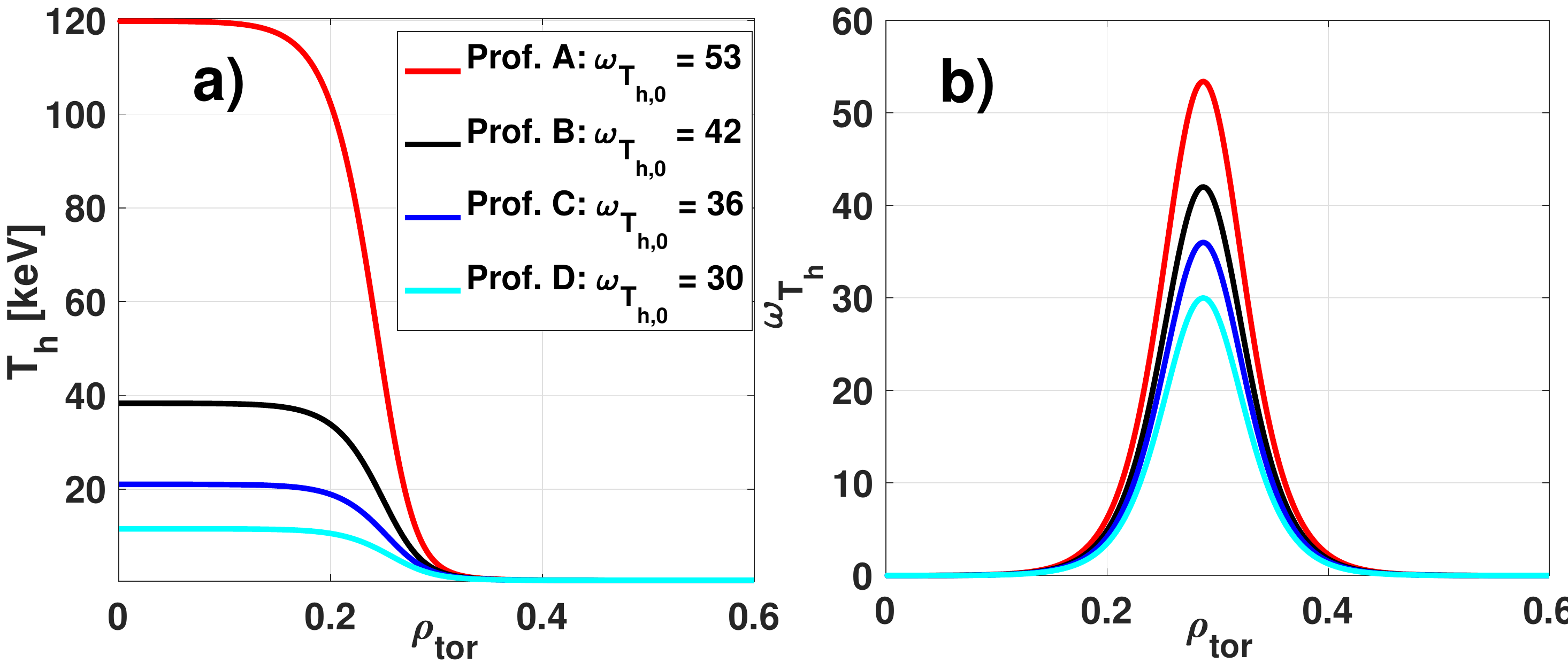}
\par\end{center}
\caption{Radial profiles of the hydrogen minority a) temperature and b) logarithmic temperature gradient corresponding to four different values of $\omega_{T_{h,0}}$.}
\label{fig:fig5}
\end{figure}

We begin by studying the effect that $\omega_{T_{h,0}}$ has on the F-ATB. This parameter defines the amplitude of the Gaussian-like logarithmic temperature gradient and thus determines the on-axis value of $T_h$. To cover a broad range of energetic particle profiles (and hence external heating powers), we have selected four different cases (labelled prof.~A to D). They are constructed from Eq.~\ref{eq:T0} by keeping $\delta_x = 0.05$, $\rho_0 = 0.287$ and varying the value of $\omega_{T_{h,0}}$ from $\omega_{T_{h,0}}=30$ to $\omega_{T_{h,0}} = 53$ (nominal value). The resulting fast particle temperature profiles and their logarithmic gradients are illustrated, respectively, in Fig.~\ref{fig:fig5}a and Fig.~\ref{fig:fig5}b.
With the numerical setup summarized in Section \ref{sec1}, we performed global GENE simulations for each of the distinct fast ion profiles shown in Fig.~\ref{fig:fig5}. The magnetic geometry and background profiles are fixed to the reference ones of Fig.~\ref{fig:fig2}.
\begin{figure*}
\begin{center}
\includegraphics[scale=0.43]{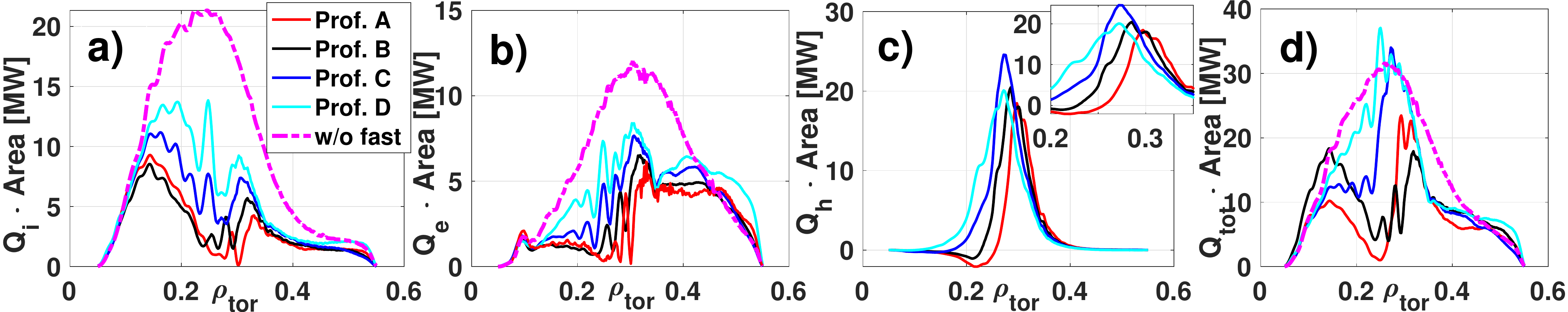}
\par\end{center}
\caption{Radial profile of the a) bulk ions, b) electrons, c) hydrogen minority species and d) overall (thermal ion + electron + fast hydrogen) heat fluxes averaged over the time domain $t [c_s / a] = [400 - 600]$ corresponding to the four different fast ion profiles of Fig.~\ref{fig:fig5}. The inlay in c) contains a zoom into the energetic particle fluxes in $\rho_{tor} = [0.2 - 0.35]$. The turbulent fluxes obtained by neglecting the energetic particles (Fig.~\ref{fig:fig3}) are shown in magenta as reference.}
\label{fig:fig6}
\end{figure*}

\begin{figure}
\begin{center}
\includegraphics[scale=0.29]{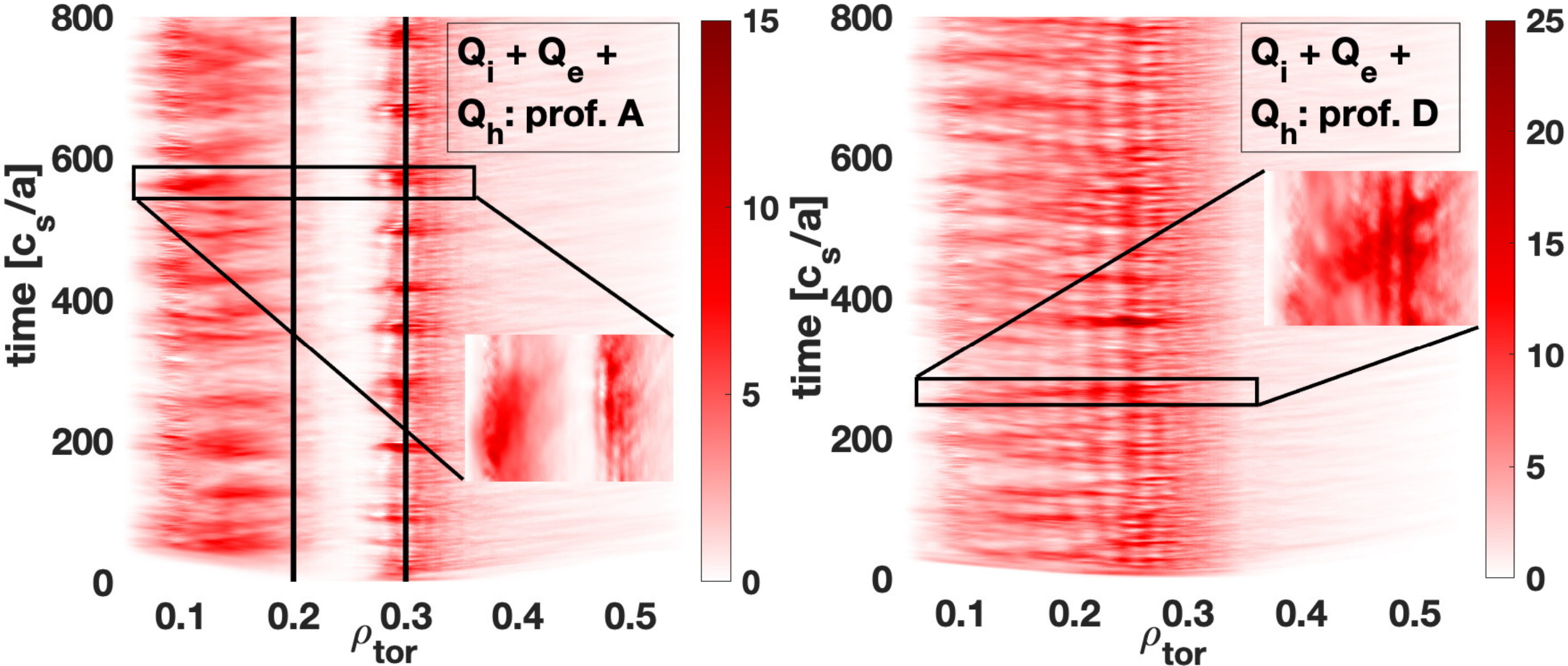}
\par\end{center}
\caption{Time evolution of the radial profile of the total heat flux (thermal ions + electrons + hydrogen) in MW corresponding to the cases labelled a) prof.~A and b) prof.~D. The vertical black lines in a) delimit the boundary positions of the transport barrier. Their radial coordinates have been computed as the position where the overall time-averaged (thermal ions + electrons + hydrogen) fluxes (see Fig.~\ref{fig:fig6}d) reach half-value before and after the transport barrier (i.e., $\rho_{tor} = [0.1 - 0.25]$ (left black line) and $\rho_{tor} = [0.3 - 0.4]$ (right black line)).}
\label{fig:fig8}
\end{figure}

The radial profiles of the time-averaged turbulent fluxes obtained from the simulations corresponding to the different energetic particle profiles are shown in Fig.~\ref{fig:fig6} for each plasma species. Furthermore, the overall (ion + electron + energetic particle) turbulent heat flux profile evolution is illustrated in Fig.~\ref{fig:fig8} for the reference fast ion temperature prof.~A and prof.~D. A first observation is that the transport barrier - localized in the radial domain $\rho_{tor} = [0.2 - 0.3]$ for the reference fast particle profile (prof.~A) - disappears as $\omega_{T_{h,0}}$ is reduced below the critical value $\omega_{T_0,c} = 42$. Without the transport barrier the overall turbulent flux profile approaches the one obtained in the absence of supra-thermal particles (see Fig.~\ref{fig:fig6}d), revealing a progressive turbulence destabilization within the radial region of the F-ATB as $\omega_{T_{h,0}}$ is reduced.

\begin{figure}
\begin{center}
\includegraphics[scale=0.29]{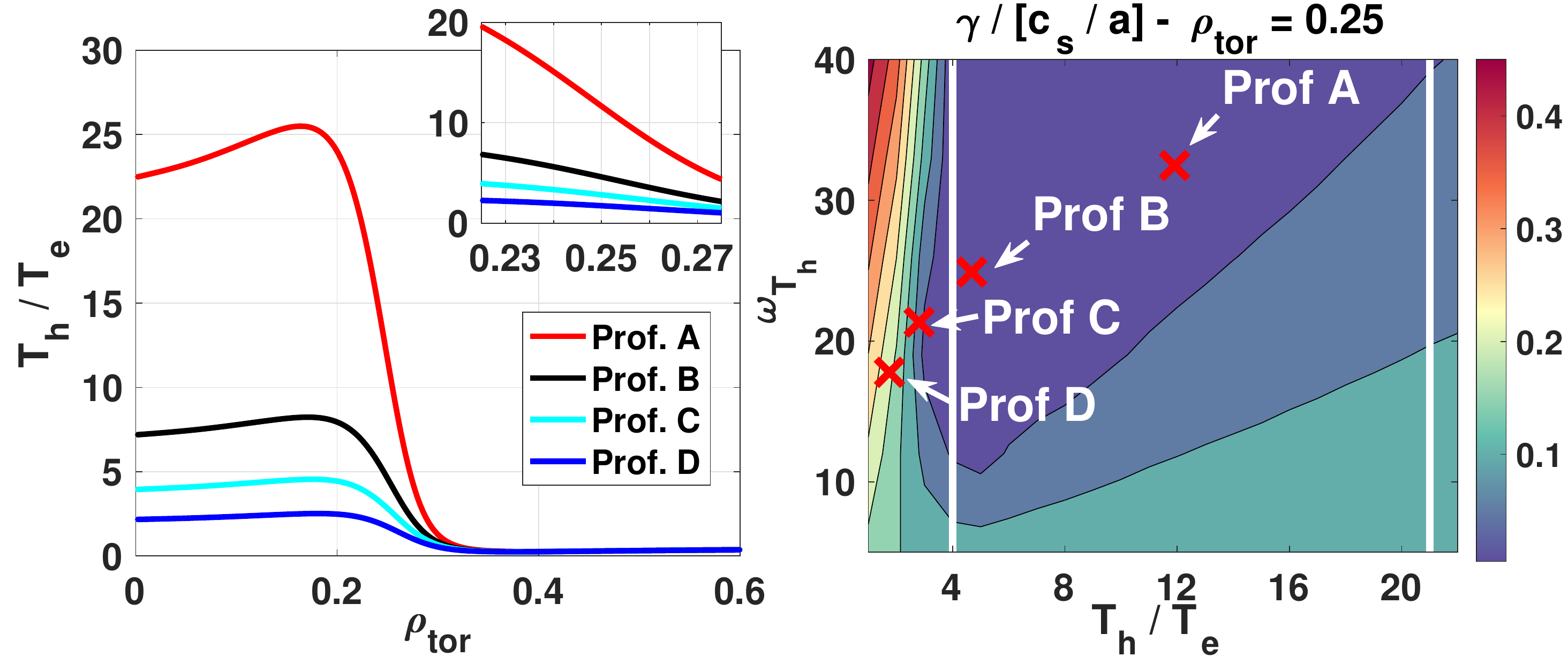}
\par\end{center}
\caption{a) Radial profile of the fast ion to electron temperature ratio $T_h / T_e$ for the four different energetic particle temperature profiles of Fig.~\ref{fig:fig5}; and b) contour plot of the most unstable linear growth rate at the toroidal mode number $n = 21$ at $\rho_{tor} = 0.25$ for different $T_h / T_e$ and $\omega_{T_{h}}$. The reference points corresponding to the values of $(T_h / T_e, \omega_{T_{h}})$ for the fast ion profiles labelled prof.~A to D are highlighted with red crosses.}
\label{fig:fig7}
\end{figure}

The energetic particle heat flux profile undergoes significant variations (see Fig.~\ref{fig:fig6}c) for $\rho_{tor} < 0.3$, where the fast ion heat flux is reduced from 20MW (prof.~D) to roughly 0MW (prof.~A) at $\rho_{tor} = 0.25$, suggesting a gradual enhancement of the beneficial fast ion effect on ITGs as $T_h / T_e$ increases with $\omega_{T_{h,0}}$. To investigate the specific role played by the simultaneous changes in the energetic particle temperature and its gradient on the underlying ITGs, linear flux-tube scans are performed at the reference position $\rho_{tor} = 0.25$ (center of the F-ATB). 
The results are illustrated in Fig.~\ref{fig:fig7}b for the toroidal-mode number $n = 21$, corresponding to the most unstable mode at the position of interest. No qualitative differences are observed if the toroidal-mode number is changed. At fixed logarithmic fast ion temperature gradients (above $\omega_{T_h} = 10$), the characteristic "sweet-spot" in the ratio $T_h / T_e$ is observed (blue region in Fig.~\ref{fig:fig7}b). It denotes the optimal fast ion temperature locating the phase-space resonance into the minimum negative drive term for the supra-thermal particles (namely $\partial_x f_0$ with $x$ the radial direction). These results are consistent with the theoretical predictions of Refs.~\cite{DiSiena_NF2018,DiSiena_PoP2019}. In particular, we note that when $T_h / T_e$ decreases, the wave-particle resonant layers move to larger ($v_\shortparallel,\mu$)-velocities resulting in a linear micro-instability de-stabilization when they are located into the positive fast ion drive regions ($v_\shortparallel^2-\mu B_0 > 3/2$). The readers are referred to Refs.~\cite{DiSiena_NF2018,DiSiena_PoP2019} for a detailed description of the specific interplay between stabilizing and de-stabilizing wave-particle resonant interactions. The energetic particle logarithmic temperature gradient effectively enhances the fast ion effects on the underlying ITGs. The operational points for the ratio $T_h / T_e$ and the fast ion logarithmic temperature gradient of the four different profiles studied in this Section are marked in Fig.~\ref{fig:fig7}b and the $T_h / T_e$ profiles are shown in Fig.~\ref{fig:fig7}a. It is clear that only prof.~A and B have fast ion temperatures exceeding the minimum value of $T_h / T_e = 4$ to effectively stabilize ITG micro-instabilities (delimited by the two white vertical lines), whilst prof.~C and D lead to a linear growth rate destabilization. These findings are consistent with the nonlinear results of Fig.~\ref{fig:fig6}.

The mild differences in the energetic particle heat flux profiles observed in Fig.~\ref{fig:fig6}c for $\rho_{tor} > 0.3$ are related to the negligible variations in the ratio $T_h / T_e$ over the four fast ions profiles employed throughout this Section.

\begin{figure*}
\begin{center}
\includegraphics[scale=0.45]{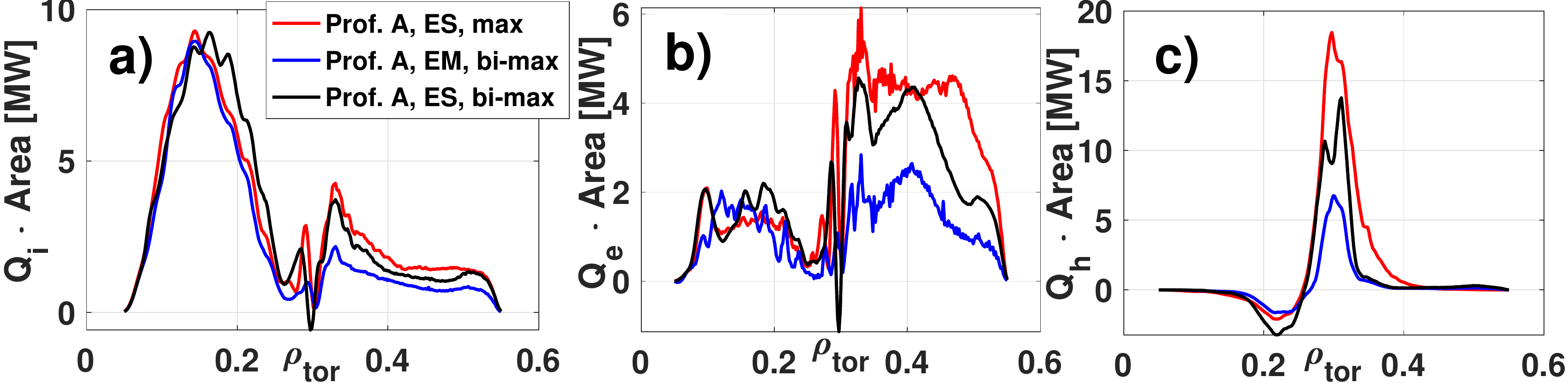}
\par\end{center}
\caption{Radial profile of the heat flux of a) bulk ions, b) electrons and c) hydrogen minority species - averaged over the time domain $t [c_s / a] = [400 - 600]$ - corresponding to the fast particle profile labelled prof.~A and different fast background distribution functions. While a Maxwellian is employed for the red line, a bi-Maxwellian background is used for the black and blue lines. Electromagnetic fluctuations are considered only for the blue line.}
\label{fig:fig9}
\end{figure*}

At the narrow radial layer where the wave-particle resonant interaction turns from stabilizing to destabilizing (i.e., $\rho_{tor} = 0.3$), a large fast ion heat flux is observed for each fast ion temperature profile. At this position, the destabilizing fast ion resonant effect on micro-turbulence is amplified by their large logarithmic temperature gradient. However, this is also the radial domain where electromagnetic fast ion effects \cite{Citrin_PRL2013,DiSiena_NF_2019,DiSiena_JPP_2021} are more pronounced possibly due to the large fast ion pressure gradient. To estimate the role of finite-$\beta_e$ on the turbulent fluxes of each plasma species, we perform, in the remaining part of this Section, global GENE simulations retaining electromagnetic fluctuations with $\beta_e (\rho_{tor} = 0.3) = 8 \pi T_e n_e / B_0^2 = 0.4\%$ for the case labeled prof.~A. The results are shown in Fig.~\ref{fig:fig9}. The electrostatic and electromagnetic GENE results of Fig.~\ref{fig:fig9} identified by the black and blue lines have been obtained by modeling the energetic particle equilibrium distribution function with a bi-Maxwellian. The parallel and perpendicular temperature profiles have been set equal to the ones computed by TORIC/SSFPQL for the experimental case summarized in Ref.~\cite{DiSiena_PRL_2021}. Consistently with Ref.~\cite{Citrin_PRL2013,DiSiena_NF_2019,DiSiena_JPP_2021}, electromagnetic effects reduce turbulent transport for each plasma species, in particular for the fast ions, which give the largest contribution at that position. This turbulence suppression is largely localized in the radial domain $\rho_{tor} = [0.3 - 0.4]$ and leads to a significant stabilization of the fast ion heat flux, which is reduced by $\sim 60\%$ (from roughly 18MW to 7MW) at $\rho_{tor} = 0.3$. No linearly unstable Alfv\'enic fast-ion driven modes are observed. These results might be a hint to a nonlinear wave-wave interplay between marginally stable fast ion modes and turbulence, proposed in Ref.~\cite{DiSiena_NF_2019,DiSiena_JPP_2021} to explain the numerical gyrokinetic simulations and experiments. A more detailed analysis assessing the role of Alfv\'en modes (stable, marginally stable or unstable) on the F-ATB is left for future studies.

To assess the relevance of correctly capturing the temperature anisotropies arising from the ICRH scheme, we add in Fig.~\ref{fig:fig9} the time-averaged (over the nonlinear saturated phase) heat flux profiles carried by the different species in the electrostatic limit by using a Maxwellian background for the supra-thermal species. By looking at the electrostatic results of Fig~\ref{fig:fig9}, only minor differences are observed for the thermal species (bulk ions and electrons). However, we note a reduction in the supra-thermal ion heat flux at $\rho_{tor} = 0.3$, which diminishes by $\sim 30\%$ (from roughly 18MW to 13MW) when the Maxwellian fast-ion distribution is replaced by a bi-Maxwellian. Interestingly, the outward fast ion flux shrinks in the radial direction. This is likely to be connected with the different parallel and perpendicular logarithmic fast particle temperature gradients, which modify the fast ion drive term, thus indirectly affecting the phase-space localization of the resonance layers. Preliminary results assessing the role of the fast ion temperature anisotropies on the wave-particle resonant interaction can be found in Refs.~\cite{Disiena_Phd}.

\section{Impact of the half-width of the logarithmic fast ion temperature gradient} \label{sec4}

\begin{figure}
\begin{center}
\includegraphics[scale=0.29]{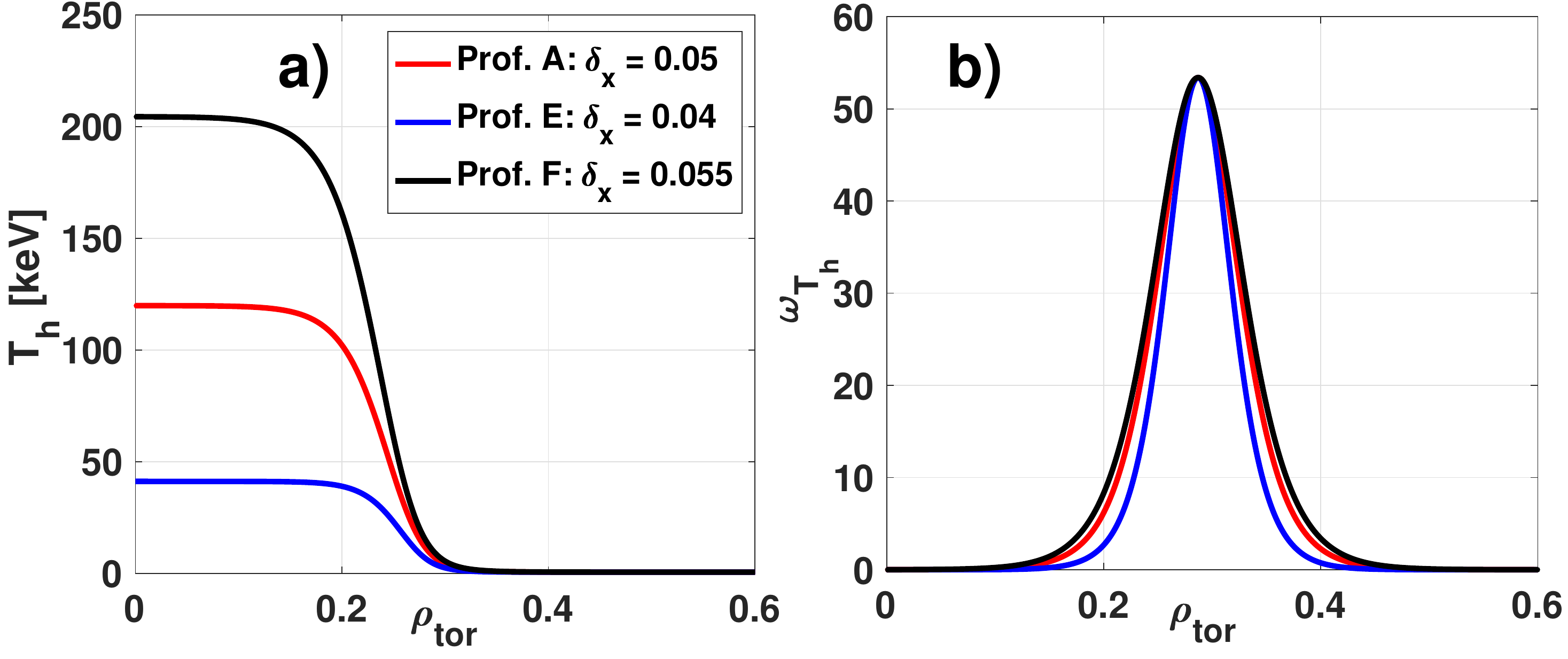}
\par\end{center}
\caption{Radial profiles of the hydrogen minority a) temperature and b) logarithmic temperature gradient corresponding to three different values of $\delta_x$.}
\label{fig:fig10}
\end{figure}

\begin{figure*}
\begin{center}
\includegraphics[scale=0.43]{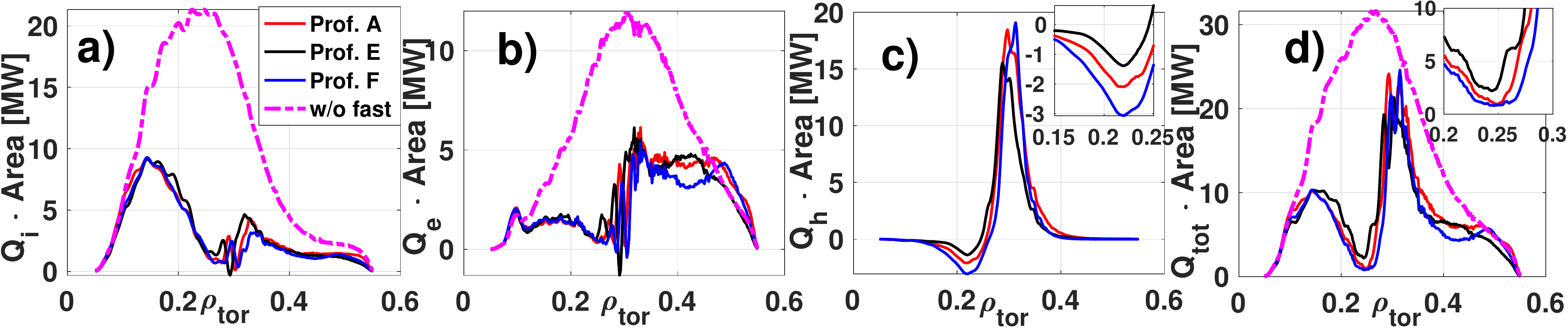}
\par\end{center}
\caption{Radial profile of the a) bulk ions, b) electrons, c) hydrogen minority species and d) overall (thermal ion + electron + fast hydrogen) heat fluxes averaged over the time domain $t [c_s / a] = [400 - 600]$ corresponding to the three different fast ion profiles of Fig.~\ref{fig:fig10}. The inlays in c) and d) contain, respectively, a zoom into the energetic particle and total fluxes in $\rho_{tor} = [0.15 - 0.25]$ and $\rho_{tor} = [0.2 - 0.3]$. The turbulent fluxes obtained by neglecting the energetic particles (Fig.~\ref{fig:fig3}) are shown in magenta as reference.}
\label{fig:fig11}
\end{figure*}

\begin{figure}
\begin{center}
\includegraphics[scale=0.40]{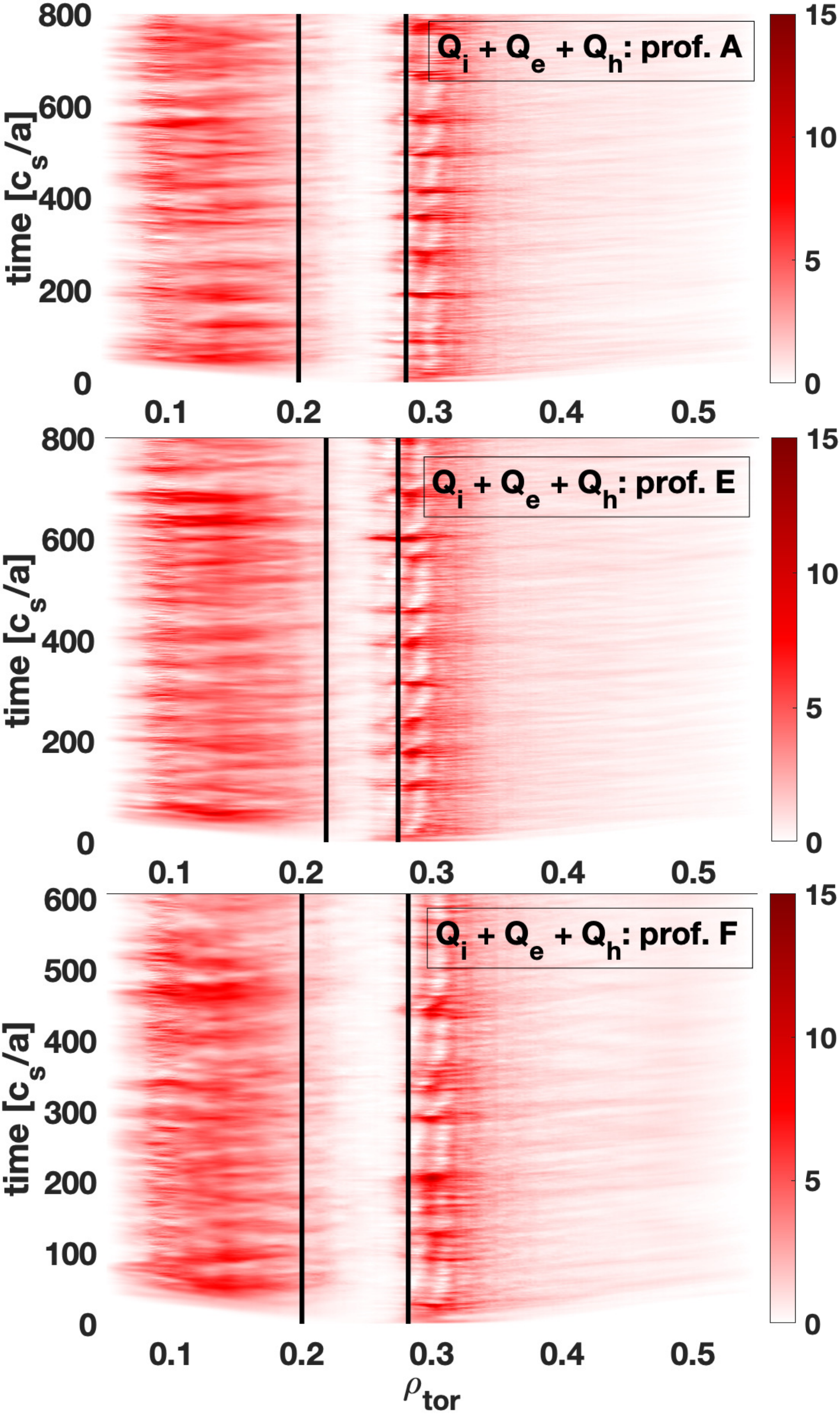}
\par\end{center}
\caption{Time evolution of the radial profile of the total heat flux (thermal ions + electrons + hydrogen) in MW corresponding to the cases labelled a) prof.~L, b) prof.~A and b) prof.~I. The vertical black lines delimit the boundary positions of the transport barrier. Their radial coordinates have been computed as the position where the overall time-averaged (thermal ions + electrons + hydrogen) fluxes (see Fig.~\ref{fig:fig11}d) reach half-value before and after the transport barrier.}
\label{fig:fig12}
\end{figure}

\begin{figure}
\begin{center}
\includegraphics[scale=0.29]{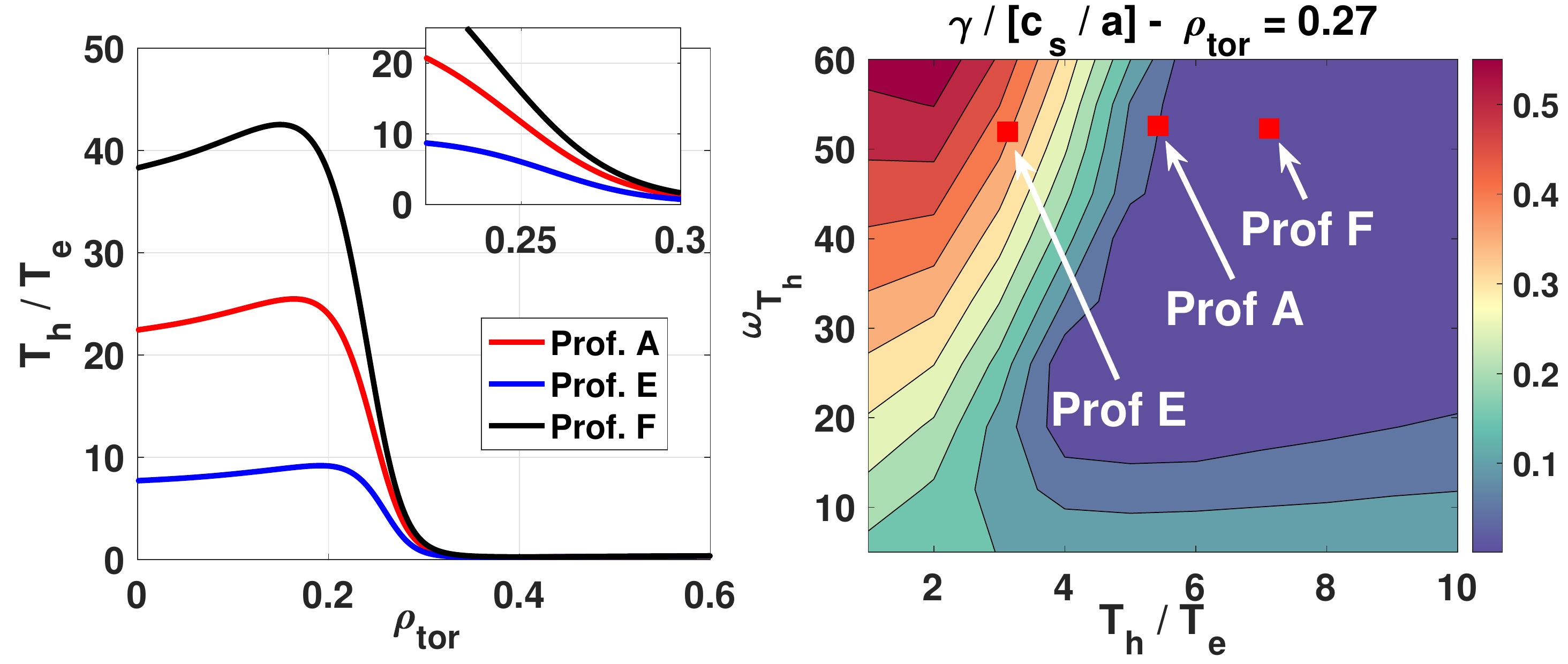}
\par\end{center}
\caption{a) Radial profile of the fast ion to electron temperature ratio $T_h / T_e$ for the three different energetic particle temperature profiles of Fig.~\ref{fig:fig10}; and b) contour plot of the most unstable linear growth rate at the toroidal mode number $n = 21$ at $\rho_{tor} = 0.27$ for different $T_h / T_e$ and $\omega_{T_h}$. The reference points corresponding to the values of $(T_h / T_e, \omega_{T_{h}})$ for the fast ion profiles labelled prof.~A, prof.~E and prof.~F are highlighted with red squares.}
\label{fig:fig13}
\end{figure}

The second parameter defining the radial profile of supra-thermal particle species is $\delta_x$. It represents the half-width of the Gaussian-like logarithmic temperature gradient and, similarly to $\omega_{T_{h,0}}$, determines also the on-axis fast ion temperature. Throughout this Section, the turbulent fluxes of the reference prof.~A ($\delta_x = 0.05$) are compared with those obtained by reducing (prof.~E, i.e. $\delta_x = 0.04$) and increasing (prof.~F, i.e., $\delta_x = 0.055$) the half-width of the fast ion temperature gradients. The energetic particles profiles are illustrated in Fig.~\ref{fig:fig10}.
Larger values of $\delta_x$ have not been explored in this work, since the resulting fast particle temperature profiles will set more demanding constraints over the radial dependent block-structured grids to fully resolve the velocity directions for each plasma species. The radial profiles of the turbulent fluxes carried by each plasma species are illustrated in Fig.~\ref{fig:fig11}. 
While no significant variations are observed in the electron heat fluxes in the whole radial region of interest, minor - but still relevant - changes are found for the bulk and supra-thermal ion heat fluxes within the transport barrier. More precisely, a reduced ion-scale turbulence stabilization is observed for the case labelled prof.~E when compared to the results obtained with the other profiles at $\rho_{tor} = 0.25$ (see Fig.~\ref{fig:fig11}a)). This result is consistent with a corresponding weakening of the wave-particle resonant interaction within the F-ATB. In particular, the region of inward fast ion heat flux - corresponding to the radial domain where the resonant interaction is most effective - is reduced in both amplitude (by roughly $66\%$) and radial extent (from $\rho_{tor} = [0.15-0.25]$ to $\rho_{tor} = [0.18-0.23]$) for the case labelled prof.~E (see Fig.~\ref{fig:fig11}c)), thus decreasing the beneficial supra-thermal ion effects on turbulent transport. This leads to more localized transport barriers, as shown in Fig.~\ref{fig:fig12}, which illustrates the time evolution of the overall radial profile of the turbulent heat fluxes. Therefore, despite the minor variations in the thermal ion and supra-thermal hydrogen fluxes, non-negligible effects on the transport barrier width are observed.
These findings are consistent with the linear flux-tube results of Fig.~\ref{fig:fig13}b showing the growth rate dependence of the toroidal-mode number $n = 21$ with the $T_h / T_e$ and the fast ion logarithmic temperature gradient at the position $\rho_{tor} = 0.27$. The corresponding $T_h / T_e$ for each profiles are illustrated in Fig.~\ref{fig:fig13}a. No qualitative differences are observed if the toroidal-mode number is changed.
Fig.~\ref{fig:fig13}b reveals that whilst the ratio $T_h / T_e$ is large enough for prof.~A and prof.~F to reach the optimal temperatures to stabilize ITG micro-instabilities, this does not happen for prof.~E, for which the resonant layers still lie in the destabilizing phase-space region of the energetic particle drive, namely at $T_h / T_e< 4$. Therefore, at $\rho_{tor} = 0.27$, the combination of fast ion temperature and temperature gradients for prof.~E leads to a linear ITG destabilization, thus reducing the radial extent of the transport barrier, consistently with Fig.~\ref{fig:fig11}d and Fig.~\ref{fig:fig12}. These results show that - for a fixed value of $\omega_{T_{h,0}}$ (amplitude of the logarithmic fast ion temperature gradient) - broader energetic particle temperature gradient profiles can trigger transport barriers more effectively. It is worth mentioning here that the supra-thermal particle temperature and its gradients are however limited by the onset of electromagnetic fast particle-driven turbulence \cite{Chen_Zonca} which is well-known to be detrimental for plasma confinement in present-day devices \cite{Citrin_PPCF2015,Biancalani_PPCF_2021}.

\section{Combined impact of the half-width and amplitude of the logarithmic fast ion temperature gradient} \label{sec5}

In the two preceding Sections, we have investigated the separate effect of variations in the (i) amplitude ($\omega_{T_{h,0}}$) and (ii) half-width ($\delta_x$) of the Gaussian-like logarithmic fast ion temperature gradient on the F-ATB formation and global properties. Consistently with the quasi-linear predictions of Ref.~\cite{DiSiena_NF2018,DiSiena_PoP2019}, we found a threshold in $\omega_{T_{h,0}}$ (or equivalently in the ratio $T_h / T_e$) to trigger the transport barrier effectively. This value corresponds to the energetic particle temperature (at fixed $T_e$) moving the resonant layers in the beneficial (negative) fast particle drive region. When this condition is fulfilled, fast particles deplete the turbulent energy content, leading to a localized transport reduction and, hence, to the F-ATB formation. For a fixed value of the amplitude of the Gaussian-like fast particle profiles, on the other hand, $\delta_x$ is found to influence the radial extent of the transport barrier. In particular, it affects the size of the radial domain reaching the "optimal" ratio of $T_h / T_e$ (i.e., $T_h / T_e \sim [5 - 25]$ see Fig.~\ref{fig:fig13}a) with large logarithmic gradients.
\begin{figure}
\begin{center}
\includegraphics[scale=0.29]{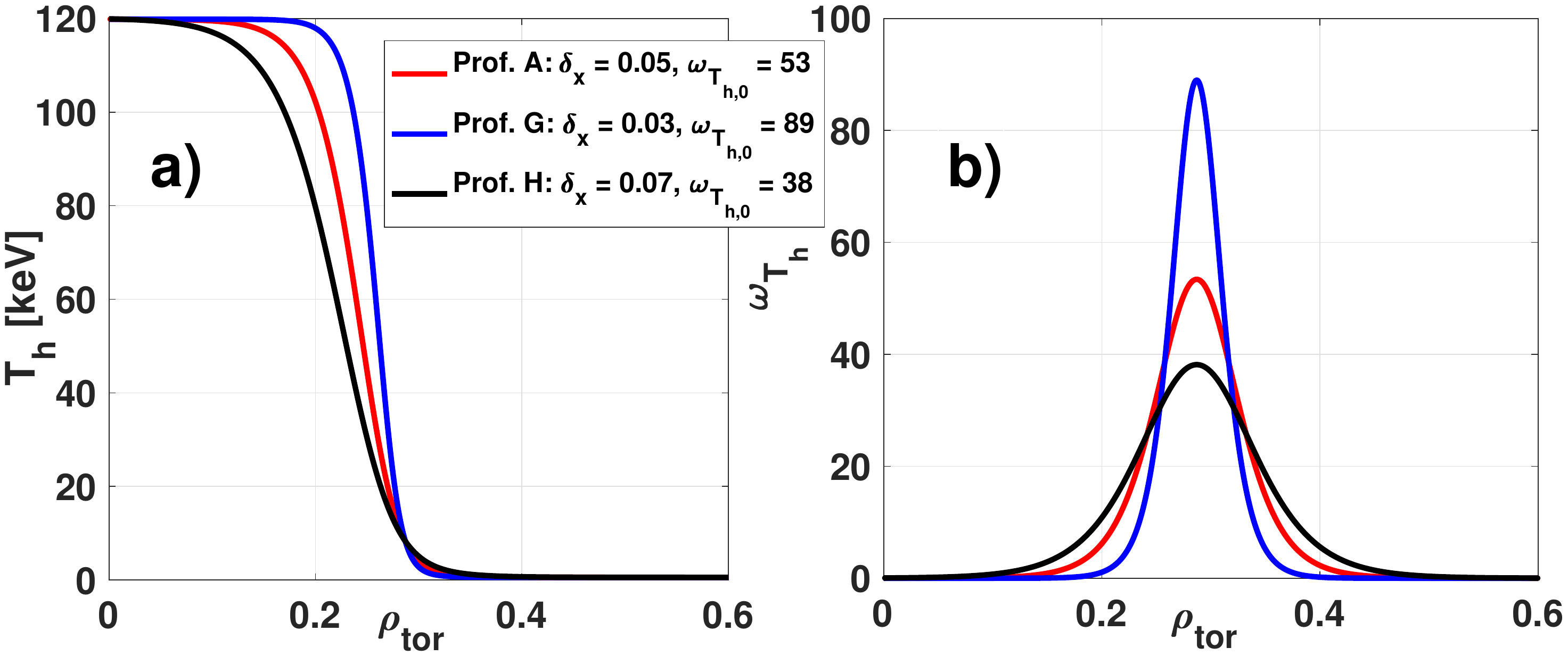}
\par\end{center}
\caption{Radial profiles of the hydrogen minority a) temperature and b) logarithmic temperature gradient corresponding to three different combinations of $\omega_{T_{h,0}}$ and $\delta_x$.}
\label{fig:fig14}
\end{figure}

In the present Section, the combined effect of $\omega_{T_{h,0}}$ and $\delta_x$ is analyzed. The energetic particle profiles selected for these studies are illustrated in Fig.~\ref{fig:fig14}. The values of $\omega_{T_{h,0}}$ and $\delta_x$ are varied simultaneously such that the same fast ion temperature on axis is obtained. By looking at the fast ion profiles depicted in Fig.~\ref{fig:fig14}, we notice that - by construction - each of them reach the optimal fast ion temperature corresponding to the "sweet-spot" observed in Fig.~\ref{fig:fig7}b. This is the value of $T_h / T_e$ maximizing the stabilizing effect of the wave-particle resonant mechanism on ITG micro-instabilities. The critical difference among the radial profiles of Fig.~\ref{fig:fig14}a, is the radial extent of the domain where this optimal range of fast ion temperatures is located. A broader, and hence, more effective ITG stabilization is expected for the supra-thermal ion profile labeled prof.~H, which is characterized by the largest $\delta_x$. Therefore, for this specific case, the beneficial wave-particle interaction is expected to be effective in the whole region $\rho_{tor} \sim [0.2-0.3]$. This domain is reduced for the case labeled prof.~G, defined by a smaller value of $\delta_x$.
\begin{figure*}
\begin{center}
\includegraphics[scale=0.43]{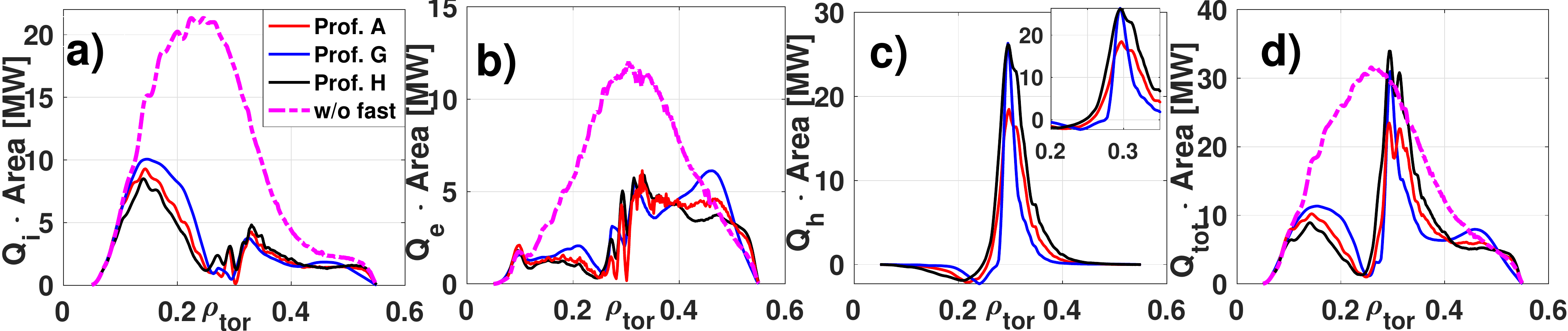}
\par\end{center}
\caption{Radial profile of the a) bulk ions, b) electrons, c) hydrogen minority species and d) overall (thermal ion + electron + fast hydrogen) heat fluxes averaged over the time domain $t [c_s / a] = [400 - 600]$ corresponding to the three different fast ion profiles of Fig.~\ref{fig:fig14}. The inlay in c) contains a zoom into the total fluxes in $\rho_{tor} = [0.2 - 0.35]$. The turbulent fluxes obtained by neglecting the energetic particles (Fig.~\ref{fig:fig3}) are shown in magenta as reference.}
\label{fig:fig15}
\end{figure*}
These predictions are confirmed by the global gyrokinetic GENE results. The time-averaged radial profiles of the turbulent fluxes are shown in Fig.~\ref{fig:fig15} for each different plasma species. The fast particle profile labeled prof.~H leads to the most effective heat flux stabilization for the bulk species (thermal ions and electrons), whilst larger fluxes are observed for prof.~G. At the radial position $\rho_{tor} = 0.2$, we notice a turbulence reduction of $\sim 50\%$ for the main ions (from $Q_i = 8.3$MW for prof.~G to $Q_i = 4.4$MW for prof.~H) and $\sim 50\%$ for the electrons (from $Q_e = 2.1$MW for prof.~G to $Q_e = 1$MW for prof.~H). The energetic particle heat flux profiles reveal a broader radial region where the fast ion heat flux is negative (inward) for prof.~H, which is consistent with our previous findings of Section \ref{sec4}. This is a signature that the wave-particle interaction is stabilizing ITG turbulence. 

It is worth mentioning that despite prof.~G leads to more localized transport barriers, the reduced half-width of its logarithmic temperature profile leads to a more narrow region where the wave-particle interaction is destabilizing ITGs (namely $\rho_{tor}> 0.3$). This is shown in Fig.~\ref{fig:fig15}c, where the radial domain of positive fast ion heat flux increases from $\rho_{tor} = [0.28 - 0.31]$ for prof.~G to $\rho_{tor} = [0.27 - 0.34]$ for prof.~H. However, as shown in Section \ref{sec3}, this outward energetic particle heat flux is strongly reduced when electromagnetic fluctuations are consistently included in the simulations.

\begin{figure}
\begin{center}
\includegraphics[scale=0.40]{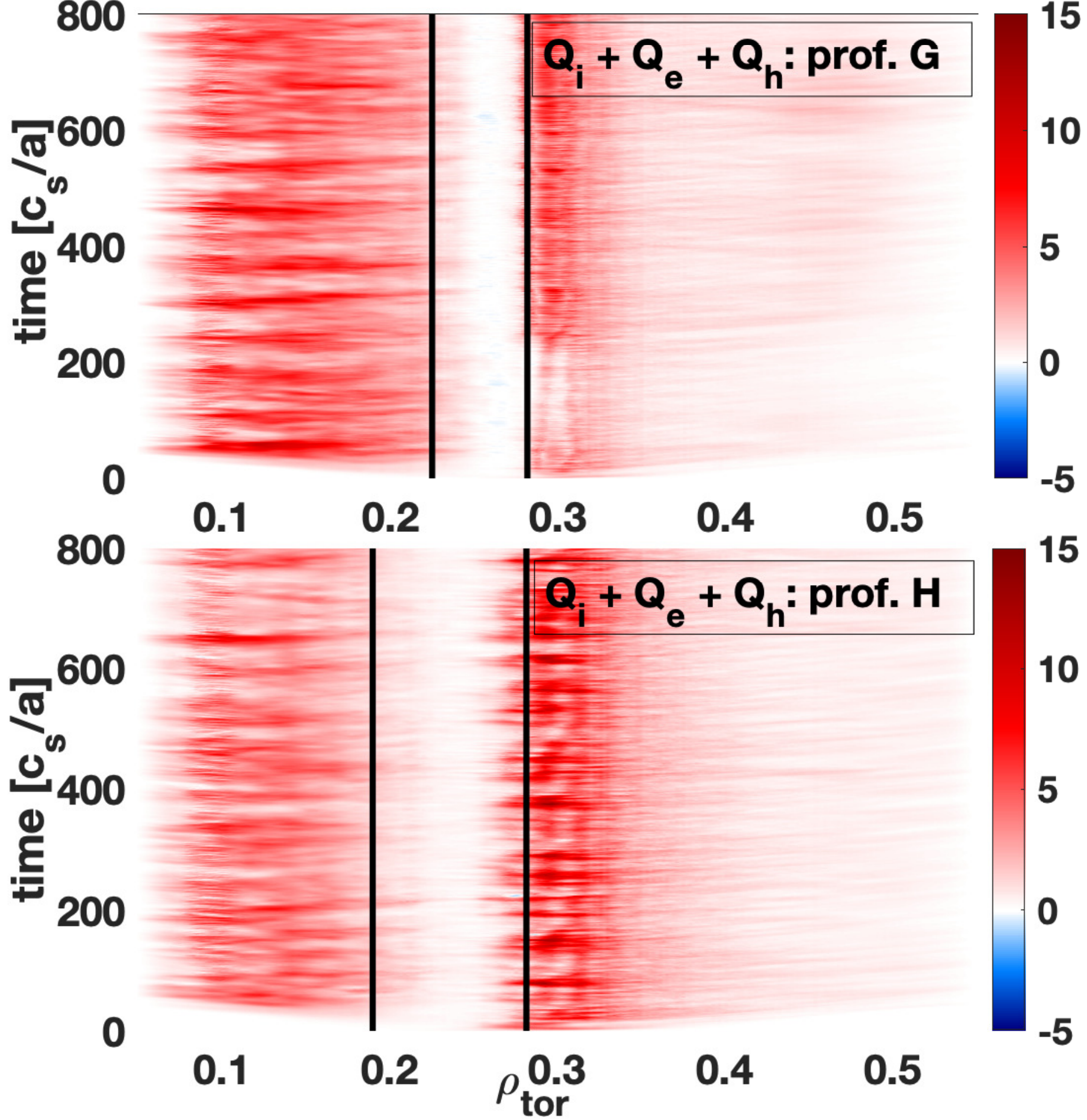}
\par\end{center}
\caption{Time evolution of the radial profile of the total heat flux (thermal ions + electrons + hydrogen) in MW corresponding to the cases labelled a) prof.~G and b) prof.~H. The vertical black lines delimit the boundary positions of the transport barrier. Their radial coordinates have been computed as the position where the overall time-averaged (thermal ions + electrons + hydrogen) fluxes (see Fig.~\ref{fig:fig15}d) reach half-value before and after the transport barrier.}
\label{fig:fig16}
\end{figure}

The time evolution of the radial profile of the overall (thermal ions + electrons + fast ions) turbulent fluxes are shown in Fig.~\ref{fig:fig16}. This figure clearly reveals that the radial extent of the transport barrier increases by roughly $65\%$ (from $\Delta_x = 0.055$ to $\Delta_x = 0.092$) when replacing the fast particle profile prof.~G with prof.~H, possibly leading to better plasma performances in experiments.
\begin{figure}
\begin{center}
\includegraphics[scale=0.29]{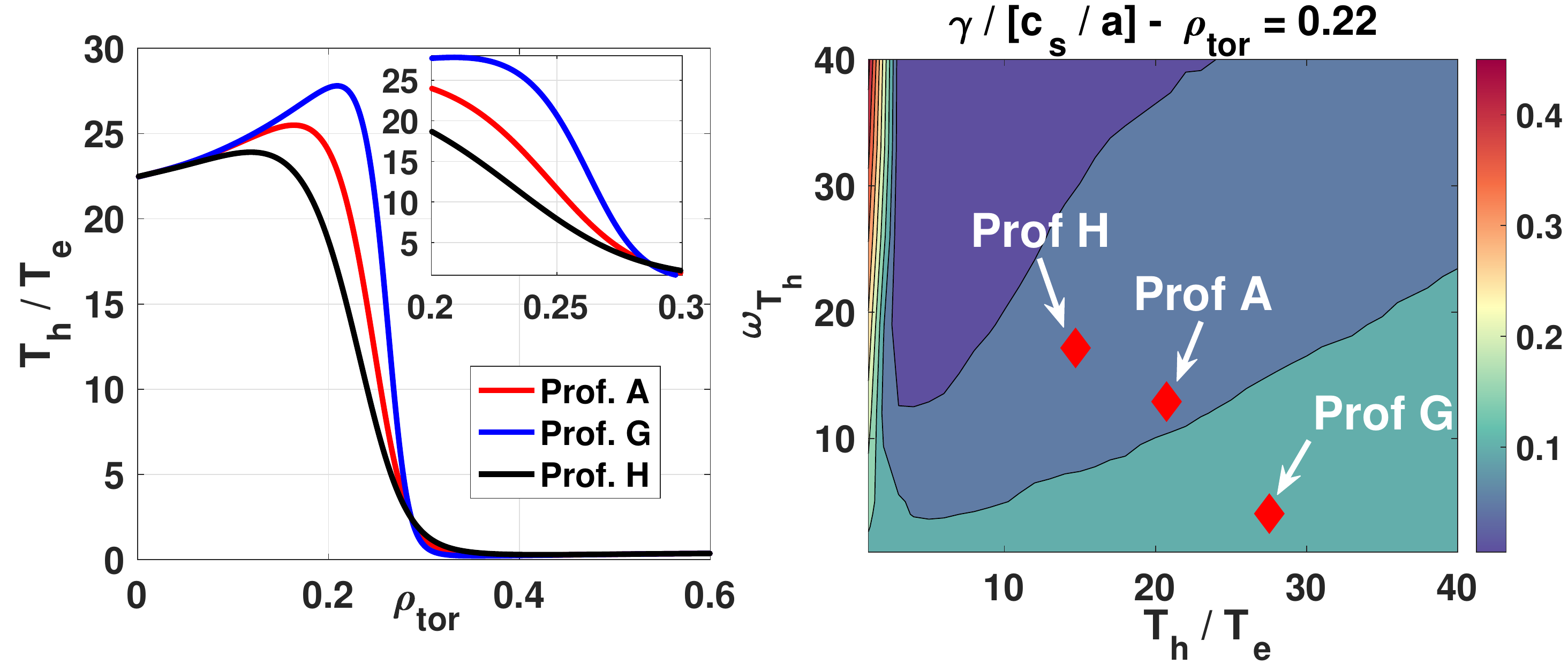}
\par\end{center}
\caption{a) Radial profile of the fast ion to electron temperature ratio $T_h / T_e$ for the three different energetic particle temperature profiles of Fig.~\ref{fig:fig14}; and b) contour plot of the most unstable linear growth rate at the toroidal mode number $n = 21$ at $\rho_{tor} = 0.22$ for different $T_h / T_e$ and $\omega_{T_{h}}$. The reference points corresponding to the values of $(T_h / T_e, \omega_{T_{h}})$ for the fast ion profiles labelled prof.~A, prof.~G and prof.~H are highlighted with red diamonds.
}
\label{fig:fig17}
\end{figure}
These results are again consistent with linear flux-tube results. They are shown in Fig.~\ref{fig:fig17}b for the toroidal mode number $n = 21$ at the radial position $\rho_{tor} = 0.22$. In particular, we notice that the strongest fast particle stabilization is observed for the case labelled prof.~H and it is progressively reduced for prof.~A and prof.~G. The latter case exhibits large values for the ratio $T_h / T_e$ and small logarithmic gradients. Therefore, the beneficial wave-particle resonant stabilization is weakened when compared to the other fast particle profiles (being it enhanced by $\omega_{T_h}$), thus leading to a reduction in the radial extend of the transport barrier. The radial profile of the ratio $T_h / T_e$ for the differences cases studied in this Section is illustrated in Fig.~\ref{fig:fig17}a.

\section{Relevance of the position of the ICRH heating} \label{sec6}

\begin{figure}
\begin{center}
\includegraphics[scale=0.29]{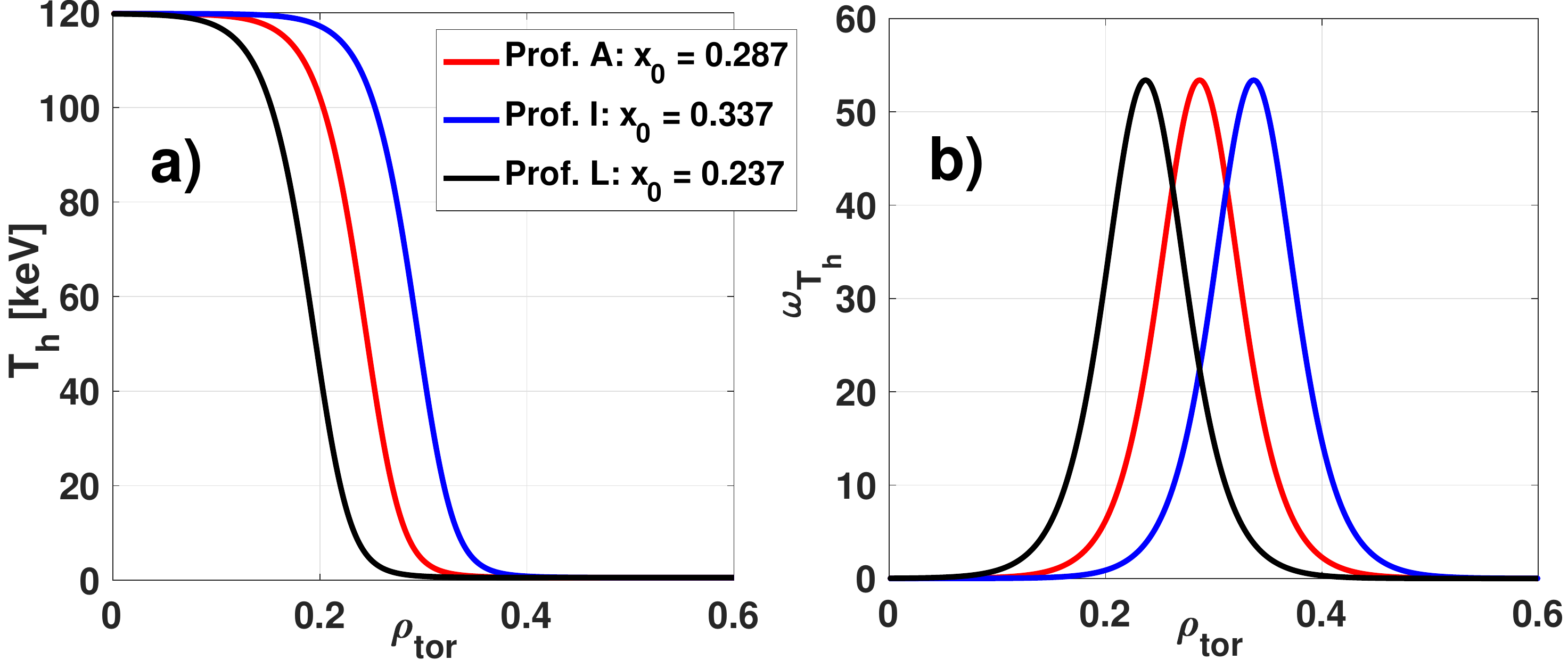}
\par\end{center}
\caption{Radial profiles of the hydrogen minority a) temperature and b) logarithmic temperature gradient corresponding to three different values of $\rho_0$.}
\label{fig:fig18}
\end{figure}

The remaining free parameter defining the energetic particle temperature and its gradient is $\rho_0$. It determines the radial location of the peak of the logarithmic fast ion temperature gradient and, hence, it affects the position of the F-ATB. For these studies, the global gyrokinetic results of the reference profile (prof.~A, i.e. $\rho_0 = 0.287$) are compared with those obtained by moving it by $\Delta \rho_0 = 0.05$ towards the magnetic axis (prof.~L, i.e. $\rho_0 = 0.237$) or the edge (prof.~I, i.e. $\rho_0 = 0.337$). These profiles are illustrated in Fig.~\ref{fig:fig18}. The amplitude and the half-width of the fast ion temperature gradients are kept constant. Although this is only a rough approximation that hardly reflects the experimental conditions, this Section aims to disentangle the impact of $\rho_0$ on the transport barrier from the other terms defining its radial shape (see Eq.~\ref{eq:T0}). 

\begin{figure*}
\begin{center}
\includegraphics[scale=0.43]{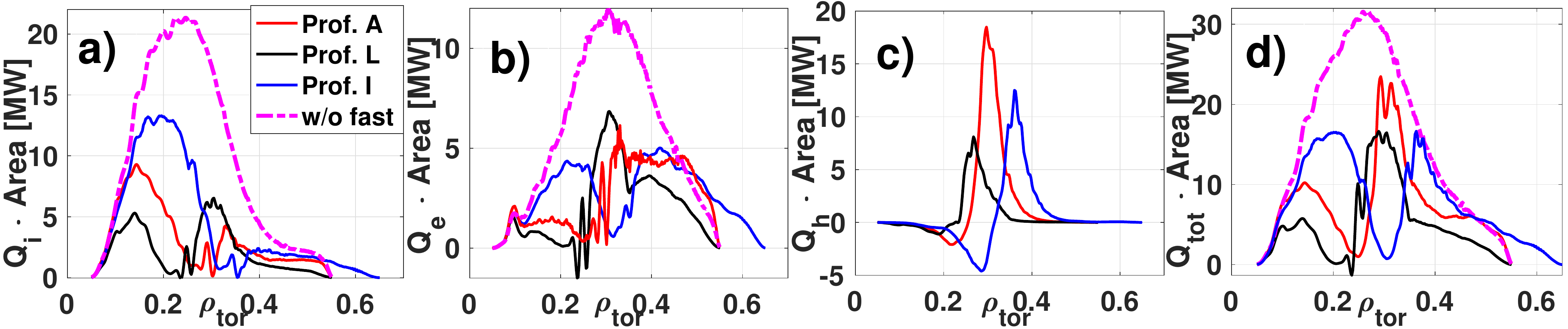}
\par\end{center}
\caption{Radial profile of the a) bulk ions, b) electrons, c) hydrogen minority species and d) overall (thermal ion + electron + fast hydrogen) heat fluxes averaged over the time domain $t [c_s / a] = [400 - 600]$ corresponding to the three different fast ion profiles of Fig.~\ref{fig:fig18}. The turbulent fluxes obtained by neglecting the energetic particles (Fig.~\ref{fig:fig3}) are shown in magenta as reference.}
\label{fig:fig19}
\end{figure*}

\begin{figure}
\begin{center}
\includegraphics[scale=0.40]{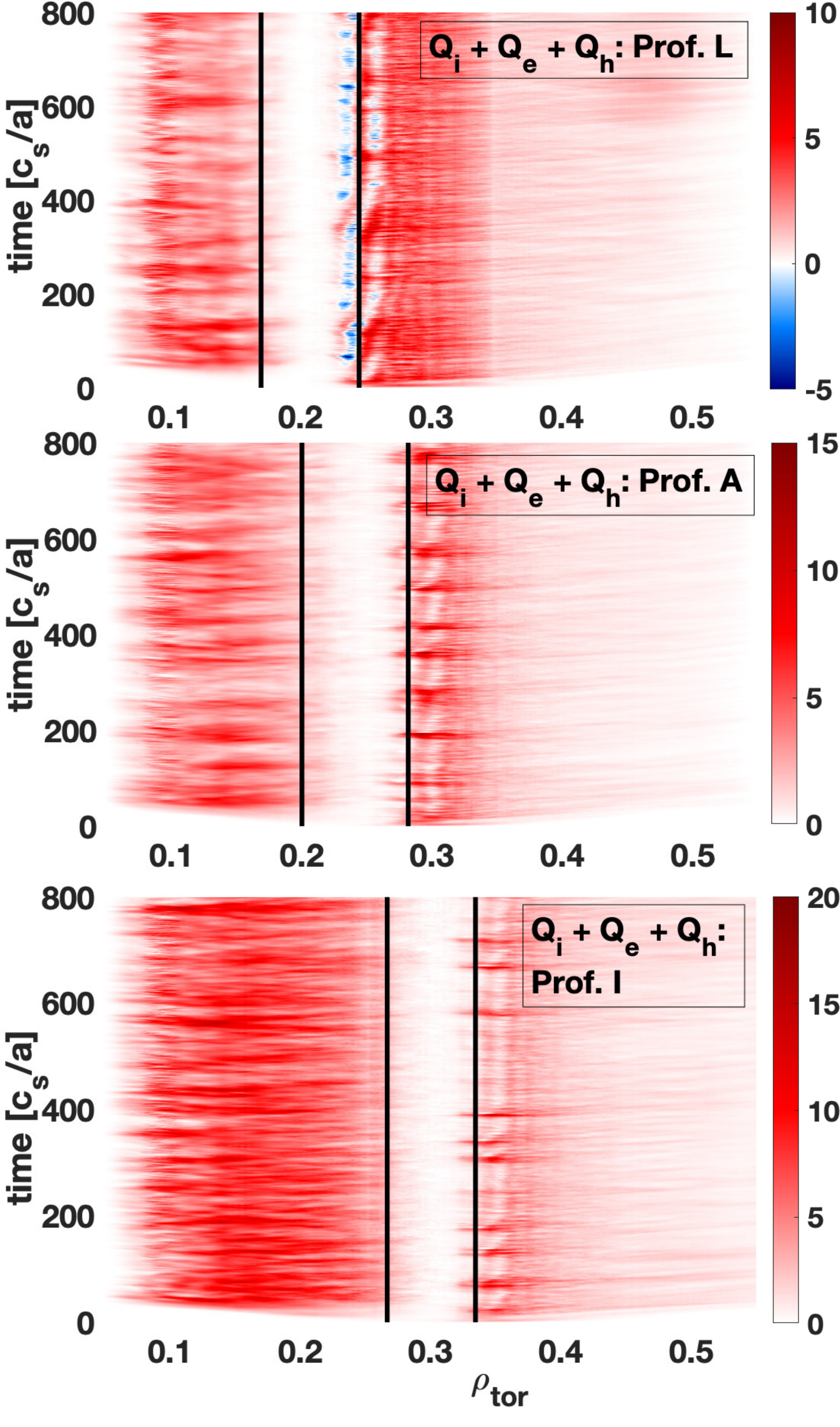}
\par\end{center}
\caption{Time evolution of the radial profile of the total heat flux (thermal ions + electrons + hydrogen) in MW corresponding to the cases labelled a) prof.~L, b) prof.~A and b) prof.~I. To compare more easily the changes in the position of the F-ATB among the different profiles, the radial domain of the case labelled prof.~I has been set equal to $\rho_{tor} = [0.05 - 0.55]$. The vertical black lines delimit the boundary positions of the transport barrier. Their radial coordinates have been computed as the position where the overall time-averaged (thermal ions + electrons + hydrogen) fluxes (see Fig.~\ref{fig:fig19}d) reach half-value before and after the transport barrier.}
\label{fig:fig20}
\end{figure}

The time-averaged heat fluxes obtained with these fast particle profiles are shown in Fig.~\ref{fig:fig19} for each different plasma species. It is worth mentioning that modifications on the radially dependent block-structured velocity grids were required to resolve the velocity directions correctly. This also led to the extension of the radial grid for the case labeled prof.~I, covering up to $\rho_{tor} = 0.65$. The other grids end at $\rho_{tor} = 0.55$. A first interesting observation by looking at Fig.~\ref{fig:fig19} is that the extension of the radial domain for the case prof.~I does not lead to significant changes in the turbulent fluxes for $\rho_{tor} > 0.4$ for the cases prof.~A and I, suggesting a negligible impact of the buffer zones on the results presented within this paper. Furthermore, we notice that the (bulk + fast) ion and electron heat fluxes are strongly affected by the changes in the energetic particle profiles. More precisely, the central region of minimum turbulent transport marking the location of the F-ATB ($\bar{\rho}_0$) follows the position of the peak of the energetic particle temperature gradient. It is $\bar{\rho}_0 \sim 0.25$ for the case labeled prof.~A, whilst it moves to $\bar{\rho}_0 \sim 0.2$ for prof.~L, and to $\bar{\rho}_0 \sim 0.28$ for prof.~I. It is worth mentioning here that the center of the F-ATB does not coincide with the position of the maximum of the fast ion logarithmic Gaussian-like temperature gradient, but it is rather determined by the ratio $T_h / T_e$. This result is again consistent with the quasi-linear predictions of Refs.~\cite{DiSiena_NF2018,DiSiena_PoP2019}. The "optimal" ratio $T_h / T_e$ (and large temperature gradient) maximizing the beneficial wave-particle resonant ITG stabilization (i.e., $T_h / T_e \sim [5 - 25]$) locates exactly in the region of reduced transport in Fig.~\ref{fig:fig20}. Similarly to the results obtained for the thermal species, also the energetic particle heat flux profiles undergoes strong modifications, which are in agreement with the changes in the resonant velocity layers, as discussed in detail in the previous Sections. The transport barrier radial dependence with the fast ion temperature profiles further corroborates the theoretical interpretation that the barrier is triggered by a wave-particle resonant interaction between supra-thermal particles and ITGs.

The time evolution of the overall turbulence fluxes are shown in Fig.~\ref{fig:fig20} for the different fast particle temperature profiles. The most effective (thermal ion + electron + fast particle) turbulence stabilization is found for prof.~L, where the thermal ion and electron fluxes are strongly reduced in the region $\rho_{tor} = [0.17 - 0.245]$. For this case, the position of the transport barrier leads to a suppression of the ion and electron heat fluxes, as shown in Fig.~\ref{fig:fig19}a-b. On the other hand, the case labeled prof.~I, moves the F-ATB outward, thus reducing its beneficial effect in the radial domain $\rho_{tor} = [0.1-0.25]$ that undergoes a large increase in both thermal ion and electron fluxes. These findings do not suggest that the "optimal" turbulence suppression is expected for transport barriers located more in the inner radii, but is a rather consequence of the specific numerical setup employed throughout this paper. More specifically, it depends on the particular radial profile of the bulk ion turbulent heat flux, which peaks around $\rho_{tor} \sim 0.2$ in the absence of fast ions. Therefore, the most effective turbulence suppression is found for the transport barriers centered at this location. 
\begin{figure}
\begin{center}
\includegraphics[scale=0.35]{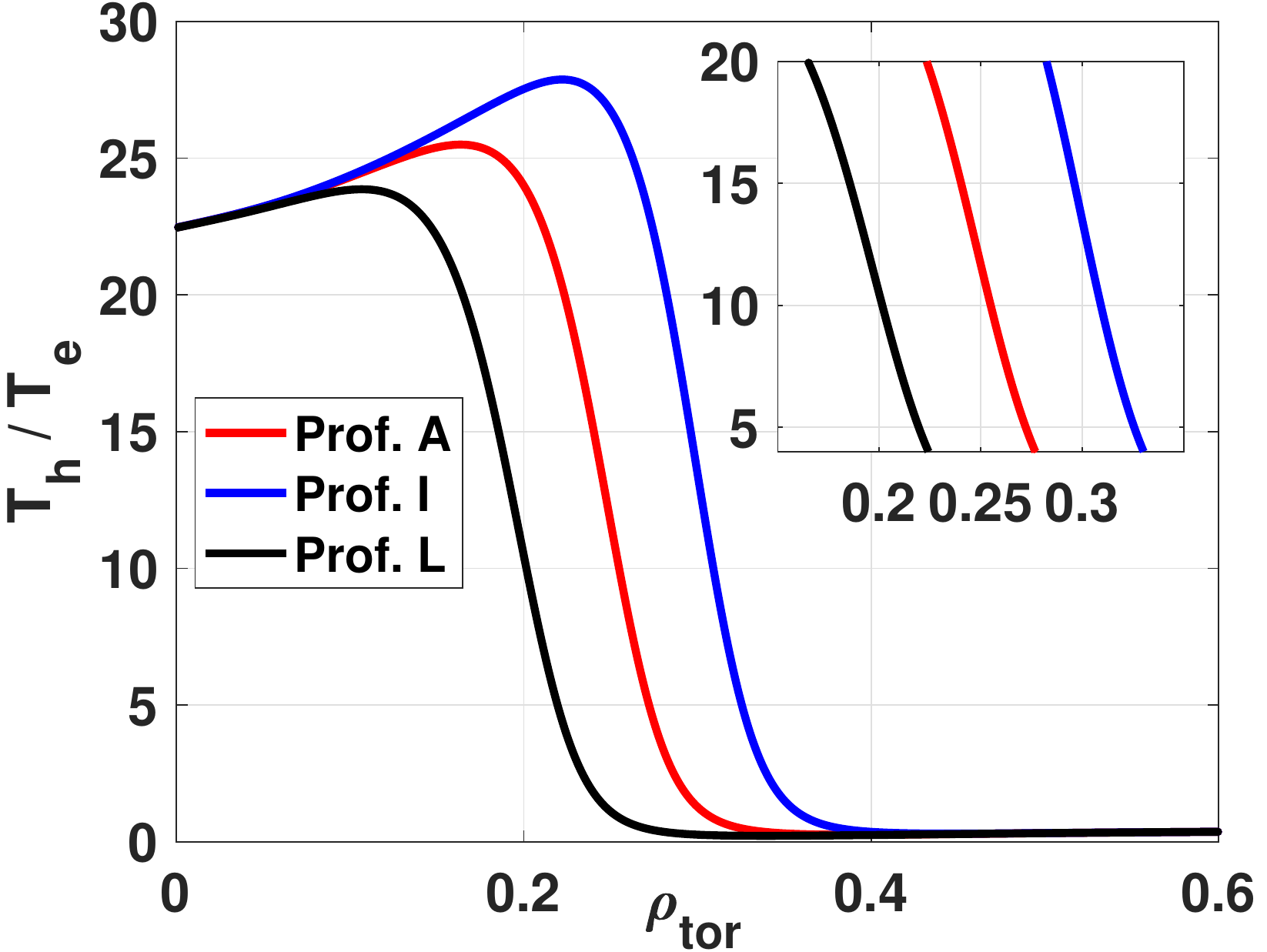}
\par\end{center}
\caption{Radial profile of the fast ion to electron temperature ratio $T_h / T_e$ for the three different energetic particle temperature profiles of Fig.~\ref{fig:fig18}.}
\label{fig:fig21}
\end{figure}

\section{Conclusions} \label{sec7}

A series of global nonlinear GENE simulations has been performed to investigate the impact of different energetic particle temperature profiles on a novel type of transport barrier called F-ATB. More specifically, we have analyzed how the particular shape of the supra-thermal ion temperature (and corresponding logarithmic gradient) determines the transport barrier width, radial localization, and its impact on turbulent transport. Realistic ASDEX Upgrade physical parameters - taken from the discharge $\# 36637$ at $t = 4.1$s - have been considered, and an analytic approximation for the fast ion temperature has been used. This is defined by a Gaussian-like logarithmic temperature gradient with essentially three free parameters, namely the amplitude ($\omega_{T_{h,0}}$), the half-width ($\delta_x$) and the radial location ($\rho_0$) of the Gaussian-like peak. Corresponding nonlinear scans over these different parameters are performed. Moreover, the role of finite $\beta_e$ (electromagnetic) effects and energetic particle temperature anisotropies - arising, e.g., from the heating schemes - on the transport barrier has been studied. It is worth mentioning that despite the relevance of the supra-thermal ion density in the dynamics of the wave-particle resonant interaction, for simplicity, it has been fixed to the value $n_h / n_e = 0.11\%$. This is the nominal fast ion density measured in the ASDEX Upgrade discharge $\# 36637$ at $t = 4.1$s inferred from neutral-particle-analyzer measurements. This holds true for the entire radial domain considered for the GENE simulations, thus fixing also the fast hydrogen logarithmic density gradient profile, which has not been varied throughout this work.

The main findings can be summarized as follows:

\renewcommand{\labelenumii}{\Roman{enumii}}
\begin{enumerate}[(i)]
    \item  We found a threshold in the amplitude of the logarithmic temperature gradient $\omega_{T_{h,0}}$ to trigger the transport barrier.
    This is related to the minimum value of $\omega_{T_{h,0}}$ required for the fast ion temperature profile to increase up to the "optimal" range of temperatures where $T_h / T_e = [5 - 25]$. When this condition is fulfilled, the magnetic fast particle drift matches the linear ITG frequencies (for the non-linearly relevant wave-numbers) in the negative fast ion drive regions ($v_\shortparallel^2 + \mu B_0 < 3/2$), thus strongly suppressing the ITG drive. This leads to a localized (only in the radial domain where such conditions are fulfilled) turbulence reduction. It is important to mention here that the threshold value is also expected to depend on the local (at the position of interest) fast ion density. A smaller fast ion density will require a larger logarithmic fast ion temperature gradient to maximize the wave-particle resonant interaction. According to Refs.~\cite{DiSiena_NF2018,DiSiena_PoP2019} the flow of the overall resonant energy exchange is proportional to both the fast ion density and its logarithmic temperature gradient. 

    \item The half-width $\delta_x$ of the Gaussian-like logarithmic fast particle temperature profile controls the radial width of the transport barrier. More specifically, it affects the size of the radial domain reaching the "optimal" ratio of $T_h / T_e$ (i.e., $T_h / T_e \sim [5- 25]$). Therefore, for a fixed value of $\omega_{T_{h,0}}$, the largest value of $\delta_x$ is found to lead to the widest transport barrier.

    \item When the on-axis fast ion temperature is fixed, the "optimal" profile has the largest half-width (despite the consequent reduction in $\omega_{T_{h,0}}$), since fixing it leads to fast ion temperature profiles always reaching the "sweet-spot" in $T_h / T_e$. This enhances the wave-particle resonant interaction. Therefore, all these profiles can efficiently trigger the F-ATB. However, the one with the largest half-width will lead to the broader transport barrier, thus amplifying its effect on transport.

    \item The central position of the transport barrier is "mainly" controlled by $\rho_0$ (position of the maximum of the Gaussian-like peak in the fast ion logarithmic temperature profile). Here, we are considering that the electron temperature profile does not vary dramatically along the radial direction. This will otherwise affect the ratio $T_h / T_e$. For the bulk profiles employed throughout this paper, the most effective overall turbulence suppression is found when the position of the peak is moved in the inner radii. This is strictly linked to the radial profile of the thermal ion turbulent flux in the absence of fast particles, which peaks around $\rho_{tor} \sim 0.2$. This result can change in other physical conditions where the bulk ion ITG turbulence is localized at a different position.

    \item Electromagnetic fluctuations do not alter qualitatively the transport barrier properties. However, finite $\beta_e$ effects lead to additional turbulence stabilization mainly in the radial region where the fast particle temperature gradient is maximum. No linearly unstable Alfv\'enic fast-ion driven modes are observed. These findings are in agreement with the results of Ref.~\cite{DiSiena_NF_2019,DiSiena_JPP_2021} where a nonlinear interplay between marginally stable fast ion modes and turbulence has been proposed to explain the electromagnetic fast particle turbulence stabilization \cite{Citrin_PRL2013}.
    
    \item Velocity space anisotropies do not lead to significant modifications on the thermal species fluxes in the electrostatic GENE results. However, non negligible changes are observed for the supra-thermal particles. This is likely to be related to the corresponding variations in the fast ion drive term (namely $\partial_x F_{0,h}$) which in turn affect the free-energy exchange between fast ions and ITG turbulence.

\end{enumerate}

The impact of such findings on plasma profiles evolution and plasma performances is still unclear. In particular, being able to predict with numerical simulations what are the critical energetic particle density, heating power and location of the ICRH absorption layer to trigger the most effective transport barrier is extremely challenging considering the numerical tools available to date. The previously mentioned quantities (energetic particle density, heating power and location of the ICRH absorption layer) depend on each others in a way which cannot be captured by gradient-driven turbulence simulations. These analyses would require the plasma profiles to freely evolve due to the combined effect of external sources and turbulent fluxes and hence a flux-driven (electromagnetic) approach, which is still not affordable numerically. An attractive alternative to flux-driven simulations is coupling the gradient-driven GENE version (acting on the short turbulence time scale) with a transport code (acting on the transport time scale). Dedicated studies pursuing this method are currently ongoing \cite{Parker_2018} and will be extended to include kinetic electrons and supra-thermal particles in the near future.

An important parameter dependence of the F-ATB which has not been investigated throughout this paper is the safety factor. It affects the energetic particle stabilization of plasma micro-turbulence and (most importantly) sets the position of rational surfaces in the radial domain. As already previously mentioned, rational surfaces in the safety factor profile can also trigger transport barriers via a fundamentally different trigger mechanism \cite{Joffrin_2002_q1,Joffrin_NF_2003} compared to the wave-particle interaction. The possible competition and/or synergy between these different effects on transport barrier formations will be analyzed in future studies.

This work has been carried out within the framework of the EUROfusion Consortium and has received funding from the Euratom research and training programme 2014-2018 and 2019-2020 under grant agreement No 633053. The views and opinions expressed herein do not necessarily reflect those of the European Commission. The simulations presented in this work were performed on the HPC-Marconi supercomputer within the framework of the ROBIN project. The authors would like to thank Philipp Lauber for fruitful discussions.

\appendix

\section{Convergence studies} \label{appendix:a}

To ensure that the transport barrier formation is not an artifact of the particular numerical setup employed, convergence studies have been performed and are illustrated throughout this Section. In particular, we analyze in details convergence over the radial and velocity (both $v_\shortparallel$ and $\mu$) resolutions together with different choices for the radially dependent block-structured velocity grid. The numerical setup and physical parameters are the same as the ones summarized in Section \ref{sec1}. Given the prohibitive computational cost for performing such studies for each of the ten different fast particle profiles employed in this manuscript, we limit our analyses only to the reference profile labelled prof.~A.

\subsection{Radial resolution}

\begin{figure*}
\begin{center}
\includegraphics[scale=0.45]{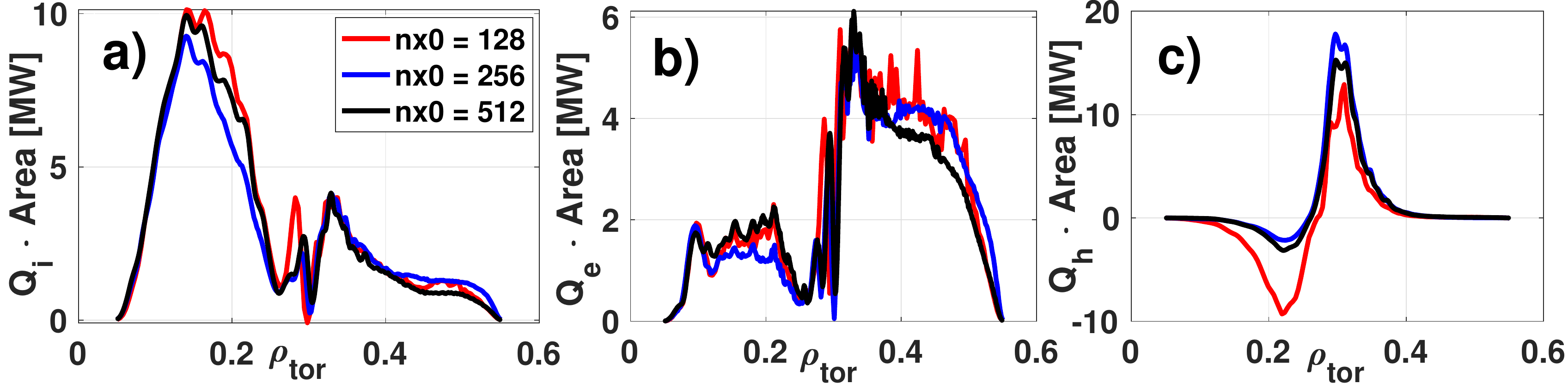}
\par\end{center}
\caption{Radial profile of the a) bulk ion, b) electron and c) hydrogen minority heat fluxes averaged over the time domain $t [c_s / a] = [200 - 300]$ corresponding to the reference profiles labelled prof.~A for different radial grid resolutions.}
\label{fig:nx0_x}
\end{figure*}

The number of radial grid points (${\rm nx0}$) employed throughout this work is ${\rm nx0} = 256$. This resolution differs from the one used in Ref.~\cite{DiSiena_PRL_2021} (namely ${\rm nx0} = 512$) by a factor of two to reduce the otherwise prohibitive computational cost required to run the ten different global simulations performed here. A comparison of the time-averaged radial profile of the heat fluxes carried by each of the plasma species is illustrated in Fig.~\ref{fig:nx0_x} for ${\rm nx0} = 128$, ${\rm nx0} = 256$ and ${\rm nx0} = 512$. The radial domain mid-radius location is $\rho_{tor} = 0.3$. It corresponds to the reference position used to normalize the different physical parameters. At this location, $1/\rho^* = a/\rho_s= 176$, with $a$ minor radius of the devise, $\rho_s = c_s / \Omega_i$, $c_s= \left(T_e/m_e\right)^{1/2}$. Therefore, the expected minimum radial resolution required to correctly resolve the radial dynamics is roughly two point per thermal ion Larmor radius. However, Fig.~\ref{fig:nx0_x} reveals only minor differences in the heat fluxes when increasing ${\rm nx0}$ from 256 to 512. This result shows that a minimum radial resolution of ${\rm nx0} = 256$ is sufficient to ensure converged heat fluxes for each plasma species within $<7 \%$ error. This relatively small minimum radial resolution may be due to the absence of low $m/n$ rational surfaces in the radial domain of interest.

\subsection{Velocity resolution}

\begin{figure*}
\begin{center}
\includegraphics[scale=0.45]{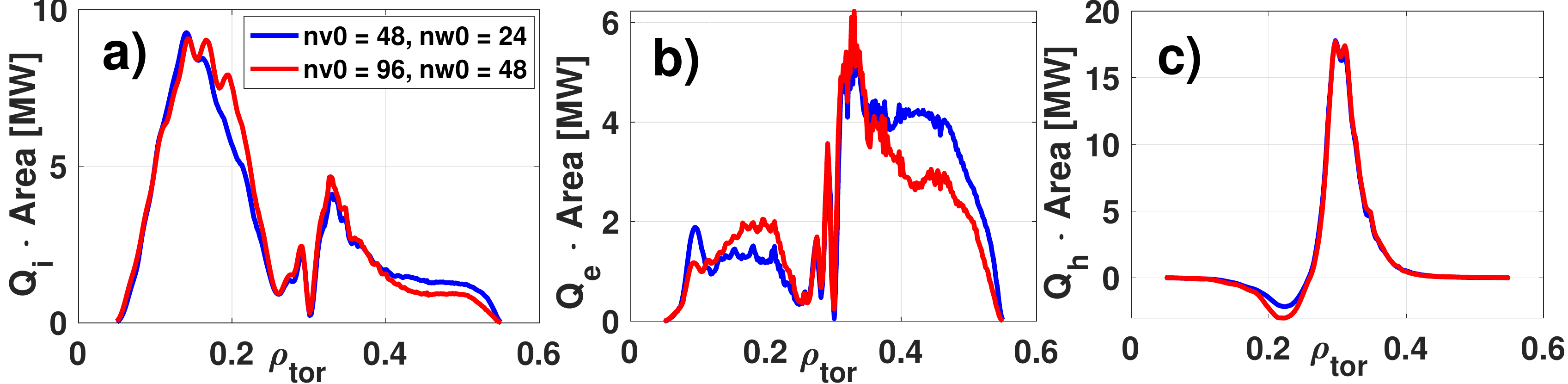}
\par\end{center}
\caption{Radial profile of the a) bulk ion, b) electron and c) hydrogen minority heat fluxes averaged over the time domain $t [c_s / a] = [200 - 300]$ corresponding to the reference profiles labelled prof.~A for different velocity grid resolutions.}
\label{fig:nv0_v}
\end{figure*}

The F-ATB trigger mechanism is the wave-particle resonant interaction discussed in Ref.~\cite{DiSiena_NF2018,DiSiena_PoP2019}. This is a complex effect involving a phase-space resonance in the velocity space. Therefore, it is particularly important to correctly resolve the $\left(v_\shortparallel,\mu\right)$ directions and ensure that the resonant layers are well captured. In the present Section, convergence studies are performed comparing the time-averaged radial heat flux profiles obtained with the reference velocity resolution employed in Sections (i.e., $(nv0,nw0) = (48,24)$), and $(nv0,nw0) = (96,48)$. The results are illustrated in Fig.~\ref{fig:nv0_v} for the different plasma species considered. The increase by a factor of two in the parallel velocity and magnetic moment resolutions leads to $\sim 12 \%$ variations in the radially-averaged turbulent fluxes, proving that the reference results are reasonably well resolved in the velocity directions.

\subsection{Block-structured grid}

\begin{figure}
\begin{center}
\includegraphics[scale=0.55]{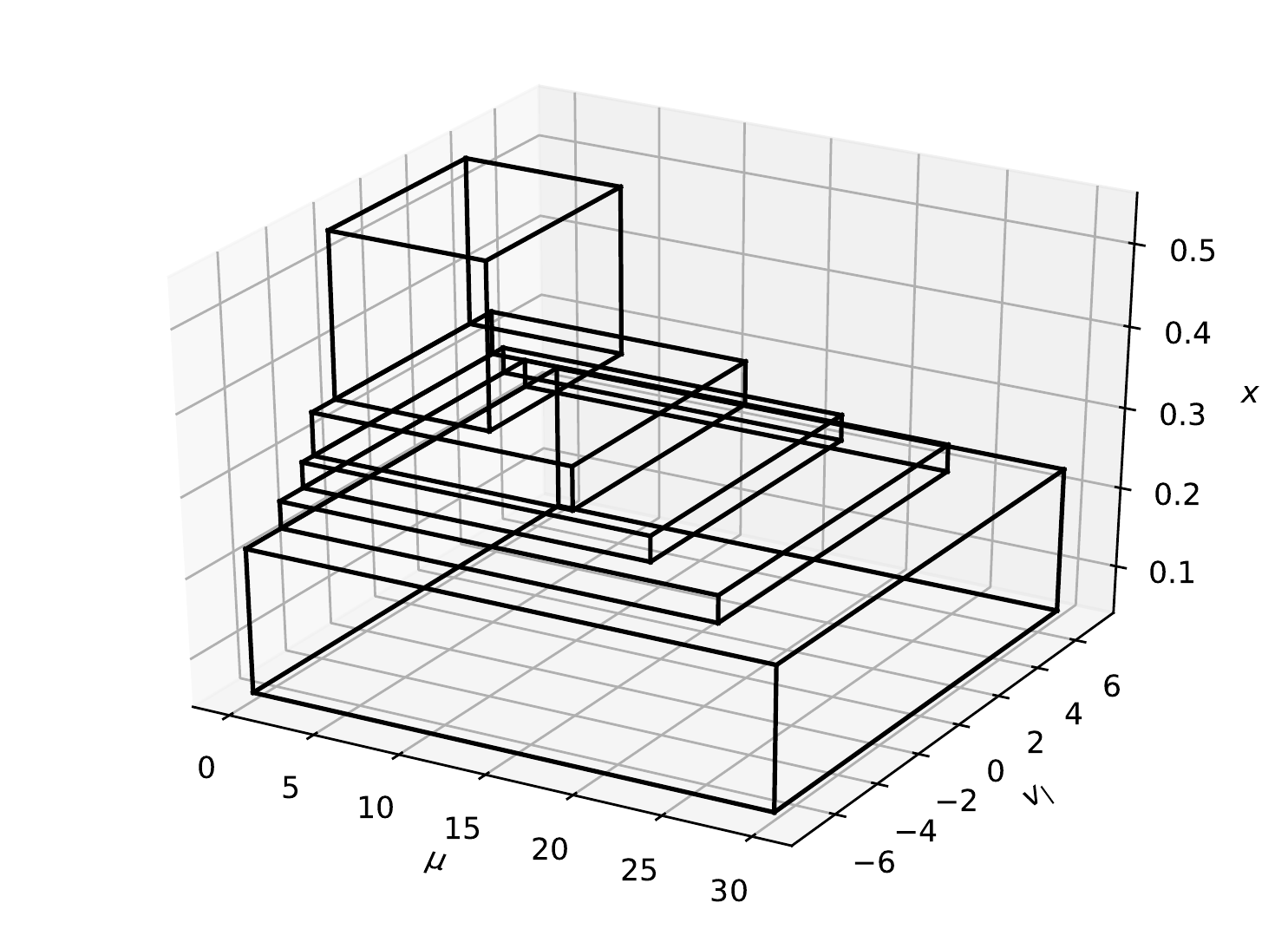}
\par\end{center}
\caption{Block-structured grid with five blocks in $(\rho_{tor}, v_\shortparallel,\mu)$ generated to check the convergence of the GENE results.}
\label{fig:bsg_x}
\end{figure}

\begin{figure*}
\begin{center}
\includegraphics[scale=0.45]{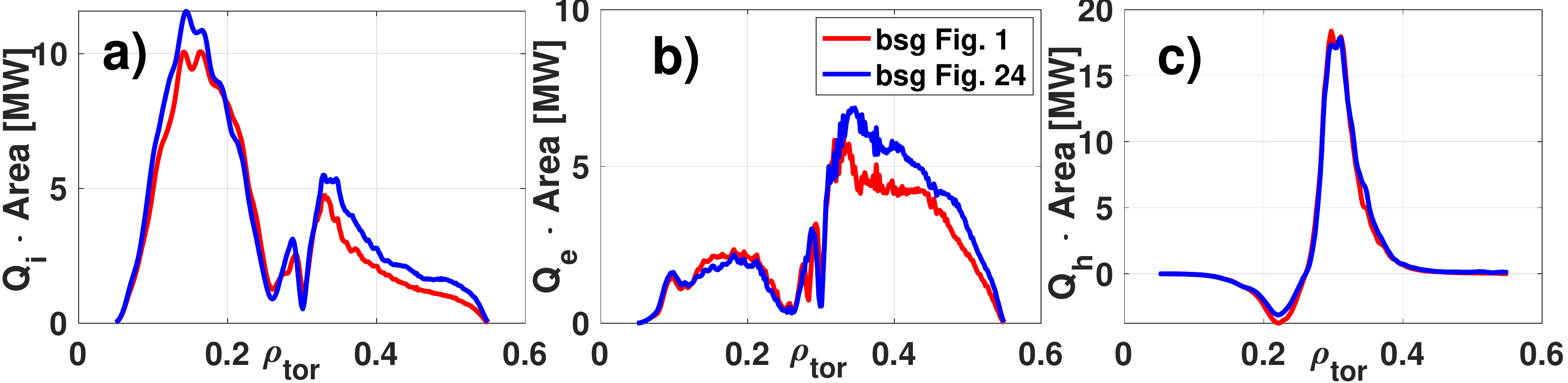}
\par\end{center}
\caption{Radial profile of the a) bulk ion, b) electron and c) hydrogen minority heat fluxes averaged over the time domain $t [c_s / a] = [200 - 300]$ corresponding to the reference profiles labelled prof.~A for different block-structured grids.}
\label{fig:bsg_x2}
\end{figure*}

The remaining convergence study discussed within this paper concerns the particular shape of the radially dependent block-structured grid employed for the global GENE simulations. As already stressed in Section \ref{sec1}, this particular feature is essential to reduce the velocity resolution requirements to resolve the dynamics of the bulk and supra-thermal species in the whole radial domain of interest. Therefore, to ensure that our numerical results are not affected by the specific choice of the block-structured grid, we present a comparison of the time-averaged turbulent fluxes obtained with the reference grid (see Fig.~\ref{fig:fig1}) and with the one introduced in Fig.~\ref{fig:bsg_x}. 
The results are shown in Fig.~\ref{fig:bsg_x2} for the different plasma species considered. The same grid resolution is employed in each directions with $(nv0,nw0) = (96,48)$. The change in the radially-dependent velocity grid does not alter the numerical results significantly and only minor differences of about $\sim 10 \%$ are observed from Fig.~\ref{fig:bsg_x2}. These findings show that the reference results are independent from the particular choice of the block-structured grid and well resolved in the radial and velocity directions.


\begin{thebibliography}{48}%
\makeatletter
\providecommand \@ifxundefined [1]{%
 \@ifx{#1\undefined}
}%
\providecommand \@ifnum [1]{%
 \ifnum #1\expandafter \@firstoftwo
 \else \expandafter \@secondoftwo
 \fi
}%
\providecommand \@ifx [1]{%
 \ifx #1\expandafter \@firstoftwo
 \else \expandafter \@secondoftwo
 \fi
}%
\providecommand \natexlab [1]{#1}%
\providecommand \enquote  [1]{``#1''}%
\providecommand \bibnamefont  [1]{#1}%
\providecommand \bibfnamefont [1]{#1}%
\providecommand \citenamefont [1]{#1}%
\providecommand \href@noop [0]{\@secondoftwo}%
\providecommand \href [0]{\begingroup \@sanitize@url \@href}%
\providecommand \@href[1]{\@@startlink{#1}\@@href}%
\providecommand \@@href[1]{\endgroup#1\@@endlink}%
\providecommand \@sanitize@url [0]{\catcode `\\12\catcode `\$12\catcode
  `\&12\catcode `\#12\catcode `\^12\catcode `\_12\catcode `\%12\relax}%
\providecommand \@@startlink[1]{}%
\providecommand \@@endlink[0]{}%
\providecommand \url  [0]{\begingroup\@sanitize@url \@url }%
\providecommand \@url [1]{\endgroup\@href {#1}{\urlprefix }}%
\providecommand \urlprefix  [0]{URL }%
\providecommand \Eprint [0]{\href }%
\providecommand \doibase [0]{http://dx.doi.org/}%
\providecommand \selectlanguage [0]{\@gobble}%
\providecommand \bibinfo  [0]{\@secondoftwo}%
\providecommand \bibfield  [0]{\@secondoftwo}%
\providecommand \translation [1]{[#1]}%
\providecommand \BibitemOpen [0]{}%
\providecommand \bibitemStop [0]{}%
\providecommand \bibitemNoStop [0]{.\EOS\space}%
\providecommand \EOS [0]{\spacefactor3000\relax}%
\providecommand \BibitemShut  [1]{\csname bibitem#1\endcsname}%
\let\auto@bib@innerbib\@empty
\bibitem [{\citenamefont {Connor}\ \emph {et~al.}(2004)\citenamefont {Connor},
  \citenamefont {Fukuda}, \citenamefont {Garbet}, \citenamefont {Gormezano},
  \citenamefont {Mukhovatov}, \citenamefont {Wakatani}, \citenamefont {the ITB
  Database~Group}, \citenamefont {the ITPA Topical Group~on Transport},\ and\
  \citenamefont {Physics}}]{Connor_NF_2004}%
  \BibitemOpen
  \bibfield  {author} {\bibinfo {author} {\bibfnamefont {J.}~\bibnamefont
  {Connor}}, \bibinfo {author} {\bibfnamefont {T.}~\bibnamefont {Fukuda}},
  \bibinfo {author} {\bibfnamefont {X.}~\bibnamefont {Garbet}}, \bibinfo
  {author} {\bibfnamefont {C.}~\bibnamefont {Gormezano}}, \bibinfo {author}
  {\bibfnamefont {V.}~\bibnamefont {Mukhovatov}}, \bibinfo {author}
  {\bibfnamefont {M.}~\bibnamefont {Wakatani}}, \bibinfo {author} {\bibnamefont
  {the ITB Database~Group}}, \bibinfo {author} {\bibnamefont {the ITPA Topical
  Group~on Transport}}, \ and\ \bibinfo {author} {\bibfnamefont {I.~B.}\
  \bibnamefont {Physics}},\ }\href {\doibase 10.1088/0029-5515/44/4/r01}
  {\bibfield  {journal} {\bibinfo  {journal} {Nuclear Fusion}\ }\textbf
  {\bibinfo {volume} {44}},\ \bibinfo {pages} {R1} (\bibinfo {year}
  {2004})}\BibitemShut {NoStop}%
\bibitem [{\citenamefont {Rice}\ \emph {et~al.}(1996)\citenamefont {Rice},
  \citenamefont {Burrell}, \citenamefont {Lao}, \citenamefont {Navratil},
  \citenamefont {Stallard}, \citenamefont {Strait}, \citenamefont {Taylor},
  \citenamefont {Austin}, \citenamefont {Casper}, \citenamefont {Chu},
  \citenamefont {Forest}, \citenamefont {Gohil}, \citenamefont {Groebner},
  \citenamefont {Heidbrink}, \citenamefont {Hyatt}, \citenamefont {Ikezi},
  \citenamefont {La~Haye}, \citenamefont {Lazarus}, \citenamefont {Lin-Liu},
  \citenamefont {Mauel}, \citenamefont {Meyer}, \citenamefont {Rettig},
  \citenamefont {Schissel}, \citenamefont {St.~John}, \citenamefont {Taylor},\
  and\ \citenamefont {Turnbull}}]{Rice_PoP_1996}%
  \BibitemOpen
  \bibfield  {author} {\bibinfo {author} {\bibfnamefont {B.~W.}\ \bibnamefont
  {Rice}}, \bibinfo {author} {\bibfnamefont {K.~H.}\ \bibnamefont {Burrell}},
  \bibinfo {author} {\bibfnamefont {L.~L.}\ \bibnamefont {Lao}}, \bibinfo
  {author} {\bibfnamefont {G.}~\bibnamefont {Navratil}}, \bibinfo {author}
  {\bibfnamefont {B.~W.}\ \bibnamefont {Stallard}}, \bibinfo {author}
  {\bibfnamefont {E.~J.}\ \bibnamefont {Strait}}, \bibinfo {author}
  {\bibfnamefont {T.~S.}\ \bibnamefont {Taylor}}, \bibinfo {author}
  {\bibfnamefont {M.~E.}\ \bibnamefont {Austin}}, \bibinfo {author}
  {\bibfnamefont {T.~A.}\ \bibnamefont {Casper}}, \bibinfo {author}
  {\bibfnamefont {M.~S.}\ \bibnamefont {Chu}}, \bibinfo {author} {\bibfnamefont
  {C.~B.}\ \bibnamefont {Forest}}, \bibinfo {author} {\bibfnamefont
  {P.}~\bibnamefont {Gohil}}, \bibinfo {author} {\bibfnamefont {R.~J.}\
  \bibnamefont {Groebner}}, \bibinfo {author} {\bibfnamefont {W.~W.}\
  \bibnamefont {Heidbrink}}, \bibinfo {author} {\bibfnamefont {A.~W.}\
  \bibnamefont {Hyatt}}, \bibinfo {author} {\bibfnamefont {H.}~\bibnamefont
  {Ikezi}}, \bibinfo {author} {\bibfnamefont {R.~J.}\ \bibnamefont {La~Haye}},
  \bibinfo {author} {\bibfnamefont {E.~A.}\ \bibnamefont {Lazarus}}, \bibinfo
  {author} {\bibfnamefont {Y.~R.}\ \bibnamefont {Lin-Liu}}, \bibinfo {author}
  {\bibfnamefont {M.~E.}\ \bibnamefont {Mauel}}, \bibinfo {author}
  {\bibfnamefont {W.~H.}\ \bibnamefont {Meyer}}, \bibinfo {author}
  {\bibfnamefont {C.~L.}\ \bibnamefont {Rettig}}, \bibinfo {author}
  {\bibfnamefont {D.~P.}\ \bibnamefont {Schissel}}, \bibinfo {author}
  {\bibfnamefont {H.~E.}\ \bibnamefont {St.~John}}, \bibinfo {author}
  {\bibfnamefont {P.~L.}\ \bibnamefont {Taylor}}, \ and\ \bibinfo {author}
  {\bibfnamefont {A.~D.}\ \bibnamefont {Turnbull}},\ }\href {\doibase
  10.1063/1.871994} {\bibfield  {journal} {\bibinfo  {journal} {Physics of
  Plasmas}\ }\textbf {\bibinfo {volume} {3}},\ \bibinfo {pages} {1983}
  (\bibinfo {year} {1996})}\BibitemShut {NoStop}%
\bibitem [{\citenamefont {Gruber}\ \emph {et~al.}(2000)\citenamefont {Gruber},
  \citenamefont {Wolf}, \citenamefont {Dux}, \citenamefont {Guenter},
  \citenamefont {Carthy}, \citenamefont {Lackner}, \citenamefont {Maraschek},
  \citenamefont {Meister}, \citenamefont {Pereverzev}, \citenamefont
  {Treutterer},\ and\ \citenamefont {the ASDEX
  Upgrade~Team}}]{Gruber_PPCF_2000}%
  \BibitemOpen
  \bibfield  {author} {\bibinfo {author} {\bibfnamefont {O.}~\bibnamefont
  {Gruber}}, \bibinfo {author} {\bibfnamefont {R.}~\bibnamefont {Wolf}},
  \bibinfo {author} {\bibfnamefont {R.}~\bibnamefont {Dux}}, \bibinfo {author}
  {\bibfnamefont {S.}~\bibnamefont {Guenter}}, \bibinfo {author} {\bibfnamefont
  {P.~J.~M.}\ \bibnamefont {Carthy}}, \bibinfo {author} {\bibfnamefont
  {K.}~\bibnamefont {Lackner}}, \bibinfo {author} {\bibfnamefont
  {M.}~\bibnamefont {Maraschek}}, \bibinfo {author} {\bibfnamefont
  {H.}~\bibnamefont {Meister}}, \bibinfo {author} {\bibfnamefont
  {G.}~\bibnamefont {Pereverzev}}, \bibinfo {author} {\bibfnamefont
  {W.}~\bibnamefont {Treutterer}}, \ and\ \bibinfo {author} {\bibnamefont {the
  ASDEX Upgrade~Team}},\ }\href {\doibase 10.1088/0741-3335/42/5a/311}
  {\bibfield  {journal} {\bibinfo  {journal} {Plasma Physics and Controlled
  Fusion}\ }\textbf {\bibinfo {volume} {42}},\ \bibinfo {pages} {A117}
  (\bibinfo {year} {2000})}\BibitemShut {NoStop}%
\bibitem [{\citenamefont {Tardini}\ \emph {et~al.}(2007)\citenamefont
  {Tardini}, \citenamefont {Hobirk}, \citenamefont {Igochine}, \citenamefont
  {Maggi}, \citenamefont {Martin}, \citenamefont {McCune}, \citenamefont
  {Peeters}, \citenamefont {Sips}, \citenamefont {St\"abler}, \citenamefont
  {Stober},\ and\ \citenamefont {the ASDEX Upgrade~Team}}]{Tardini_NF_2007}%
  \BibitemOpen
  \bibfield  {author} {\bibinfo {author} {\bibfnamefont {G.}~\bibnamefont
  {Tardini}}, \bibinfo {author} {\bibfnamefont {J.}~\bibnamefont {Hobirk}},
  \bibinfo {author} {\bibfnamefont {V.}~\bibnamefont {Igochine}}, \bibinfo
  {author} {\bibfnamefont {C.}~\bibnamefont {Maggi}}, \bibinfo {author}
  {\bibfnamefont {P.}~\bibnamefont {Martin}}, \bibinfo {author} {\bibfnamefont
  {D.}~\bibnamefont {McCune}}, \bibinfo {author} {\bibfnamefont
  {A.}~\bibnamefont {Peeters}}, \bibinfo {author} {\bibfnamefont
  {A.}~\bibnamefont {Sips}}, \bibinfo {author} {\bibfnamefont {A.}~\bibnamefont
  {St\"abler}}, \bibinfo {author} {\bibfnamefont {J.}~\bibnamefont {Stober}}, \
  and\ \bibinfo {author} {\bibnamefont {the ASDEX Upgrade~Team}},\ }\href
  {\doibase 10.1088/0029-5515/47/4/006} {\bibfield  {journal} {\bibinfo
  {journal} {Nuclear Fusion}\ }\textbf {\bibinfo {volume} {47}},\ \bibinfo
  {pages} {280} (\bibinfo {year} {2007})}\BibitemShut {NoStop}%
\bibitem [{\citenamefont {Fujita}\ \emph {et~al.}(1997)\citenamefont {Fujita},
  \citenamefont {Ide}, \citenamefont {Shirai}, \citenamefont {Kikuchi},
  \citenamefont {Naito}, \citenamefont {Koide}, \citenamefont {Takeji},
  \citenamefont {Kubo},\ and\ \citenamefont {Ishida}}]{Fujita_PRL_1997}%
  \BibitemOpen
  \bibfield  {author} {\bibinfo {author} {\bibfnamefont {T.}~\bibnamefont
  {Fujita}}, \bibinfo {author} {\bibfnamefont {S.}~\bibnamefont {Ide}},
  \bibinfo {author} {\bibfnamefont {H.}~\bibnamefont {Shirai}}, \bibinfo
  {author} {\bibfnamefont {M.}~\bibnamefont {Kikuchi}}, \bibinfo {author}
  {\bibfnamefont {O.}~\bibnamefont {Naito}}, \bibinfo {author} {\bibfnamefont
  {Y.}~\bibnamefont {Koide}}, \bibinfo {author} {\bibfnamefont
  {S.}~\bibnamefont {Takeji}}, \bibinfo {author} {\bibfnamefont
  {H.}~\bibnamefont {Kubo}}, \ and\ \bibinfo {author} {\bibfnamefont
  {S.}~\bibnamefont {Ishida}},\ }\href {\doibase 10.1103/PhysRevLett.78.2377}
  {\bibfield  {journal} {\bibinfo  {journal} {Phys. Rev. Lett.}\ }\textbf
  {\bibinfo {volume} {78}},\ \bibinfo {pages} {2377} (\bibinfo {year}
  {1997})}\BibitemShut {NoStop}%
\bibitem [{\citenamefont {Levinton}\ \emph {et~al.}(1995)\citenamefont
  {Levinton}, \citenamefont {Zarnstorff}, \citenamefont {Batha}, \citenamefont
  {Bell}, \citenamefont {Bell}, \citenamefont {Budny}, \citenamefont {Bush},
  \citenamefont {Chang}, \citenamefont {Fredrickson}, \citenamefont {Janos},
  \citenamefont {Manickam}, \citenamefont {Ramsey}, \citenamefont {Sabbagh},
  \citenamefont {Schmidt}, \citenamefont {Synakowski},\ and\ \citenamefont
  {Taylor}}]{Levinton_PRL_1995}%
  \BibitemOpen
  \bibfield  {author} {\bibinfo {author} {\bibfnamefont {F.~M.}\ \bibnamefont
  {Levinton}}, \bibinfo {author} {\bibfnamefont {M.~C.}\ \bibnamefont
  {Zarnstorff}}, \bibinfo {author} {\bibfnamefont {S.~H.}\ \bibnamefont
  {Batha}}, \bibinfo {author} {\bibfnamefont {M.}~\bibnamefont {Bell}},
  \bibinfo {author} {\bibfnamefont {R.~E.}\ \bibnamefont {Bell}}, \bibinfo
  {author} {\bibfnamefont {R.~V.}\ \bibnamefont {Budny}}, \bibinfo {author}
  {\bibfnamefont {C.}~\bibnamefont {Bush}}, \bibinfo {author} {\bibfnamefont
  {Z.}~\bibnamefont {Chang}}, \bibinfo {author} {\bibfnamefont
  {E.}~\bibnamefont {Fredrickson}}, \bibinfo {author} {\bibfnamefont
  {A.}~\bibnamefont {Janos}}, \bibinfo {author} {\bibfnamefont
  {J.}~\bibnamefont {Manickam}}, \bibinfo {author} {\bibfnamefont
  {A.}~\bibnamefont {Ramsey}}, \bibinfo {author} {\bibfnamefont {S.~A.}\
  \bibnamefont {Sabbagh}}, \bibinfo {author} {\bibfnamefont {G.~L.}\
  \bibnamefont {Schmidt}}, \bibinfo {author} {\bibfnamefont {E.~J.}\
  \bibnamefont {Synakowski}}, \ and\ \bibinfo {author} {\bibfnamefont
  {G.}~\bibnamefont {Taylor}},\ }\href {\doibase 10.1103/PhysRevLett.75.4417}
  {\bibfield  {journal} {\bibinfo  {journal} {Phys. Rev. Lett.}\ }\textbf
  {\bibinfo {volume} {75}},\ \bibinfo {pages} {4417} (\bibinfo {year}
  {1995})}\BibitemShut {NoStop}%
\bibitem [{\citenamefont {Litaudon}\ \emph {et~al.}(1999)\citenamefont
  {Litaudon}, \citenamefont {Aniel}, \citenamefont {Baranov}, \citenamefont
  {Bartlett}, \citenamefont {B{\'{e}}coulet}, \citenamefont {Challis},
  \citenamefont {Conway}, \citenamefont {Cottrell}, \citenamefont {Ekedahl},
  \citenamefont {Erba}, \citenamefont {Eriksson}, \citenamefont {Gormezano},
  \citenamefont {Hoang}, \citenamefont {Huysmans}, \citenamefont {Imbeaux},
  \citenamefont {Joffrin}, \citenamefont {Mantsinen}, \citenamefont {Parail},
  \citenamefont {Peysson}, \citenamefont {Rochard}, \citenamefont {Schild},
  \citenamefont {Sips}, \citenamefont {S{\~{a}}\"oldner}, \citenamefont
  {Tubbing}, \citenamefont {Voitsekhovitch}, \citenamefont {Ward},\ and\
  \citenamefont {Zwingmann}}]{Litaudon_PPCF_1999}%
  \BibitemOpen
  \bibfield  {author} {\bibinfo {author} {\bibfnamefont {X.}~\bibnamefont
  {Litaudon}}, \bibinfo {author} {\bibfnamefont {T.}~\bibnamefont {Aniel}},
  \bibinfo {author} {\bibfnamefont {Y.}~\bibnamefont {Baranov}}, \bibinfo
  {author} {\bibfnamefont {D.}~\bibnamefont {Bartlett}}, \bibinfo {author}
  {\bibfnamefont {A.}~\bibnamefont {B{\'{e}}coulet}}, \bibinfo {author}
  {\bibfnamefont {C.}~\bibnamefont {Challis}}, \bibinfo {author} {\bibfnamefont
  {G.~D.}\ \bibnamefont {Conway}}, \bibinfo {author} {\bibfnamefont {G.~A.}\
  \bibnamefont {Cottrell}}, \bibinfo {author} {\bibfnamefont {A.}~\bibnamefont
  {Ekedahl}}, \bibinfo {author} {\bibfnamefont {M.}~\bibnamefont {Erba}},
  \bibinfo {author} {\bibfnamefont {L.}~\bibnamefont {Eriksson}}, \bibinfo
  {author} {\bibfnamefont {C.}~\bibnamefont {Gormezano}}, \bibinfo {author}
  {\bibfnamefont {G.~T.}\ \bibnamefont {Hoang}}, \bibinfo {author}
  {\bibfnamefont {G.}~\bibnamefont {Huysmans}}, \bibinfo {author}
  {\bibfnamefont {F.}~\bibnamefont {Imbeaux}}, \bibinfo {author} {\bibfnamefont
  {E.}~\bibnamefont {Joffrin}}, \bibinfo {author} {\bibfnamefont
  {M.}~\bibnamefont {Mantsinen}}, \bibinfo {author} {\bibfnamefont
  {V.}~\bibnamefont {Parail}}, \bibinfo {author} {\bibfnamefont
  {Y.}~\bibnamefont {Peysson}}, \bibinfo {author} {\bibfnamefont
  {F.}~\bibnamefont {Rochard}}, \bibinfo {author} {\bibfnamefont
  {P.}~\bibnamefont {Schild}}, \bibinfo {author} {\bibfnamefont
  {A.}~\bibnamefont {Sips}}, \bibinfo {author} {\bibfnamefont {F.~X.}\
  \bibnamefont {S{\~{a}}\"oldner}}, \bibinfo {author} {\bibfnamefont
  {B.}~\bibnamefont {Tubbing}}, \bibinfo {author} {\bibfnamefont
  {I.}~\bibnamefont {Voitsekhovitch}}, \bibinfo {author} {\bibfnamefont
  {D.}~\bibnamefont {Ward}}, \ and\ \bibinfo {author} {\bibfnamefont
  {W.}~\bibnamefont {Zwingmann}},\ }\href {\doibase
  10.1088/0741-3335/41/3a/066} {\bibfield  {journal} {\bibinfo  {journal}
  {Plasma Physics and Controlled Fusion}\ }\textbf {\bibinfo {volume} {41}},\
  \bibinfo {pages} {A733} (\bibinfo {year} {1999})}\BibitemShut {NoStop}%
\bibitem [{\citenamefont {Ernst}\ \emph {et~al.}(2004)\citenamefont {Ernst},
  \citenamefont {Bonoli}, \citenamefont {Catto}, \citenamefont {Dorland},
  \citenamefont {Fiore}, \citenamefont {Granetz}, \citenamefont {Greenwald},
  \citenamefont {Hubbard}, \citenamefont {Porkolab}, \citenamefont {Redi},
  \citenamefont {Rice}, \citenamefont {Zhurovich},\ and\ \citenamefont
  {Group}}]{Ernst_PoP_2004}%
  \BibitemOpen
  \bibfield  {author} {\bibinfo {author} {\bibfnamefont {D.~R.}\ \bibnamefont
  {Ernst}}, \bibinfo {author} {\bibfnamefont {P.~T.}\ \bibnamefont {Bonoli}},
  \bibinfo {author} {\bibfnamefont {P.~J.}\ \bibnamefont {Catto}}, \bibinfo
  {author} {\bibfnamefont {W.}~\bibnamefont {Dorland}}, \bibinfo {author}
  {\bibfnamefont {C.~L.}\ \bibnamefont {Fiore}}, \bibinfo {author}
  {\bibfnamefont {R.~S.}\ \bibnamefont {Granetz}}, \bibinfo {author}
  {\bibfnamefont {M.}~\bibnamefont {Greenwald}}, \bibinfo {author}
  {\bibfnamefont {A.~E.}\ \bibnamefont {Hubbard}}, \bibinfo {author}
  {\bibfnamefont {M.}~\bibnamefont {Porkolab}}, \bibinfo {author}
  {\bibfnamefont {M.~H.}\ \bibnamefont {Redi}}, \bibinfo {author}
  {\bibfnamefont {J.~E.}\ \bibnamefont {Rice}}, \bibinfo {author}
  {\bibfnamefont {K.}~\bibnamefont {Zhurovich}}, \ and\ \bibinfo {author}
  {\bibfnamefont {A.~C.-M.}\ \bibnamefont {Group}},\ }\href {\doibase
  10.1063/1.1705653} {\bibfield  {journal} {\bibinfo  {journal} {Physics of
  Plasmas}\ }\textbf {\bibinfo {volume} {11}},\ \bibinfo {pages} {2637}
  (\bibinfo {year} {2004})}\BibitemShut {NoStop}%
\bibitem [{\citenamefont {Zhurovich}\ \emph {et~al.}(2007)\citenamefont
  {Zhurovich}, \citenamefont {Fiore}, \citenamefont {Ernst}, \citenamefont
  {Bonoli}, \citenamefont {Greenwald}, \citenamefont {Hubbard}, \citenamefont
  {Hughes}, \citenamefont {Marmar}, \citenamefont {Mikkelsen}, \citenamefont
  {Phillips},\ and\ \citenamefont {Rice}}]{Zhurovich_NF_2007}%
  \BibitemOpen
  \bibfield  {author} {\bibinfo {author} {\bibfnamefont {K.}~\bibnamefont
  {Zhurovich}}, \bibinfo {author} {\bibfnamefont {C.}~\bibnamefont {Fiore}},
  \bibinfo {author} {\bibfnamefont {D.}~\bibnamefont {Ernst}}, \bibinfo
  {author} {\bibfnamefont {P.}~\bibnamefont {Bonoli}}, \bibinfo {author}
  {\bibfnamefont {M.}~\bibnamefont {Greenwald}}, \bibinfo {author}
  {\bibfnamefont {A.}~\bibnamefont {Hubbard}}, \bibinfo {author} {\bibfnamefont
  {J.}~\bibnamefont {Hughes}}, \bibinfo {author} {\bibfnamefont
  {E.}~\bibnamefont {Marmar}}, \bibinfo {author} {\bibfnamefont
  {D.}~\bibnamefont {Mikkelsen}}, \bibinfo {author} {\bibfnamefont
  {P.}~\bibnamefont {Phillips}}, \ and\ \bibinfo {author} {\bibfnamefont
  {J.}~\bibnamefont {Rice}},\ }\href {\doibase 10.1088/0029-5515/47/9/019}
  {\bibfield  {journal} {\bibinfo  {journal} {Nuclear Fusion}\ }\textbf
  {\bibinfo {volume} {47}},\ \bibinfo {pages} {1220} (\bibinfo {year}
  {2007})}\BibitemShut {NoStop}%
\bibitem [{\citenamefont {Gormezano}(1999)}]{Gormezano_PPCF_1999}%
  \BibitemOpen
  \bibfield  {author} {\bibinfo {author} {\bibfnamefont {C.}~\bibnamefont
  {Gormezano}},\ }\href {\doibase 10.1088/0741-3335/41/12b/327} {\bibfield
  {journal} {\bibinfo  {journal} {Plasma Physics and Controlled Fusion}\
  }\textbf {\bibinfo {volume} {41}},\ \bibinfo {pages} {B367} (\bibinfo {year}
  {1999})}\BibitemShut {NoStop}%
\bibitem [{\citenamefont {Ida}\ \emph {et~al.}(2003)\citenamefont {Ida},
  \citenamefont {Shimozuma}, \citenamefont {Funaba}, \citenamefont {Narihara},
  \citenamefont {Kubo}, \citenamefont {Murakami}, \citenamefont {Wakasa},
  \citenamefont {Yokoyama}, \citenamefont {Takeiri}, \citenamefont {Watanabe},
  \citenamefont {Tanaka}, \citenamefont {Yoshinuma}, \citenamefont {Liang},\
  and\ \citenamefont {Ohyabu}}]{Ida_PRL_2003}%
  \BibitemOpen
  \bibfield  {author} {\bibinfo {author} {\bibfnamefont {K.}~\bibnamefont
  {Ida}}, \bibinfo {author} {\bibfnamefont {T.}~\bibnamefont {Shimozuma}},
  \bibinfo {author} {\bibfnamefont {H.}~\bibnamefont {Funaba}}, \bibinfo
  {author} {\bibfnamefont {K.}~\bibnamefont {Narihara}}, \bibinfo {author}
  {\bibfnamefont {S.}~\bibnamefont {Kubo}}, \bibinfo {author} {\bibfnamefont
  {S.}~\bibnamefont {Murakami}}, \bibinfo {author} {\bibfnamefont
  {A.}~\bibnamefont {Wakasa}}, \bibinfo {author} {\bibfnamefont
  {M.}~\bibnamefont {Yokoyama}}, \bibinfo {author} {\bibfnamefont
  {Y.}~\bibnamefont {Takeiri}}, \bibinfo {author} {\bibfnamefont {K.~Y.}\
  \bibnamefont {Watanabe}}, \bibinfo {author} {\bibfnamefont {K.}~\bibnamefont
  {Tanaka}}, \bibinfo {author} {\bibfnamefont {M.}~\bibnamefont {Yoshinuma}},
  \bibinfo {author} {\bibfnamefont {Y.}~\bibnamefont {Liang}}, \ and\ \bibinfo
  {author} {\bibfnamefont {N.}~\bibnamefont {Ohyabu}} (\bibinfo {collaboration}
  {LHD experimental group}),\ }\href {\doibase 10.1103/PhysRevLett.91.085003}
  {\bibfield  {journal} {\bibinfo  {journal} {Phys. Rev. Lett.}\ }\textbf
  {\bibinfo {volume} {91}},\ \bibinfo {pages} {085003} (\bibinfo {year}
  {2003})}\BibitemShut {NoStop}%
\bibitem [{\citenamefont {Fujisawa}\ \emph {et~al.}(1999)\citenamefont
  {Fujisawa}, \citenamefont {Iguchi}, \citenamefont {Minami}, \citenamefont
  {Yoshimura}, \citenamefont {Sanuki}, \citenamefont {Itoh}, \citenamefont
  {Lee}, \citenamefont {Tanaka}, \citenamefont {Yokoyama}, \citenamefont
  {Kojima}, \citenamefont {Itoh}, \citenamefont {Okamura}, \citenamefont
  {Akiyama}, \citenamefont {Ida}, \citenamefont {Isobe}, \citenamefont
  {Morita}, \citenamefont {Nishimura}, \citenamefont {Osakabe}, \citenamefont
  {Shimizu}, \citenamefont {Takahashi}, \citenamefont {Toi}, \citenamefont
  {Hamada}, \citenamefont {Matsuoka},\ and\ \citenamefont
  {Fujiwara}}]{Fujisawa_PRL_1999}%
  \BibitemOpen
  \bibfield  {author} {\bibinfo {author} {\bibfnamefont {A.}~\bibnamefont
  {Fujisawa}}, \bibinfo {author} {\bibfnamefont {H.}~\bibnamefont {Iguchi}},
  \bibinfo {author} {\bibfnamefont {T.}~\bibnamefont {Minami}}, \bibinfo
  {author} {\bibfnamefont {Y.}~\bibnamefont {Yoshimura}}, \bibinfo {author}
  {\bibfnamefont {H.}~\bibnamefont {Sanuki}}, \bibinfo {author} {\bibfnamefont
  {K.}~\bibnamefont {Itoh}}, \bibinfo {author} {\bibfnamefont {S.}~\bibnamefont
  {Lee}}, \bibinfo {author} {\bibfnamefont {K.}~\bibnamefont {Tanaka}},
  \bibinfo {author} {\bibfnamefont {M.}~\bibnamefont {Yokoyama}}, \bibinfo
  {author} {\bibfnamefont {M.}~\bibnamefont {Kojima}}, \bibinfo {author}
  {\bibfnamefont {S.-I.}\ \bibnamefont {Itoh}}, \bibinfo {author}
  {\bibfnamefont {S.}~\bibnamefont {Okamura}}, \bibinfo {author} {\bibfnamefont
  {R.}~\bibnamefont {Akiyama}}, \bibinfo {author} {\bibfnamefont
  {K.}~\bibnamefont {Ida}}, \bibinfo {author} {\bibfnamefont {M.}~\bibnamefont
  {Isobe}}, \bibinfo {author} {\bibfnamefont {S.}~\bibnamefont {Morita}},
  \bibinfo {author} {\bibfnamefont {S.}~\bibnamefont {Nishimura}}, \bibinfo
  {author} {\bibfnamefont {M.}~\bibnamefont {Osakabe}}, \bibinfo {author}
  {\bibfnamefont {A.}~\bibnamefont {Shimizu}}, \bibinfo {author} {\bibfnamefont
  {C.}~\bibnamefont {Takahashi}}, \bibinfo {author} {\bibfnamefont
  {K.}~\bibnamefont {Toi}}, \bibinfo {author} {\bibfnamefont {Y.}~\bibnamefont
  {Hamada}}, \bibinfo {author} {\bibfnamefont {K.}~\bibnamefont {Matsuoka}}, \
  and\ \bibinfo {author} {\bibfnamefont {M.}~\bibnamefont {Fujiwara}},\ }\href
  {\doibase 10.1103/PhysRevLett.82.2669} {\bibfield  {journal} {\bibinfo
  {journal} {Phys. Rev. Lett.}\ }\textbf {\bibinfo {volume} {82}},\ \bibinfo
  {pages} {2669} (\bibinfo {year} {1999})}\BibitemShut {NoStop}%
\bibitem [{\citenamefont {Castej{\'{o}}n}\ \emph {et~al.}(2002)\citenamefont
  {Castej{\'{o}}n}, \citenamefont {Tribaldos}, \citenamefont
  {Garc{\'{\i}}a-Cort{\'{e}}s}, \citenamefont {de~la Luna}, \citenamefont
  {Herranz}, \citenamefont {Pastor}, \citenamefont {Estrada},\ and\
  \citenamefont {{TJ-II Team}}}]{Castejon_NF_2002}%
  \BibitemOpen
  \bibfield  {author} {\bibinfo {author} {\bibfnamefont {F.}~\bibnamefont
  {Castej{\'{o}}n}}, \bibinfo {author} {\bibfnamefont {V.}~\bibnamefont
  {Tribaldos}}, \bibinfo {author} {\bibfnamefont {I.}~\bibnamefont
  {Garc{\'{\i}}a-Cort{\'{e}}s}}, \bibinfo {author} {\bibfnamefont
  {E.}~\bibnamefont {de~la Luna}}, \bibinfo {author} {\bibfnamefont
  {J.}~\bibnamefont {Herranz}}, \bibinfo {author} {\bibfnamefont
  {I.}~\bibnamefont {Pastor}}, \bibinfo {author} {\bibfnamefont
  {T.}~\bibnamefont {Estrada}}, \ and\ \bibinfo {author} {\bibnamefont {{TJ-II
  Team}}},\ }\href {\doibase 10.1088/0029-5515/42/3/307} {\bibfield  {journal}
  {\bibinfo  {journal} {Nuclear Fusion}\ }\textbf {\bibinfo {volume} {42}},\
  \bibinfo {pages} {271} (\bibinfo {year} {2002})}\BibitemShut {NoStop}%
\bibitem [{\citenamefont {Stroth}\ \emph {et~al.}(2001)\citenamefont {Stroth},
  \citenamefont {Itoh}, \citenamefont {Itoh}, \citenamefont {Hartfuss},
  \citenamefont {Laqua}, \citenamefont {the ECRH~team},\ and\ \citenamefont
  {the W7-AS~team}}]{Stroth_PRL_2001}%
  \BibitemOpen
  \bibfield  {author} {\bibinfo {author} {\bibfnamefont {U.}~\bibnamefont
  {Stroth}}, \bibinfo {author} {\bibfnamefont {K.}~\bibnamefont {Itoh}},
  \bibinfo {author} {\bibfnamefont {S.-I.}\ \bibnamefont {Itoh}}, \bibinfo
  {author} {\bibfnamefont {H.}~\bibnamefont {Hartfuss}}, \bibinfo {author}
  {\bibfnamefont {H.}~\bibnamefont {Laqua}}, \bibinfo {author} {\bibnamefont
  {the ECRH~team}}, \ and\ \bibinfo {author} {\bibnamefont {the W7-AS~team}},\
  }\href {\doibase 10.1103/PhysRevLett.86.5910} {\bibfield  {journal} {\bibinfo
   {journal} {Phys. Rev. Lett.}\ }\textbf {\bibinfo {volume} {86}},\ \bibinfo
  {pages} {5910} (\bibinfo {year} {2001})}\BibitemShut {NoStop}%
\bibitem [{\citenamefont {Joffrin}\ \emph
  {et~al.}(2002{\natexlab{a}})\citenamefont {Joffrin}, \citenamefont {Challis},
  \citenamefont {Hender}, \citenamefont {Howell},\ and\ \citenamefont
  {Huysmans}}]{Joffrin_NF_2002}%
  \BibitemOpen
  \bibfield  {author} {\bibinfo {author} {\bibfnamefont {E.}~\bibnamefont
  {Joffrin}}, \bibinfo {author} {\bibfnamefont {C.}~\bibnamefont {Challis}},
  \bibinfo {author} {\bibfnamefont {T.}~\bibnamefont {Hender}}, \bibinfo
  {author} {\bibfnamefont {D.}~\bibnamefont {Howell}}, \ and\ \bibinfo {author}
  {\bibfnamefont {G.}~\bibnamefont {Huysmans}},\ }\href {\doibase
  10.1088/0029-5515/42/3/302} {\bibfield  {journal} {\bibinfo  {journal}
  {Nuclear Fusion}\ }\textbf {\bibinfo {volume} {42}},\ \bibinfo {pages} {235}
  (\bibinfo {year} {2002}{\natexlab{a}})}\BibitemShut {NoStop}%
\bibitem [{\citenamefont {Joffrin}\ \emph
  {et~al.}(2002{\natexlab{b}})\citenamefont {Joffrin}, \citenamefont {Gorini},
  \citenamefont {Challis}, \citenamefont {Hawkes}, \citenamefont {Hender},
  \citenamefont {Howell}, \citenamefont {Maget}, \citenamefont {Mantica},
  \citenamefont {Mazon}, \citenamefont {Sharapov}, \citenamefont {Tresset},\
  and\ \citenamefont {contributors to~the
  EFDA-JET~Workprogramme}}]{Joffrin_PPCF_2002}%
  \BibitemOpen
  \bibfield  {author} {\bibinfo {author} {\bibfnamefont {E.}~\bibnamefont
  {Joffrin}}, \bibinfo {author} {\bibfnamefont {G.}~\bibnamefont {Gorini}},
  \bibinfo {author} {\bibfnamefont {C.~D.}\ \bibnamefont {Challis}}, \bibinfo
  {author} {\bibfnamefont {N.~C.}\ \bibnamefont {Hawkes}}, \bibinfo {author}
  {\bibfnamefont {T.~C.}\ \bibnamefont {Hender}}, \bibinfo {author}
  {\bibfnamefont {D.~F.}\ \bibnamefont {Howell}}, \bibinfo {author}
  {\bibfnamefont {P.}~\bibnamefont {Maget}}, \bibinfo {author} {\bibfnamefont
  {P.}~\bibnamefont {Mantica}}, \bibinfo {author} {\bibfnamefont
  {D.}~\bibnamefont {Mazon}}, \bibinfo {author} {\bibfnamefont {S.~E.}\
  \bibnamefont {Sharapov}}, \bibinfo {author} {\bibfnamefont {G.}~\bibnamefont
  {Tresset}}, \ and\ \bibinfo {author} {\bibnamefont {contributors to~the
  EFDA-JET~Workprogramme}},\ }\href {\doibase 10.1088/0741-3335/44/8/320}
  {\bibfield  {journal} {\bibinfo  {journal} {Plasma Physics and Controlled
  Fusion}\ }\textbf {\bibinfo {volume} {44}},\ \bibinfo {pages} {1739}
  (\bibinfo {year} {2002}{\natexlab{b}})}\BibitemShut {NoStop}%
\bibitem [{\citenamefont {Garbet}\ \emph {et~al.}(2001)\citenamefont {Garbet},
  \citenamefont {Bourdelle}, \citenamefont {Hoang}, \citenamefont {Maget},
  \citenamefont {Benkadda}, \citenamefont {Beyer}, \citenamefont {Figarella},
  \citenamefont {Voitsekovitch}, \citenamefont {Agullo},\ and\ \citenamefont
  {Bian}}]{Garbet_PoP_2001}%
  \BibitemOpen
  \bibfield  {author} {\bibinfo {author} {\bibfnamefont {X.}~\bibnamefont
  {Garbet}}, \bibinfo {author} {\bibfnamefont {C.}~\bibnamefont {Bourdelle}},
  \bibinfo {author} {\bibfnamefont {G.~T.}\ \bibnamefont {Hoang}}, \bibinfo
  {author} {\bibfnamefont {P.}~\bibnamefont {Maget}}, \bibinfo {author}
  {\bibfnamefont {S.}~\bibnamefont {Benkadda}}, \bibinfo {author}
  {\bibfnamefont {P.}~\bibnamefont {Beyer}}, \bibinfo {author} {\bibfnamefont
  {C.}~\bibnamefont {Figarella}}, \bibinfo {author} {\bibfnamefont
  {I.}~\bibnamefont {Voitsekovitch}}, \bibinfo {author} {\bibfnamefont
  {O.}~\bibnamefont {Agullo}}, \ and\ \bibinfo {author} {\bibfnamefont
  {N.}~\bibnamefont {Bian}},\ }\href {\doibase 10.1063/1.1367320} {\bibfield
  {journal} {\bibinfo  {journal} {Physics of Plasmas}\ }\textbf {\bibinfo
  {volume} {8}},\ \bibinfo {pages} {2793} (\bibinfo {year} {2001})}\BibitemShut
  {NoStop}%
\bibitem [{\citenamefont {Joffrin}\ \emph
  {et~al.}(2002{\natexlab{c}})\citenamefont {Joffrin}, \citenamefont {Wolf},
  \citenamefont {Alper}, \citenamefont {Baranov}, \citenamefont {Challis},
  \citenamefont {de~Baar}, \citenamefont {Giroud}, \citenamefont {Gowers},
  \citenamefont {Hawkes}, \citenamefont {Hender}, \citenamefont {Marachek},
  \citenamefont {Mazon}, \citenamefont {Parail}, \citenamefont {Peeters},
  \citenamefont {Zastrow},\ and\ \citenamefont {contributors to~the
  EFDA-JET~Workprogramme}}]{Joffrin_2002_q1}%
  \BibitemOpen
  \bibfield  {author} {\bibinfo {author} {\bibfnamefont {E.}~\bibnamefont
  {Joffrin}}, \bibinfo {author} {\bibfnamefont {R.}~\bibnamefont {Wolf}},
  \bibinfo {author} {\bibfnamefont {B.}~\bibnamefont {Alper}}, \bibinfo
  {author} {\bibfnamefont {Y.}~\bibnamefont {Baranov}}, \bibinfo {author}
  {\bibfnamefont {C.~D.}\ \bibnamefont {Challis}}, \bibinfo {author}
  {\bibfnamefont {M.}~\bibnamefont {de~Baar}}, \bibinfo {author} {\bibfnamefont
  {C.}~\bibnamefont {Giroud}}, \bibinfo {author} {\bibfnamefont {C.~W.}\
  \bibnamefont {Gowers}}, \bibinfo {author} {\bibfnamefont {N.~C.}\
  \bibnamefont {Hawkes}}, \bibinfo {author} {\bibfnamefont {T.~C.}\
  \bibnamefont {Hender}}, \bibinfo {author} {\bibfnamefont {M.}~\bibnamefont
  {Marachek}}, \bibinfo {author} {\bibfnamefont {D.}~\bibnamefont {Mazon}},
  \bibinfo {author} {\bibfnamefont {V.}~\bibnamefont {Parail}}, \bibinfo
  {author} {\bibfnamefont {A.}~\bibnamefont {Peeters}}, \bibinfo {author}
  {\bibfnamefont {K.-D.}\ \bibnamefont {Zastrow}}, \ and\ \bibinfo {author}
  {\bibnamefont {contributors to~the EFDA-JET~Workprogramme}},\ }\href
  {\doibase 10.1088/0741-3335/44/7/310} {\bibfield  {journal} {\bibinfo
  {journal} {Plasma Physics and Controlled Fusion}\ }\textbf {\bibinfo {volume}
  {44}},\ \bibinfo {pages} {1203} (\bibinfo {year}
  {2002}{\natexlab{c}})}\BibitemShut {NoStop}%
\bibitem [{\citenamefont {Joffrin}\ \emph {et~al.}(2003)\citenamefont
  {Joffrin}, \citenamefont {Challis}, \citenamefont {Conway}, \citenamefont
  {Garbet}, \citenamefont {Gude}, \citenamefont {G\"unter}, \citenamefont
  {Hawkes}, \citenamefont {Hender}, \citenamefont {Howell}, \citenamefont
  {Huysmans}, \citenamefont {Lazzaro}, \citenamefont {Maget}, \citenamefont
  {Marachek}, \citenamefont {Peeters}, \citenamefont {Pinches}, \citenamefont
  {Sharapov},\ and\ \citenamefont {{JET-EFDA contributors}}}]{Joffrin_NF_2003}%
  \BibitemOpen
  \bibfield  {author} {\bibinfo {author} {\bibfnamefont {E.}~\bibnamefont
  {Joffrin}}, \bibinfo {author} {\bibfnamefont {C.}~\bibnamefont {Challis}},
  \bibinfo {author} {\bibfnamefont {G.}~\bibnamefont {Conway}}, \bibinfo
  {author} {\bibfnamefont {X.}~\bibnamefont {Garbet}}, \bibinfo {author}
  {\bibfnamefont {A.}~\bibnamefont {Gude}}, \bibinfo {author} {\bibfnamefont
  {S.}~\bibnamefont {G\"unter}}, \bibinfo {author} {\bibfnamefont
  {N.}~\bibnamefont {Hawkes}}, \bibinfo {author} {\bibfnamefont
  {T.}~\bibnamefont {Hender}}, \bibinfo {author} {\bibfnamefont
  {D.}~\bibnamefont {Howell}}, \bibinfo {author} {\bibfnamefont
  {G.}~\bibnamefont {Huysmans}}, \bibinfo {author} {\bibfnamefont
  {E.}~\bibnamefont {Lazzaro}}, \bibinfo {author} {\bibfnamefont
  {P.}~\bibnamefont {Maget}}, \bibinfo {author} {\bibfnamefont
  {M.}~\bibnamefont {Marachek}}, \bibinfo {author} {\bibfnamefont
  {A.}~\bibnamefont {Peeters}}, \bibinfo {author} {\bibfnamefont
  {S.}~\bibnamefont {Pinches}}, \bibinfo {author} {\bibfnamefont
  {S.}~\bibnamefont {Sharapov}}, \ and\ \bibinfo {author} {\bibnamefont
  {{JET-EFDA contributors}}},\ }\href {\doibase 10.1088/0029-5515/43/10/018}
  {\bibfield  {journal} {\bibinfo  {journal} {Nuclear Fusion}\ }\textbf
  {\bibinfo {volume} {43}},\ \bibinfo {pages} {1167} (\bibinfo {year}
  {2003})}\BibitemShut {NoStop}%
\bibitem [{\citenamefont {Wong}\ \emph {et~al.}(2004)\citenamefont {Wong},
  \citenamefont {Heidbrink}, \citenamefont {Ruskov}, \citenamefont {Petty},
  \citenamefont {Greenfield}, \citenamefont {Nazikian},\ and\ \citenamefont
  {Budny}}]{Wong_NF_2004}%
  \BibitemOpen
  \bibfield  {author} {\bibinfo {author} {\bibfnamefont {K.}~\bibnamefont
  {Wong}}, \bibinfo {author} {\bibfnamefont {W.}~\bibnamefont {Heidbrink}},
  \bibinfo {author} {\bibfnamefont {E.}~\bibnamefont {Ruskov}}, \bibinfo
  {author} {\bibfnamefont {C.}~\bibnamefont {Petty}}, \bibinfo {author}
  {\bibfnamefont {C.}~\bibnamefont {Greenfield}}, \bibinfo {author}
  {\bibfnamefont {R.}~\bibnamefont {Nazikian}}, \ and\ \bibinfo {author}
  {\bibfnamefont {R.}~\bibnamefont {Budny}},\ }\href {\doibase
  10.1088/0029-5515/45/1/004} {\bibfield  {journal} {\bibinfo  {journal}
  {Nuclear Fusion}\ }\textbf {\bibinfo {volume} {45}},\ \bibinfo {pages} {30}
  (\bibinfo {year} {2004})}\BibitemShut {NoStop}%
\bibitem [{\citenamefont {Romanelli}\ \emph {et~al.}(2010)\citenamefont
  {Romanelli}, \citenamefont {Zocco}, \citenamefont {Crisanti},\ and\
  \citenamefont {{JET-EFDA Contributors}}}]{Romanelli_PPCF_2010}%
  \BibitemOpen
  \bibfield  {author} {\bibinfo {author} {\bibfnamefont {M.}~\bibnamefont
  {Romanelli}}, \bibinfo {author} {\bibfnamefont {A.}~\bibnamefont {Zocco}},
  \bibinfo {author} {\bibfnamefont {F.}~\bibnamefont {Crisanti}}, \ and\
  \bibinfo {author} {\bibnamefont {{JET-EFDA Contributors}}},\ }\href {\doibase
  10.1088/0741-3335/52/4/045007} {\bibfield  {journal} {\bibinfo  {journal}
  {Plasma Physics and Controlled Fusion}\ }\textbf {\bibinfo {volume} {52}},\
  \bibinfo {pages} {045007} (\bibinfo {year} {2010})}\BibitemShut {NoStop}%
\bibitem [{\citenamefont {Dif-Pradalier}\ \emph {et~al.}(2015)\citenamefont
  {Dif-Pradalier}, \citenamefont {Hornung}, \citenamefont {Ghendrih},
  \citenamefont {Sarazin}, \citenamefont {Clairet}, \citenamefont {Vermare},
  \citenamefont {Diamond}, \citenamefont {Abiteboul}, \citenamefont
  {Cartier-Michaud}, \citenamefont {Ehrlacher}, \citenamefont {Est\`eve},
  \citenamefont {Garbet}, \citenamefont {Grandgirard}, \citenamefont
  {G\"urcan}, \citenamefont {Hennequin}, \citenamefont {Kosuga}, \citenamefont
  {Latu}, \citenamefont {Maget}, \citenamefont {Morel}, \citenamefont
  {Norscini}, \citenamefont {Sabot},\ and\ \citenamefont
  {Storelli}}]{Dif-Pradalier_PRL_2015}%
  \BibitemOpen
  \bibfield  {author} {\bibinfo {author} {\bibfnamefont {G.}~\bibnamefont
  {Dif-Pradalier}}, \bibinfo {author} {\bibfnamefont {G.}~\bibnamefont
  {Hornung}}, \bibinfo {author} {\bibfnamefont {P.}~\bibnamefont {Ghendrih}},
  \bibinfo {author} {\bibfnamefont {Y.}~\bibnamefont {Sarazin}}, \bibinfo
  {author} {\bibfnamefont {F.}~\bibnamefont {Clairet}}, \bibinfo {author}
  {\bibfnamefont {L.}~\bibnamefont {Vermare}}, \bibinfo {author} {\bibfnamefont
  {P.~H.}\ \bibnamefont {Diamond}}, \bibinfo {author} {\bibfnamefont
  {J.}~\bibnamefont {Abiteboul}}, \bibinfo {author} {\bibfnamefont
  {T.}~\bibnamefont {Cartier-Michaud}}, \bibinfo {author} {\bibfnamefont
  {C.}~\bibnamefont {Ehrlacher}}, \bibinfo {author} {\bibfnamefont
  {D.}~\bibnamefont {Est\`eve}}, \bibinfo {author} {\bibfnamefont
  {X.}~\bibnamefont {Garbet}}, \bibinfo {author} {\bibfnamefont
  {V.}~\bibnamefont {Grandgirard}}, \bibinfo {author} {\bibfnamefont {O.~D.}\
  \bibnamefont {G\"urcan}}, \bibinfo {author} {\bibfnamefont {P.}~\bibnamefont
  {Hennequin}}, \bibinfo {author} {\bibfnamefont {Y.}~\bibnamefont {Kosuga}},
  \bibinfo {author} {\bibfnamefont {G.}~\bibnamefont {Latu}}, \bibinfo {author}
  {\bibfnamefont {P.}~\bibnamefont {Maget}}, \bibinfo {author} {\bibfnamefont
  {P.}~\bibnamefont {Morel}}, \bibinfo {author} {\bibfnamefont
  {C.}~\bibnamefont {Norscini}}, \bibinfo {author} {\bibfnamefont
  {R.}~\bibnamefont {Sabot}}, \ and\ \bibinfo {author} {\bibfnamefont
  {A.}~\bibnamefont {Storelli}},\ }\href {\doibase
  10.1103/PhysRevLett.114.085004} {\bibfield  {journal} {\bibinfo  {journal}
  {Phys. Rev. Lett.}\ }\textbf {\bibinfo {volume} {114}},\ \bibinfo {pages}
  {085004} (\bibinfo {year} {2015})}\BibitemShut {NoStop}%
\bibitem [{\citenamefont {Strugarek}\ \emph {et~al.}(2013)\citenamefont
  {Strugarek}, \citenamefont {Sarazin}, \citenamefont {Zarzoso}, \citenamefont
  {Abiteboul}, \citenamefont {Brun}, \citenamefont {Cartier-Michaud},
  \citenamefont {Dif-Pradalier}, \citenamefont {Garbet}, \citenamefont
  {Ghendrih}, \citenamefont {Grandgirard}, \citenamefont {Latu}, \citenamefont
  {Passeron},\ and\ \citenamefont {Thomine}}]{Strugarek_PPCF_2013}%
  \BibitemOpen
  \bibfield  {author} {\bibinfo {author} {\bibfnamefont {A.}~\bibnamefont
  {Strugarek}}, \bibinfo {author} {\bibfnamefont {Y.}~\bibnamefont {Sarazin}},
  \bibinfo {author} {\bibfnamefont {D.}~\bibnamefont {Zarzoso}}, \bibinfo
  {author} {\bibfnamefont {J.}~\bibnamefont {Abiteboul}}, \bibinfo {author}
  {\bibfnamefont {A.~S.}\ \bibnamefont {Brun}}, \bibinfo {author}
  {\bibfnamefont {T.}~\bibnamefont {Cartier-Michaud}}, \bibinfo {author}
  {\bibfnamefont {G.}~\bibnamefont {Dif-Pradalier}}, \bibinfo {author}
  {\bibfnamefont {X.}~\bibnamefont {Garbet}}, \bibinfo {author} {\bibfnamefont
  {P.}~\bibnamefont {Ghendrih}}, \bibinfo {author} {\bibfnamefont
  {V.}~\bibnamefont {Grandgirard}}, \bibinfo {author} {\bibfnamefont
  {G.}~\bibnamefont {Latu}}, \bibinfo {author} {\bibfnamefont {C.}~\bibnamefont
  {Passeron}}, \ and\ \bibinfo {author} {\bibfnamefont {O.}~\bibnamefont
  {Thomine}},\ }\href {\doibase 10.1088/0741-3335/55/7/074013} {\bibfield
  {journal} {\bibinfo  {journal} {Plasma Physics and Controlled Fusion}\
  }\textbf {\bibinfo {volume} {55}},\ \bibinfo {pages} {074013} (\bibinfo
  {year} {2013})}\BibitemShut {NoStop}%
\bibitem [{\citenamefont {{Jenko}}\ \emph {et~al.}(2000)\citenamefont
  {{Jenko}}, \citenamefont {{Dorland}}, \citenamefont {{Kotschenreuther}},\
  and\ \citenamefont {{Rogers}}}]{Jenko_PoP2000}%
  \BibitemOpen
  \bibfield  {author} {\bibinfo {author} {\bibfnamefont {F.}~\bibnamefont
  {{Jenko}}}, \bibinfo {author} {\bibfnamefont {W.}~\bibnamefont {{Dorland}}},
  \bibinfo {author} {\bibfnamefont {M.}~\bibnamefont {{Kotschenreuther}}}, \
  and\ \bibinfo {author} {\bibfnamefont {B.~N.}\ \bibnamefont {{Rogers}}},\
  }\href {\doibase 10.1063/1.874014} {\bibfield  {journal} {\bibinfo  {journal}
  {Phys.~Plasmas}\ }\textbf {\bibinfo {volume} {7}},\ \bibinfo {pages} {1904}
  (\bibinfo {year} {2000})}\BibitemShut {NoStop}%
\bibitem [{\citenamefont {{G{\"o}rler}}\ \emph {et~al.}(2011)\citenamefont
  {{G{\"o}rler}}, \citenamefont {{Lapillonne}}, \citenamefont {{Brunner}},
  \citenamefont {{Dannert}}, \citenamefont {{Jenko}}, \citenamefont {{Merz}},\
  and\ \citenamefont {{Told}}}]{Goerler_JCP2011}%
  \BibitemOpen
  \bibfield  {author} {\bibinfo {author} {\bibfnamefont {T.}~\bibnamefont
  {{G{\"o}rler}}}, \bibinfo {author} {\bibfnamefont {X.}~\bibnamefont
  {{Lapillonne}}}, \bibinfo {author} {\bibfnamefont {S.}~\bibnamefont
  {{Brunner}}}, \bibinfo {author} {\bibfnamefont {T.}~\bibnamefont
  {{Dannert}}}, \bibinfo {author} {\bibfnamefont {F.}~\bibnamefont {{Jenko}}},
  \bibinfo {author} {\bibfnamefont {F.}~\bibnamefont {{Merz}}}, \ and\ \bibinfo
  {author} {\bibfnamefont {D.}~\bibnamefont {{Told}}},\ }\href {\doibase
  10.1016/j.jcp.2011.05.034} {\bibfield  {journal} {\bibinfo  {journal}
  {Journal of Computational Physics}\ }\textbf {\bibinfo {volume} {230}},\
  \bibinfo {pages} {7053} (\bibinfo {year} {2011})}\BibitemShut {NoStop}%
\bibitem [{\citenamefont {Di~Siena}\ \emph
  {et~al.}(2021{\natexlab{a}})\citenamefont {Di~Siena}, \citenamefont {Bilato},
  \citenamefont {G\"orler}, \citenamefont {Ba\~n\'on Navarro}, \citenamefont
  {Poli}, \citenamefont {Bobkov}, \citenamefont {Jarema}, \citenamefont
  {Fable}, \citenamefont {Angioni}, \citenamefont {Kazakov}, \citenamefont
  {Ochoukov}, \citenamefont {Schneider}, \citenamefont {Weiland}, \citenamefont
  {Jenko},\ and\ \citenamefont {the ASDEX Upgrade~Team}}]{DiSiena_PRL_2021}%
  \BibitemOpen
  \bibfield  {author} {\bibinfo {author} {\bibfnamefont {A.}~\bibnamefont
  {Di~Siena}}, \bibinfo {author} {\bibfnamefont {R.}~\bibnamefont {Bilato}},
  \bibinfo {author} {\bibfnamefont {T.}~\bibnamefont {G\"orler}}, \bibinfo
  {author} {\bibfnamefont {A.}~\bibnamefont {Ba\~n\'on Navarro}}, \bibinfo
  {author} {\bibfnamefont {E.}~\bibnamefont {Poli}}, \bibinfo {author}
  {\bibfnamefont {V.}~\bibnamefont {Bobkov}}, \bibinfo {author} {\bibfnamefont
  {D.}~\bibnamefont {Jarema}}, \bibinfo {author} {\bibfnamefont
  {E.}~\bibnamefont {Fable}}, \bibinfo {author} {\bibfnamefont
  {C.}~\bibnamefont {Angioni}}, \bibinfo {author} {\bibfnamefont {Y.~O.}\
  \bibnamefont {Kazakov}}, \bibinfo {author} {\bibfnamefont {R.}~\bibnamefont
  {Ochoukov}}, \bibinfo {author} {\bibfnamefont {P.}~\bibnamefont {Schneider}},
  \bibinfo {author} {\bibfnamefont {M.}~\bibnamefont {Weiland}}, \bibinfo
  {author} {\bibfnamefont {F.}~\bibnamefont {Jenko}}, \ and\ \bibinfo {author}
  {\bibnamefont {the ASDEX Upgrade~Team}},\ }\href {\doibase
  10.1103/PhysRevLett.127.025002} {\bibfield  {journal} {\bibinfo  {journal}
  {Phys. Rev. Lett.}\ }\textbf {\bibinfo {volume} {127}},\ \bibinfo {pages}
  {025002} (\bibinfo {year} {2021}{\natexlab{a}})}\BibitemShut {NoStop}%
\bibitem [{\citenamefont {{Di Siena}}\ \emph {et~al.}(2018)\citenamefont {{Di
  Siena}}, \citenamefont {{G{\"o}rler}}, \citenamefont {{Doerk}}, \citenamefont
  {{Poli}},\ and\ \citenamefont {{Bilato}}}]{DiSiena_NF2018}%
  \BibitemOpen
  \bibfield  {author} {\bibinfo {author} {\bibfnamefont {A.}~\bibnamefont {{Di
  Siena}}}, \bibinfo {author} {\bibfnamefont {T.}~\bibnamefont {{G{\"o}rler}}},
  \bibinfo {author} {\bibfnamefont {H.}~\bibnamefont {{Doerk}}}, \bibinfo
  {author} {\bibfnamefont {E.}~\bibnamefont {{Poli}}}, \ and\ \bibinfo {author}
  {\bibfnamefont {R.}~\bibnamefont {{Bilato}}},\ }\href {\doibase
  10.1088/1741-4326/aaaf26} {\bibfield  {journal} {\bibinfo  {journal}
  {Nucl.~Fusion}\ }\textbf {\bibinfo {volume} {58}},\ \bibinfo {pages} {054002}
  (\bibinfo {year} {2018})}\BibitemShut {NoStop}%
\bibitem [{\citenamefont {Di~Siena}\ \emph {et~al.}(2019)\citenamefont
  {Di~Siena}, \citenamefont {G\"orler}, \citenamefont {Poli}, \citenamefont
  {Bilato}, \citenamefont {Doerk},\ and\ \citenamefont
  {Zocco}}]{DiSiena_PoP2019}%
  \BibitemOpen
  \bibfield  {author} {\bibinfo {author} {\bibfnamefont {A.}~\bibnamefont
  {Di~Siena}}, \bibinfo {author} {\bibfnamefont {T.}~\bibnamefont {G\"orler}},
  \bibinfo {author} {\bibfnamefont {E.}~\bibnamefont {Poli}}, \bibinfo {author}
  {\bibfnamefont {R.}~\bibnamefont {Bilato}}, \bibinfo {author} {\bibfnamefont
  {H.}~\bibnamefont {Doerk}}, \ and\ \bibinfo {author} {\bibfnamefont
  {A.}~\bibnamefont {Zocco}},\ }\href {\doibase 10.1063/1.5087203} {\bibfield
  {journal} {\bibinfo  {journal} {Physics of Plasmas}\ }\textbf {\bibinfo
  {volume} {26}},\ \bibinfo {pages} {052504} (\bibinfo {year}
  {2019})}\BibitemShut {NoStop}%
\bibitem [{\citenamefont {Di~Siena}\ \emph {et~al.}(2020)\citenamefont
  {Di~Siena}, \citenamefont {Ba\~n\'on Navarro},\ and\ \citenamefont
  {Jenko}}]{W7X}%
  \BibitemOpen
  \bibfield  {author} {\bibinfo {author} {\bibfnamefont {A.}~\bibnamefont
  {Di~Siena}}, \bibinfo {author} {\bibfnamefont {A.}~\bibnamefont {Ba\~n\'on
  Navarro}}, \ and\ \bibinfo {author} {\bibfnamefont {F.}~\bibnamefont
  {Jenko}},\ }\href {\doibase 10.1103/PhysRevLett.125.105002} {\bibfield
  {journal} {\bibinfo  {journal} {Phys. Rev. Lett.}\ }\textbf {\bibinfo
  {volume} {125}},\ \bibinfo {pages} {105002} (\bibinfo {year}
  {2020})}\BibitemShut {NoStop}%
\bibitem [{\citenamefont {Bonanomi}\ \emph {et~al.}(2018)\citenamefont
  {Bonanomi}, \citenamefont {Mantica}, \citenamefont {Siena}, \citenamefont
  {Delabie}, \citenamefont {Giroud}, \citenamefont {Johnson}, \citenamefont
  {Lerche}, \citenamefont {Menmuir}, \citenamefont {Tsalas},\ and\
  \citenamefont {and}}]{Bonanomi_2018}%
  \BibitemOpen
  \bibfield  {author} {\bibinfo {author} {\bibfnamefont {N.}~\bibnamefont
  {Bonanomi}}, \bibinfo {author} {\bibfnamefont {P.}~\bibnamefont {Mantica}},
  \bibinfo {author} {\bibfnamefont {A.~D.}\ \bibnamefont {Siena}}, \bibinfo
  {author} {\bibfnamefont {E.}~\bibnamefont {Delabie}}, \bibinfo {author}
  {\bibfnamefont {C.}~\bibnamefont {Giroud}}, \bibinfo {author} {\bibfnamefont
  {T.}~\bibnamefont {Johnson}}, \bibinfo {author} {\bibfnamefont
  {E.}~\bibnamefont {Lerche}}, \bibinfo {author} {\bibfnamefont
  {S.}~\bibnamefont {Menmuir}}, \bibinfo {author} {\bibfnamefont
  {M.}~\bibnamefont {Tsalas}}, \ and\ \bibinfo {author} {\bibfnamefont
  {D.~V.~E.}\ \bibnamefont {and}},\ }\href {\doibase 10.1088/1741-4326/aab733}
  {\bibfield  {journal} {\bibinfo  {journal} {Nuclear Fusion}\ }\textbf
  {\bibinfo {volume} {58}},\ \bibinfo {pages} {056025} (\bibinfo {year}
  {2018})}\BibitemShut {NoStop}%
\bibitem [{\citenamefont {{Di Siena}}\ \emph {et~al.}(2016)\citenamefont {{Di
  Siena}}, \citenamefont {{G{\"o}rler}}, \citenamefont {{Doerk}}, \citenamefont
  {{Citrin}}, \citenamefont {{Johnson}}, \citenamefont {{Schneider}},
  \citenamefont {{Poli}},\ and\ \citenamefont {{JET
  Contributors}}}]{DiSiena_2016}%
  \BibitemOpen
  \bibfield  {author} {\bibinfo {author} {\bibfnamefont {A.}~\bibnamefont {{Di
  Siena}}}, \bibinfo {author} {\bibfnamefont {T.}~\bibnamefont {{G{\"o}rler}}},
  \bibinfo {author} {\bibfnamefont {H.}~\bibnamefont {{Doerk}}}, \bibinfo
  {author} {\bibfnamefont {J.}~\bibnamefont {{Citrin}}}, \bibinfo {author}
  {\bibfnamefont {T.}~\bibnamefont {{Johnson}}}, \bibinfo {author}
  {\bibfnamefont {M.}~\bibnamefont {{Schneider}}}, \bibinfo {author}
  {\bibfnamefont {E.}~\bibnamefont {{Poli}}}, \ and\ \bibinfo {author}
  {\bibnamefont {{JET Contributors}}},\ }\href {\doibase
  10.1088/1742-6596/775/1/012003} {\bibfield  {journal} {\bibinfo  {journal}
  {J.~Phys.~Conf.~Ser.}\ }\textbf {\bibinfo {volume} {775}},\ \bibinfo {pages}
  {012003} (\bibinfo {year} {2016})}\BibitemShut {NoStop}%
\bibitem [{\citenamefont {Di~Siena}\ \emph {et~al.}(2018)\citenamefont
  {Di~Siena}, \citenamefont {G\"orler}, \citenamefont {Doerk}, \citenamefont
  {Bilato}, \citenamefont {Citrin}, \citenamefont {Johnson}, \citenamefont
  {Schneider}, \citenamefont {Poli},\ and\ \citenamefont {{JET
  Contributors}}}]{DiSiena_PoP_2018}%
  \BibitemOpen
  \bibfield  {author} {\bibinfo {author} {\bibfnamefont {A.}~\bibnamefont
  {Di~Siena}}, \bibinfo {author} {\bibfnamefont {T.}~\bibnamefont {G\"orler}},
  \bibinfo {author} {\bibfnamefont {H.}~\bibnamefont {Doerk}}, \bibinfo
  {author} {\bibfnamefont {R.}~\bibnamefont {Bilato}}, \bibinfo {author}
  {\bibfnamefont {J.}~\bibnamefont {Citrin}}, \bibinfo {author} {\bibfnamefont
  {T.}~\bibnamefont {Johnson}}, \bibinfo {author} {\bibfnamefont
  {M.}~\bibnamefont {Schneider}}, \bibinfo {author} {\bibfnamefont
  {E.}~\bibnamefont {Poli}}, \ and\ \bibinfo {author} {\bibnamefont {{JET
  Contributors}}},\ }\href {\doibase 10.1063/1.5020122} {\bibfield  {journal}
  {\bibinfo  {journal} {Physics of Plasmas}\ }\textbf {\bibinfo {volume}
  {25}},\ \bibinfo {pages} {042304} (\bibinfo {year} {2018})}\BibitemShut
  {NoStop}%
\bibitem [{\citenamefont {Siena}\ \emph {et~al.}(2018)\citenamefont {Siena},
  \citenamefont {Biancalani}, \citenamefont {G\"orler}, \citenamefont {Doerk},
  \citenamefont {Novikau}, \citenamefont {Lauber}, \citenamefont {Bottino},
  \citenamefont {Poli},\ and\ \citenamefont {{The ASDEX Upgrade
  Team}}}]{DiSiena_NF_2018_egam}%
  \BibitemOpen
  \bibfield  {author} {\bibinfo {author} {\bibfnamefont {A.~D.}\ \bibnamefont
  {Siena}}, \bibinfo {author} {\bibfnamefont {A.}~\bibnamefont {Biancalani}},
  \bibinfo {author} {\bibfnamefont {T.}~\bibnamefont {G\"orler}}, \bibinfo
  {author} {\bibfnamefont {H.}~\bibnamefont {Doerk}}, \bibinfo {author}
  {\bibfnamefont {I.}~\bibnamefont {Novikau}}, \bibinfo {author} {\bibfnamefont
  {P.}~\bibnamefont {Lauber}}, \bibinfo {author} {\bibfnamefont
  {A.}~\bibnamefont {Bottino}}, \bibinfo {author} {\bibfnamefont
  {E.}~\bibnamefont {Poli}}, \ and\ \bibinfo {author} {\bibnamefont {{The ASDEX
  Upgrade Team}}},\ }\href {\doibase 10.1088/1741-4326/aad51d} {\bibfield
  {journal} {\bibinfo  {journal} {Nuclear Fusion}\ }\textbf {\bibinfo {volume}
  {58}},\ \bibinfo {pages} {106014} (\bibinfo {year} {2018})}\BibitemShut
  {NoStop}%
\bibitem [{\citenamefont {Crandall}\ \emph {et~al.}(2020)\citenamefont
  {Crandall}, \citenamefont {Jarema}, \citenamefont {Doerk}, \citenamefont
  {Pan}, \citenamefont {Merlo}, \citenamefont {G\"orler}, \citenamefont
  {{Ba\~n\'on Navarro}}, \citenamefont {Told}, \citenamefont {Maurer},\ and\
  \citenamefont {Jenko}}]{Crandall_CPC_2020}%
  \BibitemOpen
  \bibfield  {author} {\bibinfo {author} {\bibfnamefont {P.}~\bibnamefont
  {Crandall}}, \bibinfo {author} {\bibfnamefont {D.}~\bibnamefont {Jarema}},
  \bibinfo {author} {\bibfnamefont {H.}~\bibnamefont {Doerk}}, \bibinfo
  {author} {\bibfnamefont {Q.}~\bibnamefont {Pan}}, \bibinfo {author}
  {\bibfnamefont {G.}~\bibnamefont {Merlo}}, \bibinfo {author} {\bibfnamefont
  {T.}~\bibnamefont {G\"orler}}, \bibinfo {author} {\bibfnamefont
  {A.}~\bibnamefont {{Ba\~n\'on Navarro}}}, \bibinfo {author} {\bibfnamefont
  {D.}~\bibnamefont {Told}}, \bibinfo {author} {\bibfnamefont {M.}~\bibnamefont
  {Maurer}}, \ and\ \bibinfo {author} {\bibfnamefont {F.}~\bibnamefont
  {Jenko}},\ }\href {\doibase https://doi.org/10.1016/j.cpc.2020.107360}
  {\bibfield  {journal} {\bibinfo  {journal} {Computer Physics Communications}\
  ,\ \bibinfo {pages} {107360}} (\bibinfo {year} {2020})}\BibitemShut {NoStop}%
\bibitem [{\citenamefont {Jarema}\ \emph {et~al.}(2016)\citenamefont {Jarema},
  \citenamefont {Bungartz}, \citenamefont {G\"orler}, \citenamefont {Jenko},
  \citenamefont {Neckel},\ and\ \citenamefont {Told}}]{Jarema_CPC_2016}%
  \BibitemOpen
  \bibfield  {author} {\bibinfo {author} {\bibfnamefont {D.}~\bibnamefont
  {Jarema}}, \bibinfo {author} {\bibfnamefont {H.}~\bibnamefont {Bungartz}},
  \bibinfo {author} {\bibfnamefont {T.}~\bibnamefont {G\"orler}}, \bibinfo
  {author} {\bibfnamefont {F.}~\bibnamefont {Jenko}}, \bibinfo {author}
  {\bibfnamefont {T.}~\bibnamefont {Neckel}}, \ and\ \bibinfo {author}
  {\bibfnamefont {D.}~\bibnamefont {Told}},\ }\href {\doibase
  https://doi.org/10.1016/j.cpc.2015.09.007} {\bibfield  {journal} {\bibinfo
  {journal} {Computer Physics Communications}\ }\textbf {\bibinfo {volume}
  {198}},\ \bibinfo {pages} {105 } (\bibinfo {year} {2016})}\BibitemShut
  {NoStop}%
\bibitem [{\citenamefont {G\"orler}\ \emph {et~al.}(2011)\citenamefont
  {G\"orler}, \citenamefont {Lapillonne}, \citenamefont {Brunner},
  \citenamefont {Dannert}, \citenamefont {Jenko}, \citenamefont {Aghdam},
  \citenamefont {Marcus}, \citenamefont {McMillan}, \citenamefont {Merz},
  \citenamefont {Sauter}, \citenamefont {Told},\ and\ \citenamefont
  {Villard}}]{Goerler_PoP_2011}%
  \BibitemOpen
  \bibfield  {author} {\bibinfo {author} {\bibfnamefont {T.}~\bibnamefont
  {G\"orler}}, \bibinfo {author} {\bibfnamefont {X.}~\bibnamefont
  {Lapillonne}}, \bibinfo {author} {\bibfnamefont {S.}~\bibnamefont {Brunner}},
  \bibinfo {author} {\bibfnamefont {T.}~\bibnamefont {Dannert}}, \bibinfo
  {author} {\bibfnamefont {F.}~\bibnamefont {Jenko}}, \bibinfo {author}
  {\bibfnamefont {S.~K.}\ \bibnamefont {Aghdam}}, \bibinfo {author}
  {\bibfnamefont {P.}~\bibnamefont {Marcus}}, \bibinfo {author} {\bibfnamefont
  {B.~F.}\ \bibnamefont {McMillan}}, \bibinfo {author} {\bibfnamefont
  {F.}~\bibnamefont {Merz}}, \bibinfo {author} {\bibfnamefont {O.}~\bibnamefont
  {Sauter}}, \bibinfo {author} {\bibfnamefont {D.}~\bibnamefont {Told}}, \ and\
  \bibinfo {author} {\bibfnamefont {L.}~\bibnamefont {Villard}},\ }\href
  {\doibase 10.1063/1.3567484} {\bibfield  {journal} {\bibinfo  {journal}
  {Physics of Plasmas}\ }\textbf {\bibinfo {volume} {18}},\ \bibinfo {pages}
  {056103} (\bibinfo {year} {2011})}\BibitemShut {NoStop}%
\bibitem [{\citenamefont {Mc~Carthy}(1999)}]{Carthy_PoP_1999}%
  \BibitemOpen
  \bibfield  {author} {\bibinfo {author} {\bibfnamefont {P.~J.}\ \bibnamefont
  {Mc~Carthy}},\ }\href {\doibase 10.1063/1.873630} {\bibfield  {journal}
  {\bibinfo  {journal} {Physics of Plasmas}\ }\textbf {\bibinfo {volume} {6}},\
  \bibinfo {pages} {3554} (\bibinfo {year} {1999})}\BibitemShut {NoStop}%
\bibitem [{\citenamefont {Xanthopoulos}\ \emph {et~al.}(2009)\citenamefont
  {Xanthopoulos}, \citenamefont {Cooper}, \citenamefont {Jenko}, \citenamefont
  {Turkin}, \citenamefont {Runov},\ and\ \citenamefont
  {Geiger}}]{Xanthopoulos_PoP_2009}%
  \BibitemOpen
  \bibfield  {author} {\bibinfo {author} {\bibfnamefont {P.}~\bibnamefont
  {Xanthopoulos}}, \bibinfo {author} {\bibfnamefont {W.~A.}\ \bibnamefont
  {Cooper}}, \bibinfo {author} {\bibfnamefont {F.}~\bibnamefont {Jenko}},
  \bibinfo {author} {\bibfnamefont {Y.}~\bibnamefont {Turkin}}, \bibinfo
  {author} {\bibfnamefont {A.}~\bibnamefont {Runov}}, \ and\ \bibinfo {author}
  {\bibfnamefont {J.}~\bibnamefont {Geiger}},\ }\href {\doibase
  10.1063/1.3187907} {\bibfield  {journal} {\bibinfo  {journal} {Physics of
  Plasmas}\ }\textbf {\bibinfo {volume} {16}},\ \bibinfo {pages} {082303}
  (\bibinfo {year} {2009})}\BibitemShut {NoStop}%
\bibitem [{\citenamefont {Brambilla}\ and\ \citenamefont
  {Bilato}(2009)}]{Brambilla_NF2009}%
  \BibitemOpen
  \bibfield  {author} {\bibinfo {author} {\bibfnamefont {M.}~\bibnamefont
  {Brambilla}}\ and\ \bibinfo {author} {\bibfnamefont {R.}~\bibnamefont
  {Bilato}},\ }\href {http://stacks.iop.org/0029-5515/49/i=8/a=085004}
  {\bibfield  {journal} {\bibinfo  {journal} {Nucl.~Fusion}\ }\textbf {\bibinfo
  {volume} {49}},\ \bibinfo {pages} {085004} (\bibinfo {year}
  {2009})}\BibitemShut {NoStop}%
\bibitem [{\citenamefont {Bilato}\ \emph {et~al.}(2011)\citenamefont {Bilato},
  \citenamefont {Brambilla}, \citenamefont {Maj}, \citenamefont {Horton},
  \citenamefont {Maggi},\ and\ \citenamefont {Stober}}]{Bilato_2011}%
  \BibitemOpen
  \bibfield  {author} {\bibinfo {author} {\bibfnamefont {R.}~\bibnamefont
  {Bilato}}, \bibinfo {author} {\bibfnamefont {M.}~\bibnamefont {Brambilla}},
  \bibinfo {author} {\bibfnamefont {O.}~\bibnamefont {Maj}}, \bibinfo {author}
  {\bibfnamefont {L.}~\bibnamefont {Horton}}, \bibinfo {author} {\bibfnamefont
  {C.}~\bibnamefont {Maggi}}, \ and\ \bibinfo {author} {\bibfnamefont
  {J.}~\bibnamefont {Stober}},\ }\href {\doibase
  10.1088/0029-5515/51/10/103034} {\bibfield  {journal} {\bibinfo  {journal}
  {Nuclear Fusion}\ }\textbf {\bibinfo {volume} {51}},\ \bibinfo {pages}
  {103034} (\bibinfo {year} {2011})}\BibitemShut {NoStop}%
\bibitem [{\citenamefont {{Citrin}}\ \emph {et~al.}(2013)\citenamefont
  {{Citrin}}, \citenamefont {{Jenko}}, \citenamefont {{Mantica}}, \citenamefont
  {{Told}}, \citenamefont {{Bourdelle}}, \citenamefont {{Garcia}},
  \citenamefont {{Haverkort}}, \citenamefont {{Hogeweij}}, \citenamefont
  {{Johnson}},\ and\ \citenamefont {{Pueschel}}}]{Citrin_PRL2013}%
  \BibitemOpen
  \bibfield  {author} {\bibinfo {author} {\bibfnamefont {J.}~\bibnamefont
  {{Citrin}}}, \bibinfo {author} {\bibfnamefont {F.}~\bibnamefont {{Jenko}}},
  \bibinfo {author} {\bibfnamefont {P.}~\bibnamefont {{Mantica}}}, \bibinfo
  {author} {\bibfnamefont {D.}~\bibnamefont {{Told}}}, \bibinfo {author}
  {\bibfnamefont {C.}~\bibnamefont {{Bourdelle}}}, \bibinfo {author}
  {\bibfnamefont {J.}~\bibnamefont {{Garcia}}}, \bibinfo {author}
  {\bibfnamefont {J.~W.}\ \bibnamefont {{Haverkort}}}, \bibinfo {author}
  {\bibfnamefont {G.~M.~D.}\ \bibnamefont {{Hogeweij}}}, \bibinfo {author}
  {\bibfnamefont {T.}~\bibnamefont {{Johnson}}}, \ and\ \bibinfo {author}
  {\bibfnamefont {M.~J.}\ \bibnamefont {{Pueschel}}},\ }\href {\doibase
  10.1103/PhysRevLett.111.155001} {\bibfield  {journal} {\bibinfo  {journal}
  {Phys.~Rev.~Lett.}\ }\textbf {\bibinfo {volume} {111}},\ \bibinfo {eid}
  {155001} (\bibinfo {year} {2013})}\BibitemShut {NoStop}%
\bibitem [{\citenamefont {Siena}\ \emph {et~al.}(2019)\citenamefont {Siena},
  \citenamefont {G{\"{o}}rler}, \citenamefont {Poli}, \citenamefont {{Ba\~n\'on
  Navarro}}, \citenamefont {Biancalani},\ and\ \citenamefont
  {Jenko}}]{DiSiena_NF_2019}%
  \BibitemOpen
  \bibfield  {author} {\bibinfo {author} {\bibfnamefont {A.~D.}\ \bibnamefont
  {Siena}}, \bibinfo {author} {\bibfnamefont {T.}~\bibnamefont {G{\"{o}}rler}},
  \bibinfo {author} {\bibfnamefont {E.}~\bibnamefont {Poli}}, \bibinfo {author}
  {\bibfnamefont {A.}~\bibnamefont {{Ba\~n\'on Navarro}}}, \bibinfo {author}
  {\bibfnamefont {A.}~\bibnamefont {Biancalani}}, \ and\ \bibinfo {author}
  {\bibfnamefont {F.}~\bibnamefont {Jenko}},\ }\href {\doibase
  10.1088/1741-4326/ab4088} {\bibfield  {journal} {\bibinfo  {journal} {Nuclear
  Fusion}\ }\textbf {\bibinfo {volume} {59}},\ \bibinfo {pages} {124001}
  (\bibinfo {year} {2019})}\BibitemShut {NoStop}%
\bibitem [{\citenamefont {Di~Siena}\ \emph
  {et~al.}(2021{\natexlab{b}})\citenamefont {Di~Siena}, \citenamefont
  {G\"orler}, \citenamefont {Poli}, \citenamefont {Ba\~n\'on Navarro},
  \citenamefont {Biancalani}, \citenamefont {Bilato}, \citenamefont {Bonanomi},
  \citenamefont {Novikau}, \citenamefont {Vannini},\ and\ \citenamefont
  {Jenko}}]{DiSiena_JPP_2021}%
  \BibitemOpen
  \bibfield  {author} {\bibinfo {author} {\bibfnamefont {A.}~\bibnamefont
  {Di~Siena}}, \bibinfo {author} {\bibfnamefont {T.}~\bibnamefont {G\"orler}},
  \bibinfo {author} {\bibfnamefont {E.}~\bibnamefont {Poli}}, \bibinfo {author}
  {\bibfnamefont {A.}~\bibnamefont {Ba\~n\'on Navarro}}, \bibinfo {author}
  {\bibfnamefont {A.}~\bibnamefont {Biancalani}}, \bibinfo {author}
  {\bibfnamefont {R.}~\bibnamefont {Bilato}}, \bibinfo {author} {\bibfnamefont
  {N.}~\bibnamefont {Bonanomi}}, \bibinfo {author} {\bibfnamefont
  {I.}~\bibnamefont {Novikau}}, \bibinfo {author} {\bibfnamefont
  {F.}~\bibnamefont {Vannini}}, \ and\ \bibinfo {author} {\bibfnamefont
  {F.}~\bibnamefont {Jenko}},\ }\href {\doibase 10.1017/S0022377821000362}
  {\bibfield  {journal} {\bibinfo  {journal} {Journal of Plasma Physics}\
  }\textbf {\bibinfo {volume} {87}},\ \bibinfo {pages} {555870201} (\bibinfo
  {year} {2021}{\natexlab{b}})}\BibitemShut {NoStop}%
\bibitem [{\citenamefont {{Di Siena}}(2019)}]{Disiena_Phd}%
  \BibitemOpen
  \bibfield  {author} {\bibinfo {author} {\bibfnamefont {A.}~\bibnamefont {{Di
  Siena}}},\ }\href {\doibase 10.18725/OPARU-29528} {\bibfield  {journal}
  {\bibinfo  {journal} {PhD thesis, ULM University, Germany}\ } (\bibinfo
  {year} {2019}),\ 10.18725/OPARU-29528}\BibitemShut {NoStop}%
\bibitem [{\citenamefont {Chen}\ and\ \citenamefont
  {Zonca}(2016)}]{Chen_Zonca}%
  \BibitemOpen
  \bibfield  {author} {\bibinfo {author} {\bibfnamefont {L.}~\bibnamefont
  {Chen}}\ and\ \bibinfo {author} {\bibfnamefont {F.}~\bibnamefont {Zonca}},\
  }\href {\doibase 10.1103/RevModPhys.88.015008} {\bibfield  {journal}
  {\bibinfo  {journal} {Rev. Mod. Phys.}\ }\textbf {\bibinfo {volume} {88}},\
  \bibinfo {pages} {015008} (\bibinfo {year} {2016})}\BibitemShut {NoStop}%
\bibitem [{\citenamefont {{Citrin}}\ \emph {et~al.}(2015)\citenamefont
  {{Citrin}}, \citenamefont {{Garcia}}, \citenamefont {{G{\"o}rler}},
  \citenamefont {{Jenko}}, \citenamefont {{Mantica}}, \citenamefont {{Told}},
  \citenamefont {{Bourdelle}}, \citenamefont {{Hatch}}, \citenamefont
  {{Hogeweij}}, \citenamefont {{Johnson}}, \citenamefont {{Pueschel}},\ and\
  \citenamefont {{Schneider}}}]{Citrin_PPCF2015}%
  \BibitemOpen
  \bibfield  {author} {\bibinfo {author} {\bibfnamefont {J.}~\bibnamefont
  {{Citrin}}}, \bibinfo {author} {\bibfnamefont {J.}~\bibnamefont {{Garcia}}},
  \bibinfo {author} {\bibfnamefont {T.}~\bibnamefont {{G{\"o}rler}}}, \bibinfo
  {author} {\bibfnamefont {F.}~\bibnamefont {{Jenko}}}, \bibinfo {author}
  {\bibfnamefont {P.}~\bibnamefont {{Mantica}}}, \bibinfo {author}
  {\bibfnamefont {D.}~\bibnamefont {{Told}}}, \bibinfo {author} {\bibfnamefont
  {C.}~\bibnamefont {{Bourdelle}}}, \bibinfo {author} {\bibfnamefont {D.~R.}\
  \bibnamefont {{Hatch}}}, \bibinfo {author} {\bibfnamefont {G.~M.~D.}\
  \bibnamefont {{Hogeweij}}}, \bibinfo {author} {\bibfnamefont
  {T.}~\bibnamefont {{Johnson}}}, \bibinfo {author} {\bibfnamefont {M.~J.}\
  \bibnamefont {{Pueschel}}}, \ and\ \bibinfo {author} {\bibfnamefont
  {M.}~\bibnamefont {{Schneider}}},\ }\href {\doibase
  10.1088/0741-3335/57/1/014032} {\bibfield  {journal} {\bibinfo  {journal}
  {Plasma Phys.~Controlled Fusion}\ }\textbf {\bibinfo {volume} {57}},\
  \bibinfo {eid} {014032} (\bibinfo {year} {2015})}\BibitemShut {NoStop}%
\bibitem [{\citenamefont {Biancalani}\ \emph {et~al.}(2021)\citenamefont
  {Biancalani}, \citenamefont {Bottino}, \citenamefont {Siena}, \citenamefont
  {G\"urcan}, \citenamefont {Hayward-Schneider}, \citenamefont {Jenko},
  \citenamefont {Lauber}, \citenamefont {Mishchenko}, \citenamefont {Morel},
  \citenamefont {Novikau}, \citenamefont {Vannini}, \citenamefont {Villard},\
  and\ \citenamefont {Zocco}}]{Biancalani_PPCF_2021}%
  \BibitemOpen
  \bibfield  {author} {\bibinfo {author} {\bibfnamefont {A.}~\bibnamefont
  {Biancalani}}, \bibinfo {author} {\bibfnamefont {A.}~\bibnamefont {Bottino}},
  \bibinfo {author} {\bibfnamefont {A.~D.}\ \bibnamefont {Siena}}, \bibinfo
  {author} {\bibfnamefont {O.}~\bibnamefont {G\"urcan}}, \bibinfo {author}
  {\bibfnamefont {T.}~\bibnamefont {Hayward-Schneider}}, \bibinfo {author}
  {\bibfnamefont {F.}~\bibnamefont {Jenko}}, \bibinfo {author} {\bibfnamefont
  {P.}~\bibnamefont {Lauber}}, \bibinfo {author} {\bibfnamefont
  {A.}~\bibnamefont {Mishchenko}}, \bibinfo {author} {\bibfnamefont
  {P.}~\bibnamefont {Morel}}, \bibinfo {author} {\bibfnamefont
  {I.}~\bibnamefont {Novikau}}, \bibinfo {author} {\bibfnamefont
  {F.}~\bibnamefont {Vannini}}, \bibinfo {author} {\bibfnamefont
  {L.}~\bibnamefont {Villard}}, \ and\ \bibinfo {author} {\bibfnamefont
  {A.}~\bibnamefont {Zocco}},\ }\href {\doibase 10.1088/1361-6587/abf256}
  {\bibfield  {journal} {\bibinfo  {journal} {Plasma Physics and Controlled
  Fusion}\ }\textbf {\bibinfo {volume} {63}},\ \bibinfo {pages} {065009}
  (\bibinfo {year} {2021})}\BibitemShut {NoStop}%
\bibitem [{\citenamefont {Parker}\ \emph {et~al.}(2018)\citenamefont {Parker},
  \citenamefont {LoDestro}, \citenamefont {Told}, \citenamefont {Merlo},
  \citenamefont {Ricketson}, \citenamefont {Campos}, \citenamefont {Jenko},\
  and\ \citenamefont {Hittinger}}]{Parker_2018}%
  \BibitemOpen
  \bibfield  {author} {\bibinfo {author} {\bibfnamefont {J.~B.}\ \bibnamefont
  {Parker}}, \bibinfo {author} {\bibfnamefont {L.~L.}\ \bibnamefont
  {LoDestro}}, \bibinfo {author} {\bibfnamefont {D.}~\bibnamefont {Told}},
  \bibinfo {author} {\bibfnamefont {G.}~\bibnamefont {Merlo}}, \bibinfo
  {author} {\bibfnamefont {L.~F.}\ \bibnamefont {Ricketson}}, \bibinfo {author}
  {\bibfnamefont {A.}~\bibnamefont {Campos}}, \bibinfo {author} {\bibfnamefont
  {F.}~\bibnamefont {Jenko}}, \ and\ \bibinfo {author} {\bibfnamefont {J.~A.}\
  \bibnamefont {Hittinger}},\ }\href {\doibase 10.1088/1741-4326/aab5c8}
  {\bibfield  {journal} {\bibinfo  {journal} {Nuclear Fusion}\ }\textbf
  {\bibinfo {volume} {58}},\ \bibinfo {pages} {054004} (\bibinfo {year}
  {2018})}\BibitemShut {NoStop}%
\end{thebibliography}%
\end{document}